\documentclass{sig-alternate}
\usepackage{graphicx}
\usepackage{amsmath}
\usepackage{algorithm}
\usepackage{algorithmic}
\usepackage{defpaper2e}
\usepackage{subcaption}
\usepackage[justification=centering]{caption}
\usepackage{multirow}
\usepackage{epstopdf}
\usepackage{hyperref}

\title{Efficient Authentication of Outsourced String Similarity Search}

\author{
\alignauthor
Boxiang Dong, Wendy Wang\\
       \affaddr{Department of Computer Science}\\
       \affaddr{Stevens Institute of Technology}\\
       \affaddr{Hoboken, NJ}\\
       \email{bdong, hwang4@stevens.edu}
}

\begin{document}
\maketitle

\begin{abstract}
%outsourcing
Cloud computing enables the outsourcing of big data analytics, where a third-party server is responsible for data storage and processing. 
In this paper, we consider the outsourcing model that provides string similarity search as the service. In particular, given a similarity search query, the service provider returns all strings from the outsourced dataset that are similar to the query string.
A major security concern of the outsourcing paradigm is to authenticate whether the service provider returns {\em sound} and {\em complete} search results. 
In this paper, we design $AutoS^3$, an authentication mechanism of outsourced string similarity search. The key idea of $AutoS^3$ is that the server returns a {\em verification object} ($VO$) to prove the result correctness.
First, we design an authenticated string indexing structure named $MB$-tree for $VO$ construction. Second, we design two lightweight authentication methods named $VS^2$ and E-$VS^2$ that can catch the service provider's various cheating behaviors with cheap verification cost.  Moreover, we generalize our solution for top-k string similarity search. 
We perform an extensive set of experiment results on real-world datasets to demonstrate the efficiency of our approach.
\end{abstract}

\section{Introduction}
%cloud computing, big data

%With the rapid growth of Internet and computer technology, we have witnessed the birth of Big data era. 
Big data analytics offers the promise of providing valuable insights. 
%For example, in a single month, the US Library of Congress has collected 525 terabytes of web archived data. {\bf CITATION?}
However, many companies, especially the small- and medium-sized organizations lack the computational resources, in-house knowledge and experience of big data analytics.
A practical solution to this dilemma is outsourcing, where the data owner outsources the data to a computational powerful third-party service provider (e.g., the cloud) for cost-effective solutions of data storage, processing, and analysis. 

In this paper, we consider {\em string similarity search}, an important data analytics operation that have been used in a broad range of applications, as the outsourced computations. 
Generally speaking, the data owner outsources a string database $D$ to a third-party service provider (server). The server provides the storage and processing of similarity search queries as services. The search queries ask for the strings in $D$ that are similar to a number of given strings, where the similarity is measured by a specific similarity function and a user-defined threshold. 

For all the benefits of outsourcing and cloud computing, though, the outsourcing paradigm deprives the data owner  of direct control over her data. This poses numerous security challenges. One of the challenges is that the server may cheat on the similarity search results. 
 For example, the server is incentivized to improve its revenue by computing with less resources (e.g., only search a portion of $D$) while charging for more. 
Therefore, it is important to authenticate whether the service provider has performed the search faithfully, and returned the correct results to the client. A naive method is to  execute the search queries locally, and compare the results with the outcome from the server. Apparently this method is prohibitively costly.
We aim to design efficient methods that enable the client to authenticate that the server returned {\em sound} and {\em complete} similar strings. By soundness we mean that the returned strings are indeed similar. By completeness we mean that all similar strings are returned. In this paper, we focus on {\em edit distance}, a commonly-used string similarity function. 

Most existing work (e.g. \cite{cheng2009query,hu2013spatial}) solve the authentication 
problem for spatial queries in the Euclidean space. To our best knowledge, ours is the first to consider the authentication of outsourced string similarity search. Intuitively, the strings can be mapped to the Euclidean space via a similarity-preserving embedding function (e.g. \cite{faloutsos1995fastmap,hjaltason2003properties}). However, such embedding functions cannot guarantee 100\% precision (i.e., the embedded points of some dissimilar strings become similar in the Euclidean space). This disables the direct use of the existing Euclidean distance based authentication approaches on string similarity queries.

%contribution
In this paper, we design $AutoS^3$, an  \underline{Aut}hentication mechanism of  \underline{O}utsourced \underline{S}tring  \underline{S}imilarity \underline{S}earch. 
The key idea of $AutoS^3$ is that besides returning the similar strings, the server returns a verification object ($VO$) that can prove the soundness and completeness of returned strings. In particular, we make the following contributions.

First, we design an authentication tree structure named {\em MB-tree}. MB-tree is constructed by integrating Merkle hash tree \cite{merkle-1980}, a popularly-used authenticated data structure, with $B^{ed}$-tree \cite{zhang2010bed}, a compact index for efficient string similarity search based on edit distance. 

Second, we design the basic verification method named $VS^2$ for the search queries that consist of a single query string. 
$VS^2$ constructs $VO$ from the MB-tree, requiring to include {\em false hits} into $VO$, where false hits refer to the strings that are not returned in the result, but are necessary for the result authentication. 
We prove that $VS^2$ is able to catch the server's cheating behaviors such as tampered values, soundness violation, and completeness violation.

A large amounts of false hits can impose a significant burden to the  client for verification. Therefore, our third contribution is the design of the {\em E-$VS^2$} algorithm that reduces the VO verification cost at the client side. E-$VS^2$ applies a similarity-preserving embedding function to map strings to the Euclidean space in the way that similar strings are mapped to close Euclidean points. Then $VO$ is constructed from both the $MB$-tree and the embedded Euclidean space.
Compared with $VS^2$, E-$VS^2$ dramatically saves the verification cost by replacing a large amounts of expensive string edit distance calculation with a small number of cheap Euclidean distance computation.
%We prove that E-$VS^2$ can catch incorrect and incomplete search results. 

Fourth, we extend to the authentication of: (1) similarity search queries that consists of multiple query strings,  and (2) top-k similarity search. We design efficient optimization methods that reduce verification cost for both cases.

Last but not least, we complement the theoretical investigation with a rich set of experiment study on real datasets. The experiment results demonstrate the efficiency of our approaches. It shows that E-$VS^2$ can save 25\% verification cost of the $VS^2$ approach.  

The rest of the paper is organized as follows. Sections  \ref{sc:related} and \ref{sc:pre} discuss the related work and preliminaries. Section \ref{sc:problem}  formally defines the problem. Section \ref{sc:single}  presents our $VS^2$ and E-$VS^2$ approaches for single-string search queries. 
Section \ref{sc:mstring} discusses the authentication of multi-string search queries.  
Section \ref{sc:ranking} extends to top-k similarity search. 
The experiment results are shown in Section \ref{sc:exp}. Section \ref{sc:concl} concludes the paper.

\section{Related Work}
\label{sc:related}
The problem of authentication of outsourced computations caught much attention from the research community in recent years. Based on the type of the outsourced data and the type of queries that are executed on the data, we classify these techniques into the following types: (1) authentication of outsourced SQL query evaluation, (2) authentication of keyword search, and (3) authentication of outsourced spatial query evaluation. None of these work considered string similarity search queries. 
%To best of our knowledge, we are the first to consider the problem of authenticating outsourced string similarity search queries. 

\noindent{\bf Authentication of outsourced SQL query evaluation.} 
The issue of providing authenticity for outsourced database was initially raised in the database-as-a-service ($DaS$) paradigm \cite{Hacigumus-sigmod02}. The aim is to assure the correctness of SQL query evaluation over the outsourced databases. 
The proposed solutions include Merkle hash trees \cite{li-sigmod06,Mykletun-2006}, signatures on a chain of paired tuples~\cite{Pang:2009:SVO:1687627.1687718}, and authenticated B-tree and R-tree structures for aggregated queries \cite{Li:2010:AIS:1880022.1880026}. 
The key idea of these techniques is that the data owner outsources not only data but also the endorsements of the data being outsourced. These endorsements are signed by the data owner against tampering with by the service provider. For the cleaning results, the service provider returns both the results and a proof, called the {\em verification object} ($VO$), which is an auxiliary data structure to store the processing traces such as index traversals.
%In the verification phase, 
The client uses the $VO$, together with the answers, to reconstruct the
endorsements and thus verify the authenticity of the results. 
An efficient authentication technique should minimize the size of $VO$, while requiring lightweight authentication at the client side.
%A possible solution of $VO$ construction is to construct an index structure that uses digital signatures at the granularity of individual tuples, which can be used for the verification of result authentication. 
In this paper, we follow the same VO-based strategy to design our authentication method. 
%Its main disadvantage is the relatively high overhead associated with building, storing and updating the index structure. To address this disadvantage, other techniques such as using challenge tokens~\cite{sion-vldb05} and adding  counterfeit records for SQL queries \cite{integrity-vldb07} are used. 

\noindent{\bf Authentication of keyword search.} 
Pang et al. \cite{pang2008authenticating} 
 targeted at search engines that perform similarity-based document retrieval, and designed a novel authentication mechanism for the search results. The key idea of the authentication is to build the Merkle hash tree (MHT) on the inverted index, and use the MHT for $VO$ construction. Though effective, it has several limitations, e.g., the MHT cannot deal data updates efficiently \cite{goodrich2012efficient}. To address these limitations, Goodrich et al. \cite{goodrich2012efficient} designed a new model that considered conjunctive keyword searches as equivalent with a set intersection on the underlying inverted index data structure. They use the authenticated data structure in \cite{papamanthou2011optimal} to  verify the correctness of set operations.
 %\cite{sun2014verifiable} considered cosine similarity measurement, and design encrypted data search functionality that support multi-keyword queries, result ranking and result verification. Its  search index is built based on the vector space model.  

 \noindent{\bf Authentication of outsourced spatial query evaluation.}
%State-of-the-art location-based services (LBSs) involve data owners, requesting clients, and service providers. The issue of how to verify the genuineness of service results has attracted much attention. 
A number of work investigated the problem of authentication of spatial query evaluation in the location-based service model. 
%The main difference between these work is the underlying authenticated data structure. 
Yang et al. \cite{yang2008spatial,yang2009authenticated} integrated an R-tree with the MHT (which is called Merkle R-tree or MR-tree) for authenticating multi-dimensional range queries.  
Yiu et al. \cite{yiu2011authentication} focused on the moving KNN queries that continuously reports the k nearest neighbors of a moving query point. They designed 
%two authentication methods, namely the vertex-based approach (VA) that that exploits the relationship between the vertices of an order-k Voronoi cell and the  materialization approach (MA) that builds 
the Voronoi MR-tree as the authenticated data  structure for $VO$ construction and authentication. 
Hu et al. \cite{hu2013spatial} also utilized neighborhood information derived from the Voronoi diagram of the underlying spatial dataset for authentication. 
%Their approach can deal with basic KNN queries and advanced spatial queries, such as reverse kNNs, k aggregate NNs and spatial skylines. 
Wu et al. \cite{wu2015authentication} 
%considers the authentication of moving top-k spatial keyword (MkSK) queries, which return the top-k spatial web objects that best match a query with respect to location and text relevance. In general, MkSK queries can be evaluted efficiently by leveraging IR-tree \cite{cong2009efficient}, in which each entry summarizes the spatial distances and text relevancies of the entries in its child node.
 designed a novel authenticated data structure named {\em Merkle-IR-tree} (MIR-tree) for moving top-k spatial keyword (MkSK) queries. MIR-tree builds a series of digests in each node of the IR-tree \cite{cong2009efficient}, in which each entry summarizes the spatial distances and text relevance of the entries in its child node. 
%Chen et al. \cite{chen2013authenticating} integrated data privacy with authentication into a unified framework. They designed the authentication methods based on MR-trees and Power Voronoi Diagram \cite{okabe2009spatial}  to enable the client to verify the soundness of privacy-preserving top-k spatial queries. 

\section{Preliminaries}
\label{sc:pre}
%edit distance computation

\noindent{\bf String similarity function.}
String similarity search is a fundamental problem in many research areas, e.g., information retrieval, database joins, and more.  
 In the literature, there are a number of string similarity functions, e.g., Hamming distance, n-grams, and edit distance (See \cite{koudas2006record} for a good tutorial.)
 In this paper, we mainly consider {\em edit distance}, one of the most popular string similarity measurement that has been used in a wide spectrum of applications. Informally, the  edit distance of two string values $s_1$ and $s_2$, denoted as $DST(s_1, s_2)$, measures the minimum number of insertion, deletion and substitution operations to transform $s_1$ to $s_2$. We say two strings $s_1$ and $s_2$ are {\em $\theta$-similar}, denoted as $s_1\approx s_2$, if $DST(s_1, s_2)\leq\theta$, where $\theta$ is a user-specified similarity threshold. Otherwise, we say $s_1$ and $s_2$ are {\em $\theta$-dissimilar} (denoted as $s_1\not\approx s_2$). 
 \nop{One efficient way to compute the edit distance is dynamic programming. Let $s[i]$ denote the $i$-th character of string $s$, then the edit distance between $s_1$ and $s_2$ is $DST(s_1, s_2)=d_{\ell_1, \ell_2}$, where $\ell_1$ ($\ell_2$) is the length of $s_1$ ($s_2$), and $d_{i,j}$ is defined in Equation \ref{eq:edit}. 
%The space and time complexity is $O(\ell_1 \ell_2)$.  
\begin{equation}
d_{i,j} = \left\{
	\begin{array}{lr}
		0 & : i=0\\
		0 & : j=0\\
		d_{i-1, j-1} & : s_1[i]=s_2[j]\\
		d_{i-1, j-1}+1 & : s_1[i]\neq s_2[j]
	\end{array}
	\right.
\label{eq:edit}
\end{equation}
}

%For any given string $s$, its $\theta$-similar strings can be found by evaluating a range query with a radius of $\theta$ with respect to $s$. 

%embedding
\label{sc:strmapping}
\noindent{\bf Mapping Strings to Euclidean Space.} Given two strings $s_1$ and $s_2$, normally the complexity of computing edit distance is $O(|s_1||s_2|)$, where $|s_1|$ and $|s_2|$ are the lengths of $s_1$ and $s_2$. 
%More efficient algorithms can achieve space complexity $O(max{|s_1|, |s_2|})$ \cite{cormen2001introduction}. 
One way to reduce the complexity of similarity measurement  is to map the strings into a multi-dimensional Euclidean space, such that the similar strings are mapped to close Euclidean points. 
The main reason of the embedding is that the computation of Euclidean distance is much cheaper than string edit distance. 
A few string embedding techniques (e.g.,  \cite{faloutsos1995fastmap,jin2003efficient,hjaltason2003properties}) exist in the literature. 
These algorithms have different properties in terms of their efficiency and distortion rate (See \cite{hjaltasoncontractive} for a good survey).  
In this paper, we consider an important property named {\em contractiveness} property of the embedding methods, which requires that for any pair of strings $(s_i, s_j)$ and their embedded Euclidean points $(p_i, p_j)$, 
$dst(p_i, p_j) \leq DST(s_i, s_j)$, 
where $dst()$ and $DST()$ are the distance function in the Euclidean space and string space respectively. In this paper, we use $dst()$ and $DST()$ to denote the Euclidean distance and edit distance. 
%The contractiveness property is important as for any two embedded points $p_i$ and $p_j$ that are $\theta$-dissimilar (i.e. $dst(p_i, p_j) > \theta$), their original strings $s_i$ and $s_j$ must be  $\theta$-dissimilar. 
In this paper, we use the {\em SparseMap} method \cite{hjaltason2003properties} for the string embedding. SparseMap preserves the contractiveness property. 
We will show how to leverage the contractiveness property to improve the performance of verification in Section \ref{sc:cvss}. 
%Thus, for any similarity query w.r.t. a string $s$, we can safely prune all strings $s'$ from the search for which $dst(p, p')>\theta$, where $p$ and $p'$ are the embedded points of $s$ and $s'$. 
%The time complexity of {\em SparseMap} embedding is $O(\sigma kn)$, where $k=[\log_2 n]^2$, $\sigma$ is a constant value, and $n$ is the number of strings. 
Note that the embedding methods may introduce false positives, i.e. the embedding points of dissimilar strings may become close in the Euclidean space. 
%we need to provide a new threshold value $\theta '$ to determine if two points are similar. 

%\noindent{\bf Euclidean distance VS. Edit distance} Our experiment results show that the Euclidean distance calculation is 10 times faster than edit distance computation.

\noindent{\bf Authenticated data structure.}
To enable the client to verify the correctness of the query results, the server returns the results along with some supplementary information that permits result verification. 
Normally the supplementary information takes the format of {\em verification
objects} ($VOs$). 
In the literature, VO generation is usually performed by an authenticated data structure (e.g., \cite{li2006dynamic,yang2009authenticated,papadopoulos2011authenticated}). One of the popular authenticated data structures is {\em Merkle tree} \cite{merkle-1980}. 
%Merkle tree is an authenticated data structure that allows data integrity verification. 
%In particular, given a Merkel tree $T$, each leaf $l$ of $T$ is assigned a digest of a record, calculated by using a one-way, collision-resistant hash function (e.g. SHA-1 \cite{menezes1996handbook}).
In particular, a Merkle tree is a tree $T$ in which each leaf node $N$ stores the digest of a record $r$: $h_{N} = h(r)$, where $h()$ is a one-way, collision-resistant hash function (e.g. SHA-1 \cite{menezes1996handbook}). 
%Each non-leaf node of $\cal T$ stores a digest, which is computed as the concatenation of the digests of its children. 
For each non-leaf node $N$ of $T$, it is assigned the value $h_{N} = h(h_{C_1}||\dots||h_{C_k})$, where $C_1, \dots, C_k$ are the children of $N$. 
%In a Merkle hash tree $\mathcal{T}$ built from $D$, the leaf node $\ell$ is assigned the value $h_\ell = hash(\ell||\mathcal{T}[\ell])$, while each internal node $v$ of children $a$ and $b$ is assigned with $h_v =hash(h_a||h_b)$. 
The root signature $sig$ is generated by signing the digest $h_{root}$ of the root node using the private key of a trusted party (e.g., the data owner). The VO enables the client to re-construct the root signature $sig$. 
%More details of the VO-based verification will be given in Section \ref{sc:single}.
%Otherwise, the client will conclude that the serve fails the verification. 
%{\bf NEED AN EXAMPLE OF THE TREE.}

%record linkage similarity
%\subsection{Record Linkage}
%Given a dataset containing $n$ records and $m$ attributes, record linkage aims to detect all record pairs $(r_i, r_j)$ such that $r_i$ is similar to $r_j$ ($r_i \sim r_j$). In this paper, we consider the similarity measurement based on edit distance. To evaluate the record pair-wise similarity across multiple attributes, there are two options. The first rule is that the client specifies $m$ threshold values $\theta_1, \dots, \theta_m$. $r_i \sim r_j$ if and only if $DST(r_i[k], r_j[k]) \leq \theta_k$ for all $k \in [1, m]$. The other policy is that the client only provides one similarity threshold value $\theta$. We concatenate all the attribute values into one string. Two records are similar if the edit distance of their concatenated strings is no larger than $\theta$. From now on, we focus the case that there is only one attribute. Our approach can be easily extended to the multiple attributes problem.

%Bed tree
\noindent{\bf $B^{ed}$-Tree for string similarity search.}
%Similarity search based on edit distance is not well supported in existing database systems due to the high dimensionality of the string domain.
%There are a number of work \cite{alborzi2007execution,arge2004priority,robinson1981kdb} that present compact data structures to speed up string similarity search. 
A number of compact data structures (e.g., \cite{arasu2006efficient,chaudhuri2003robust, li2008efficient}) are designed to handle edit distance based similarity measurement. In this paper, we consider 
{\em $B^{ed}$-tree} \cite{zhang2010bed} due to its simplicity and efficiency. $B^{ed}$-tree is a $B^+$-tree based index structure that can handle arbitrary edit distance thresholds. The tree is built upon a {\em string ordering} scheme which is a mapping function $\varphi$ to map each string to an integer value.  To simplify notation, we say that $s\in [s_i,s_j]$ if $\varphi(s_i)\leq\varphi(s)\leq\varphi(s_j)$. 
%The mapping $\varphi$ satisfies that for any given string $s$, $s\in [s_i, s_j]$ if $\varphi(s_i)\leq\varphi(s)\leq\varphi(s_j)$. 
Based on the string ordering,  each $B^{ed}$-tree node $N$ is associated with a string range $[N_{b}, N_{e}]$. Each leaf node contains $f$ strings $\{s_1, \dots, s_f\}$, where $s_i\in[N_b, N_e]$, for each $i\in[1, f]$. Each intermediate node contains  multiple children nodes, where for each child of $N$, its range $[N_b', N_e'] \subseteq [N_b, N_e]$, as $[N_b, N_e]$ is the range of $N$. We say these strings that  stored in the
sub-tree rooted at $N$ as the strings that are {\em covered} by $N$. 

We use $DST_{min} (s_q, N)$ to denote the minimal edit distance between a query string $s_q$ and any $s$ that is covered by a $B^{ed}$-tree node $N$.  
A nice property of the $B^{ed}$-tree is that, for any string $s_q$ and node $N$, the string ordering $\varphi$ enables to compute $DST_{min} (s_q, N)$ efficiently by computing $DST(s_q, N_b)$ and $DST(s_q, N_e)$ only, where $N_b, N_e$ refer to the string range values of $N$. Then:
\vspace{-0.05in}
\begin{definition}
\label{def:bedtree}
{\em
Given a query string $s_q$ and a similarity threshold $\theta$, we say a $B^{ed}$-tree node $N$ is a {\em candidate} if $DST_{min}(s_q, N)\leq\theta$.  Otherwise, $N$ is a {\em non-candidate}. \qed
}
\end{definition}
\vspace{-0.05in}
$B^{ed}$-tree has an important monotone property. 
Given a node $N_i$ in the $B^{ed}$-tree and any child node $N_j$ of $N_i$, for any query string $s_q$, it must be true that $DST_{min}(s_q, N_i)\leq DST_{min}(s_q, N_j)$. Therefore, for any non-candidate node, all of its children must be non-candidates. This monotone property enables early termination of search on the branches that contain non-candidate nodes.

For any given query string $s_q$, the string similarity search algorithm starts from the root of the $B^{ed}$-tree, and iteratively 
visits the candidate nodes, until all candidate  nodes are visited. The algorithm does not visit those non-candidate nodes as well as their descendants.
An important note is that for each candidate node, some of its covered strings may still be dissimilar to the query string. Therefore, given a query string $s_q$ and a candidate $B^{ed}$-tree node $N$ that is associated with a range $[N_b, N_e]$, it is necessary to compute $DST(s_q, s)$, for each $s\in[N_b, N_e]$.

\section{Problem Formulation}
\label{sc:problem}

In this section, we describe the authentication problem that we plan to study in this paper.

\noindent{\bf System Model.} 
We consider the outsourcing model that involves three parties - a data owner who possesses a dataset $D$ that contains $n$ string values, the user (client) who requests for the similarity search on $D$, and a third-party service provider (server) that executes the similarity search on $D$. The data owner outsources $D$ to the server. The server provides storage and similarity search as services. 
%To provide authenticated similarity searches, the data provider generates some auxiliary information of $D$, and sends $D$ together with the auxiliary information, as well as the similarity search software to the server.  {\bf WHY THE SEARCH SOFTWARE IS NEEDED TO BE OUTSOURCED?}
The client can be the data owner or a  trusted party.  
Given the fact that the client may not possess $D$, we require that the availability of $D$ is not necessary for the authentication procedure.

\noindent{\bf Similarity search queries.}
The server accepts similarity search queries from the clients. We consider the similarity search queries that take the format $(S, \theta)$, where $S$ is a set of query strings, and $\theta$ is the similarity threshold. 
We consider two types of similarity search search queries: 
(1) {\em Single-string similarity search}: the query $Q$ contains a single search string $s_q$ (i.e., $S=\{s_q\}$). The server returns $R$ that contains all similar strings of $s_q$ in $D$; and 
(2) {\em Multi-string similarity search}: $Q$ contains multiple unique search strings $S = \{s_1, \dots, s_\ell\}$. The server returns the search results in the format of $\{\{s_1$, $R_{s_1}\}$, $\dots$, $\{s_\ell$, $R_{s_\ell}\}\}$, where $R_{s_i} (1\leq i\leq \ell)$ is the set of similar strings of $s_i$ in $D$.

Note that the query strings may not necessarily exist in $D$. For each search string $s_q$, we consider two types of similarity outputs: (1) {\em Un-ranked results}: all the similar strings of $s_q$ are returned without any ranking; and (2) 
{\em Top-$k$ results}: the strings that are of the top-$k$ smallest distances to $s_q$ are returned by their distance to $s_q$ in an  ascending order. 
%Apparently when $k$ is sufficiently large, all similar strings of $s$ are ranked. 
We consider both un-ranked and top-k ranking cases in the paper. 
We assume that the client sends a large number of  similarity search queries to the server.

Our query model can be easily extended to support other types of string queries, e.g., range queries, KNN queries, and all-pairs join queries \cite{zhang2010bed} that find all similar string pairs in two given string datasets. 

\noindent{\bf Result correctness.} 
We define the result correctness for two different types of query outputs. 

{\bf Un-ranked search results.}
Given a query string $s_q$, the un-ranked search result $R$ of $s_q$ is {\em correct} if and only if it satisfies the following two conditions: (1) {\em soundness}: for any string $s'\in R$, $s'$ must reside in $D$, and $s_q\approx s'$; and (2) {\em completeness}: for any string $s'\in D$ such that $s_q\approx s'$, $s'$ must be included in $R$. In other words, for any string $s'\not\in R$, it must be true that $s_q\not\approx s'$. 

{\bf Top-k results.}
Given a query string $s_q$, let $R$ be the ranked search result, in which the strings are ranked by their distance to $s_q$ in an ascending order. We use $R[i]\ (1\leq i\leq k)$ to denote the $i$-th string of $R$. Then $R$ is {\em correct} if it satisfies the following two conditions: (1) {\em soundness}:  $\forall 1\leq i < j \leq k$, both $R[i$] and $R[j]$ must exist in $D$, and $DST(s_q, R[i]) \leq DST(s_q, R[j])$;
and (2) {\em completeness}: for any string $s' \not\in R$, it must be true that $DST(s_q, s') \geq DST(s_q, R[k])$.

\noindent{\bf Threat model.} 
In this paper, we assume that both the data owner and the client are fully trusted. However, the third-party server is not fully trusted as it could be compromised by the attacker (either inside or outside). The server 
may alter the received dataset $D$ and return any search result that does not exist in $D$. It also may
tamper with the search results. For instance, the server may return {\em incomplete} results that omit some legitimate documents in the similarity search results \cite{pang2008authenticating}, or alter the ranking orders of the top-k results. 
%Given the importance of the similarity search results, it is vital to provide efficient authentication mechanisms that enable the client to verify the correctness of the search results.
Note that we do not consider privacy protection for the user queries. This 
issue can be addressed by private information retrieval (PIR) \cite{chor1998private} and is beyond the scope of this paper.

\vspace{-0.1in}
\section{Single-string Similarity Search}
\label{sc:single}
In this section, we consider the single-string  similarity search queries. We first present our basic verification approach named $VS^2$ (Section \ref{sc:vss}). Then we present our E-$VS^2$ method with improved verification cost (Section \ref{sc:cvss}). 

\subsection{Basic Approach: $VS^2$}
\label{sc:vss}
Given the dataset $D$, the query string $s_q$, and the distance threshold $\theta$, the server returns all strings that are $\theta$-similar to $s_q$. Besides the similar strings, the server also returns a  {\em proof} of result correctness. 
In this section, we explain the details of our basic \underline{v}erification method of \underline{s}imilarity \underline{s}earch ($VS^2$). $VS^2$ consists of three phases: (1) the {\em pre-processing} phase in which the data owner constructs the authenticated data  structure $T$ of the dataset $D$. Both  $D$ and the root signature $sig$ of $T$ are outsourced to the server; (2) the {\em query processing} phase in which the server executes the similarity search query on $D$, and constructs the {\em verification object} ($VO$) of the search results $R$.  The server returns both $R$ and $VO$  to the client; and (3) {\em  verification} phase in which the client verifies the correctness of $R$ by leveraging $VO$. 
Next we explain the details of these three phases. 

%client preparation
\vspace{-0.1in}
\subsubsection{Pre-Processing}
\label{sc:preprocess}

In this one-time phase, the data owner constructs the authenticated data structure of the dataset $D$ before outsourcing $D$ to the server. We design a new authenticated data structure named the {\em Merkle $B^{ed}$ tree} ($MB$-tree). Next, we explain the details of $MB$-tree. 

\begin{figure}
	\centering
		\includegraphics[scale=0.2]{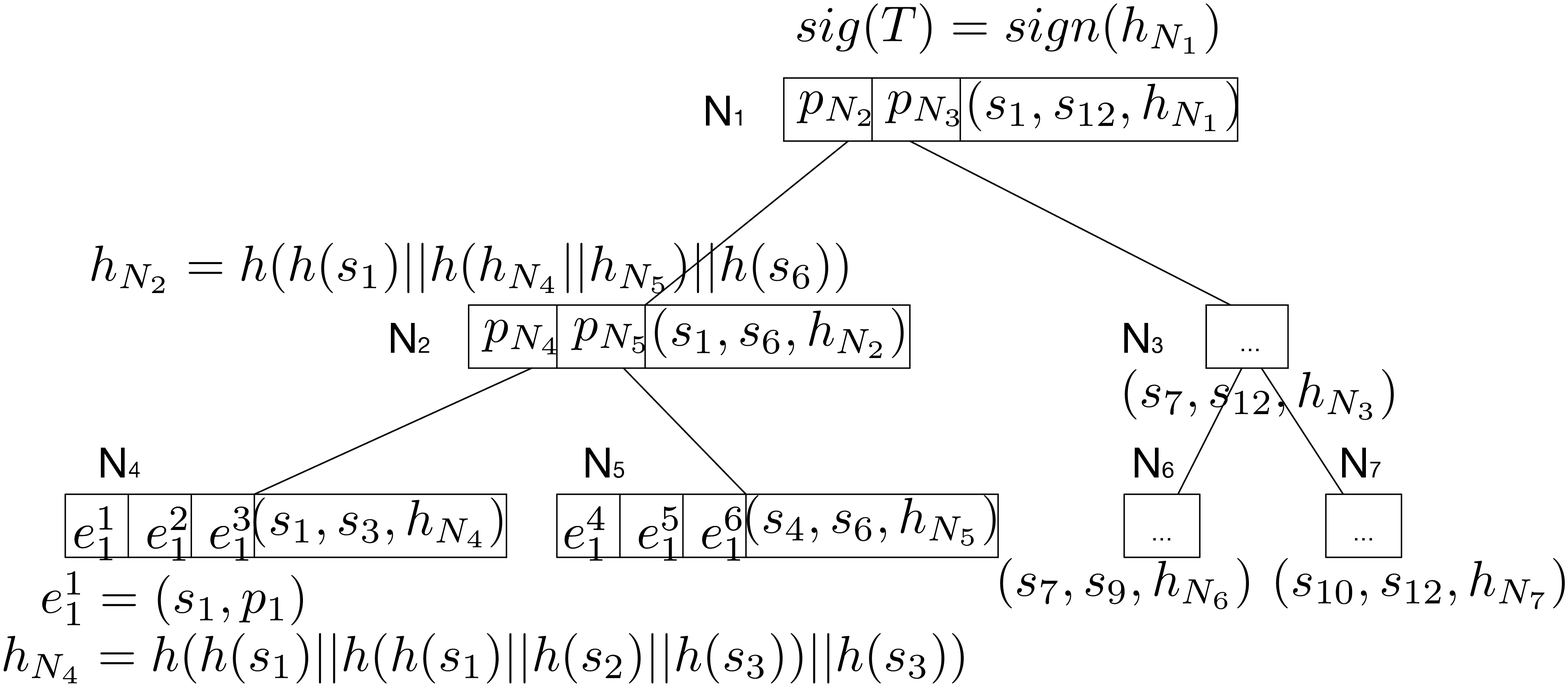}
		\vspace{-0.2in}
	\caption{An example of MB-tree}
	\label{fig:mbtree-example}
\vspace{-0.2in}
\end{figure}

The $MB$-tree is constructed on top of the $B^{ed}$ tree by assigning the digests to each $B^{ed}$ node. 
In particular, every $MB$-tree node contains a triple $(N_b, N_e, h_N)$, where 
 $N_b, N_e$ correspond to the string range values associated with $N$, and $h_N$ is the digest value computed as $h_N=h(h(N_b)||h(N_e)||h^{1\rightarrow f})$, where $h^{1\rightarrow f}=h(h_{C_1}||\dots||h_{C_f})$, with $C_1, \dots, C_f$ being the children of $N$. If $N$ is a leaf node, then $C_1, \dots, C_f$ are the strings $s_1, \dots, s_f$ covered by $N$. 
Besides the triple, each $MB$-tree node contains multiple entries. In particular, for any leaf node $N$, assume it covers $f$ strings. Then it contains $f$ entries, each of the format $(s, p)$, where $s$ is a string that is covered by $N$, and $p$ is the pointer to the disk block that stores $s$. 
%The digest value $h_N$ is computed as  $h_N=h(h(N_b)||h^{1\rightarrow f}||h(N_e))$, where $h^{1\rightarrow f}=h(h(s_1)||\dots||h(s_f))$, with $s_1, \dots, s_f$ the strings covered by $N$, and $h()$ a one-way, collision-resistant hash function (e.g. SHA-1 \cite{menezes1996handbook}).
For any intermediate node, assume that it has $f$ children nodes. Then it contains $f$ entries, each entry consisting of a pointer to one of its children nodes. 
%Suppose the internal node $N$ has $f$ children nodes $\{C_1, \dots, C_f\}$, then  $h_N=h(h(N_b)||h^{1\rightarrow f}||h(N_e))$, where $h^{1\rightarrow f}=h(h(C_1)||\dots||h(C_f))$.

%Let $N_R$ denote the root node. The client takes the hash value from the concatenation of $N_b$, $N_e$ and the hash value of every entry in the root node and takes the it as the $MB$-tree's signature. 
The digests of the $MB$-tree $T$ can be constructed in the bottom-up fashion, starting from the leaf nodes. 
After all nodes of $T$ are associated with the digest values, the data owner signs the root with her private key. The signature can be created by using a public-key cryptosystem (e.g., RSA).
%The constructed Merkle tree $T$ delivers the $MB$-tree. 
An example of the $MB$-tree structure is presented in Figure \ref{fig:mbtree-example}.
The data owner sends both $D$ and $T$ to the server. The data owner keeps the root signature of $T$ locally, and sends it to any client who requests for it for authentication purpose.

%The maximum number of entries in a leaf node is $f_1=\frac{P}{|p|+|s|+|h|}$. 
Following \cite{comer1979ubiquitous, li-sigmod06}, we assume that each node of the $MB$-tree occupies a disk page.
For the constructed $MB$-tree $T$, each entry in the leaf node occupies $|s|+|p|$ space, where $|p|$ is the size of a pointer, and $|s|$ is the maximum length of a string value. The triple $(N_b, N_e, h_N)$ takes the space of $2|s|+ h$, where $|h|$ is the size of a hash value. Therefore, a leaf node can have $f_1 =  [\frac{P-2|s|-|h|}{|p|+|s|}]$ entries at most, where $P$ is the page size. Given $n$ unique strings in the dataset, there are $[\frac{n}{f_1}]$ leaf nodes in $T$. Similarly, for the internal nodes, each entry takes the space of $|p|$. Thus each internal node can have at most $f_2=[\frac{P-2|s|-|h|}{|p|}]$ entries (i.e., [$\frac{P-2|s|-|h|}{|p|}]$ children nodes). Therefore, the height $h$ of $T$ $h \geq log_{f_2}[\frac{n}{f_1}]$. 
\nop{
\begin{equation}
f_2=\frac{P}{2|s|+|h|+|p|},
\label{eq:f2}
\end{equation}
where $|h|$ is the length of a hash value, and $|p|$ is the point's size.
}

The construction complexity of $MB$-tree is $O(n)$, where $n$ is the number of unique strings in $D$.
%and $C_H$ is the complexity to hash a value. The complexity mainly comes from two parts. The first part is to hash all the stringvalues to fill in the entries in the leaf nodes, whose complexity is $O(nC_H)$. The second part is to compute the hash values of the internal nodes in a bottom-up fashion with complexity $O(n_IC_H)$, where $n_I$ is the number of internal nodes. As $n_I<n$, the total construction complexity is $O(nC_H)$}.
It is cheaper than the complexity of pairwise similarity search over $D$. 
%It is comparable to the complexity of similarity search over $D$. This naturally fits into the amortized model for outsourced computation \cite{gennaro2010non}: the data owner performs a one-time computationally expensive phase (in our case constructing $MB$-tree and the root signature), whose cost is amortized over the authentication of all the future query executions. 

%server verification objects
\vspace{-0.1in}
\subsubsection{$VO$ Construction}
\label{sc:vo}

Upon receiving the similarity search query $Q(s_q, \theta)$ from the client, the server calculates the edit distance between all the string pairs and distills the similar pairs. For the similarity search result $R$, the server constructs a verification object $VO$ to show that $R$ are both sound and complete.

First, we define {\em false hits}. 
Given a query string $s_q$ and a similarity threshold $\theta$, the {\em false hits} of $s_q$, denoted as $F$, are all the strings that are dissimilar to $s_q$. In other words,  
$F =\{s|s\in D, s_q\not\approx s\}$. 
Intuitively, to verify that $R$ is sound and complete, the $VO$ includes both similar strings $R$ and false hits $F$. Apparently including all false hits may lead to a large $VO$, and thus high network communication cost and the verification cost at the client side. Therefore, we aim to reduce the $VO$ size of $F$. 

\nop{
Then we define {\em pure-negative} and {\em partial-negative} nodes based on the covered strings. 

\begin{definition}
Given a $MB$-tree $T$ and a search string $s_q$, a node $N$ of $T$ is {\em pure-negative} if $cover(N)$ only contains dissimilar strings of $s_q$. Otherwise we say $N$ is {\em partial-negative}.
\end{definition}

For example, consider the MB-tree in Figure \ref{fig:example} (a), and the query string $s_1$. Assume the  similar strings $R=\{s_3, s_5\}$. Then nodes $N_4$ and $N_5$ are partial-negative, while $N_6$ and $N_7$ are pure-negative. Note that the candidate nodes (Def.  \ref{def:bedtree}) can be either pure-negative or  partial-negative, while the non-candidate nodes must be pure-negative. 
}

Before we explain how to reduce $VO$ size, we first define C-strings and NC-strings. 
Apparently, each false hit string is covered by a leaf node of the $MB$-tree $T$. 
Based on whether a leaf node in $MB$-tree is a candidate, the false hits $F$ are classified into two types: 
(1) {\bf $C$-strings}: the strings that are covered by {\em candidate} leaf  nodes; and 
%(i.e., $C = \cup_{\forall\ leaf\ candidate\ node N}cover(N)}$.)
(2) {\bf $NC$-strings}: the strings that are covered by {\em non-candidate} leaf nodes. 
%(i.e., $NC = \cup_{\forall\ leaf\ non-candidate\ node N}cover(N)}$.)

Our key idea to reduce $VO$ size of $F$ is to include {\em representatives} of NC-strings instead of individual NC-strings. 
The representatives of NC-strings take the format of {\em maximal false hit subtrees} ($MFs$). 
Formally, given a $MB$-tree $T$, we say a subtree $T^N$ that is rooted at node $N$ is a {\em false hit subtree} if $N$ is a {\em non-candidate} node. We say the false hit subtree $T^N$ rooted at $N$ is {\em maximal} if the parent of $N$ is a candidate node.
The $MFs$ can root at leaf nodes. Apparently, all strings covered by the $MFs$ must be NC-strings. And each NC-string must be covered by a $MF$ node. Furthermore, $MFs$ are disjoint (i.e., no two $MFs$ cover the same string). Therefore, 
instead of including individual NC-strings into the $VO$, we include their $MFs$. As the number of $MFs$ is normally much smaller than the number of NC-strings, this can effectively reduce  $VO$ size. We are ready to define the $VO$. 
%We say the maximal false hit tree $T^N$ is a  {\em leaf maximal false hit tree} ($LF$) if $N$ is a leaf node of $T$.
%We say the maximal false hit tree $T^N$ is {\em pure-negative} if $N$ is a leaf node of $T$.

%In the MB-tree $T$, for an internal node $v$, we use $S(v)$ to denote the set of strings whose leaf node is a descendant of $v$. If $s \in S(v)$, we say the subtree at $v$ covers string $s$. A subtree $t(v)$ rooted at $v$ is said to be a {\em false hit subtree} of string $s$ if it meets the following two conditions.

%Condition 1. $cover(N)\subseteq F$, i.e., all the strings covered by $N$ are false hits. {\bf WHY CONDITION 1? IS IT POSSIBLE THAT THE NODES SATISFY CONDITION 2 BUT NOT CONDITION 1?}

%Condition 2. $DST_{min} > \theta$, i.e., the lower bound of the edit distance between $s$ and any string of node $N$ is greater than $\theta$. Therefore, none of the strings in $cover(N)$ can be similar to $s$. 

%For any query string $s_q$, the server performs a depth-first traversal of the $MB$-tree $T$, and assign the strings to  two sets: (1) the result set $R$ that includes all strings similar to $s_q$, and (2) the false hits $F$. Then the server constructs the maximal false hit subtrees $MFs$ of $F$. Finally, the server generates the $VO$ that is formally defined below. The server uses the same hash function as the one that the data owner uses to construct the $MB$-tree. 
\vspace{-0.1in}
\begin{definition}
{\em 
Given a dataset $D$, a query string $s_q$, let $R$ be the returned similar strings of $s_q$.  Let $T$ be the  $MB$-tree of $D$, and $NC$ be the strings that are covered by non-candidate nodes of $T$. Let $\cal M$ be a set of $MFs$ of $NC$. Then the $VO$ of $s_q$ consists of: 
(i) string $s$, for each $s\in D - NC$; and 
(ii) a pair $(N, h^{1\rightarrow f})$ for each $MF\in\cal M$ that is rooted at node $N$, where $N$ is represented as  $[N_b, N_e]$, with $[N_b, N_e]$ the string range associated with $N$, and $h^{1\rightarrow f}=h(h_{C_1}||\dots||h_{C_f})$, with $C_1, \dots, C_f$ being the children of $N$.  If $N$ is a leaf node, then $C_1, \dots, C_f$ are the strings $s_1, \dots, s_f$ covered by $N$. Furthermore, in $VO$, a pair of brackets is added around the strings and/or the pairs that share the same parent in $T$.\qed
}
\end{definition}
\vspace{-0.1in}
Intuitively, in $VO$,  the similar strings and C-strings are present in the original string format, while NC-strings are represented by the $MFs$ (i.e., in the format of $([N_b, N_e], h_N)$). 
\vspace{-0.15in}
\begin{example}
\label{exp:vo}
{\em
Consider the MB-tree $T$ in Figure \ref{fig:mbtree-example}, and the query string $s_1$. Note that $s_1\in D$ (but in general $s_q$ may not  be present in $D$). Assume the similar strings are $R=\{s_1, s_3, s_5\}$. Also assume that node $N_6$ of $T$ is the only non-candidate node. Then NC-strings are $NC = \{s_7, s_8, s_9\}$, and C-strings are $\{s_2, s_4, s_6, s_{10}, s_{11}, s_{12}\}$.
The set of $MFs$ $\cal M$ =$\{N_6\}$. Therefore, 
\begin{eqnarray*}
VO(s_1) = \{(((s_1, s_2, s_3), (s_4, s_5, s_6)), (([s_7, s_9], h^{7\rightarrow 9}),\\\nonumber
(s_{10}, s_{11}, s_{12})))\}, where\ h^{7\rightarrow 9}=h(h(s_7)||h(s_8)||h(s_9)) \qed\nonumber.
\end{eqnarray*}
}
\end{example}
\vspace{-0.15in}
For each $MF$, we do not require that $h(N_b)$ and $h(N_e)$ appear in $VO$. However, the server must include both $N_b$ and $N_e$ in $VO$. This is to prevent the server to cheat on the non-candidate nodes by including incorrect $[N_b, N_e]$ in $VO$. More details of the robustness of our authentication procedure can be found in Section \ref{sc:vssrobust}.  

\nop{
More details of the verification objects can be found in Algorithm \ref{alg:vo}. {\bf PSEUDO CODE FOR VERIFICATION OBJECT CREATION. ADD DESCRIPTION OF ALGORITHM}
In Algorithm \ref{alg:vss_vo_construction}, we build the VO starting from the $MB^{ed}$-tree's root. In Algorithm \ref{alg:dfs_vo_construction}, for each non-leaf $MF$ node, we add its triple to $VO$ (Line 2 to 7). However, for each $LF$, we only need to add the string to $VO$ as it only contains one string (Line 10 to 12).

\begin{algorithm}
\begin{small}
\caption{ $VSS\_VO\_Construction(s_q, T)$)
\label{alg:vss_vo_construction}
}
\begin{algorithmic}[1]
\REQUIRE{the query string $s_q$, the $MB^{ed}-tree$ $T$}
\ENSURE{The verification object to authenticate the query result $R(s_q)$.}

\STATE{set $ht$ to be the root's height in $T$}
%\STATE{initialize $VO=\{R(s_q)\}$}
\STATE{$VO=DFS\_VO(s_q, T.root, ht)$}
\RETURN{$VO$}
\end{algorithmic}
\end{small}
\end{algorithm}

\begin{algorithm}
\begin{small}
\caption{ $DFS\_VO(s_q, N, ht)$)
\label{alg:dfs_vo_construction}
}
\begin{algorithmic}[1]
\REQUIRE{the query string $s_q$, a node $N$ in $MB^{ed}-tree$ $T$ and its height $ht$}

\STATE{$VO_N='['$}
\IF{$ht \neq 0$}
	\FORALL{entry $e=(N_b, N_e, p, h_e) \in N$}
		\IF{$DST_{min}(s, N_b, N_e) > \theta$}
			\STATE{$VO_N = VO_N + (N_b, N_e, h_e)$}		
		\ELSE
			\STATE{$VO_N = VO_N + DFS\_VO(s_q, *p, ht-1)$}
		\ENDIF
	\ENDFOR
\ELSE
	\FORALL{entry $(s, p, h_e) \in N$}
		\STATE{$VO_N=VO_N+s$ }
	\ENDFOR
\ENDIF
\STATE{$VO_N=VO_N+']'$}
\RETURN{$VO_N$}
\end{algorithmic}
\end{small}
\end{algorithm}
}

%The server basically returns all the entries of the leave nodes except for the $MF$ node rooted at $N_4$, on which the server returns the triple at $N_4$.

%client verification
\vspace{-0.1in}
\subsubsection{Authentication Phase}
\label{sc:vss-vp}

For a given query string $s_q$, the server returns $\{R, VO\}$ to the client. Before result authentication, the client obtains the data owner's public key from a certificate authority (e.g., VeriSign \cite{rfc2459}). The client also obtains the hash function, the root signature $sig$ of the $MB$-tree, and the string ordering scheme and the $B^{ed}$-tree construction procedure from the data owner. The verification procedure consists of three steps. In Step 1, the client re-constructs the $MB$-tree from $VO$. In Step 2, the client re-computes the root signature $sig'$, and compares  $sig'$ with $sig$. In Step 3, the client re-computes the edit distance between $s_q$ and a {\em subset} of strings in $VO$. Next, we explain the details of these steps.  

\noindent{\bf Step 1: Re-construction of $MB$-tree}: First, the client  sorts the strings and string ranges (in the format of $[N_b, N_e]$) in $VO$ by their mapping values according to the string ordering scheme. String $s$ is put ahead the range $[N_b, N_e]$ if $s<N_b$. It should return a total order of strings and string ranges. If there exists any two ranges $[N_b, N_e]$ and $[N_b', N_e']$ that overlap, the client concludes that the $VO$ is not correct. If there exists a string $s\in R$ and a range $[N_b, N_e]\in VO$ such that $s\in[N_b, N_e]$, the client concludes that $R$ is not sound, as $s$ indeed is a dissimilar string (i.e., it is included in a non-candidate node). Second, the client maps each string $s\in R$ to an entry in a leaf node in $T$, and each pair $([N_b, N_e], h_N) \in VO$ to an internal node in $T$. The client re-constructs the parent-children relationships between these nodes by following the matching brackets () in $VO$. 

\noindent{\bf Step 2: Re-computation of root signature}: After the $MB$-tree $T$ is re-constructed, the client computes the root signature of $T$.
For each string value $s$, the client calculates $h(s)$, where $h()$ is the same hash function used by the data owner for the construction of the  $MB^{ed}$-tree.  For each internal node that corresponds to a pair $([N_b, N_e], h^{1\rightarrow f})$ in $VO$, the client computes the hash $h_N$ of $N$ as $h_N = h(h(N_b)||h(N_e)||h^{1\rightarrow f})$. Finally, the client re-computes the hash value of the root node, and rebuilds the root signature $sig'$ by signing $h_{root}$ using the data owner's public key. The client then compares $sig'$ with $sig$. If $sig' \neq sig$, the client concludes that the server's results are not correct. 

\noindent{\bf Step 3: Re-computation of necessary edit distance}: First, for each string $s\in R$, the client re-computes the edit distance $DST(s_q, s)$, and verifies whether $DST(s_q, s)\leq\theta$. If all strings $s\in R$ pass the verification, then the client concludes that $R$ is {\em sound}. Second, for each C-string $s\in VO$ (i.e., those strings appear in $VO$ but not $R$), the client verifies whether $DST(s_q, s)>\theta$. If it is not (i.e., $s$ is a similar string indeed), the client concludes that the server fails the completeness verification. Third, for each range $[N_b, N_e]\in VO$, the client verifies whether $DST_{min}(s_q, N)>\theta$, where $N$ is the corresponding $MB$-tree node associated with the range $[N_b, N_e]$. If it is not (i.e., node $N$ is indeed a candidate node), the client concludes that the server fails the soundness verification. 
\vspace{-0.05in}
\begin{example}
\label{exp:vs}
{\em 
Consider the $MB$-tree in Figure \ref{fig:mbtree-example} as an example, and the query string $s_1$. Assume the  similar strings $R=\{s_1, s_3, s_5\}$. Consider the VO shown in Example \ref{exp:vo}. 
The C-strings are $C = \{s_2, s_4, s_6, s_{10}, s_{11}, s_{12}\}$.
After the client re-constructs the $MB$-tree, it re-computes the hash values of strings $R\cup C$. It also computes the digest $h_{N_6} = (h(s_7)||h(s_9)||h^{7\rightarrow 9})$. Then it computes the root signature $sig'$ from these hash values. It also performs the following distance computations: 
(1) for $R=\{s_1, s_3, s_5\}$, compute the edit distance between $s_1$ and string in $R$, 
(2) for $C = \{s_2, s_4, s_6, s_{10}, s_{11}, s_{12}\}$, compute the edit distance between $s_1$ and any C-string in $C$, and 
(3) for the pair $([s_7, s_9], h_{N_6})\in VO$, compute $DST_{min}(s_q, N_6)$. \qed
}
\end{example}
\vspace{-0.1in}
\subsubsection{Security Analysis}
\label{sc:vssrobust}
Given a query string $s_q$ and a similarity threshold $\theta$, let $R$ ($F$, resp.) be the similar strings (false hits, resp.) of $s_q$. An untrusted server may  perform the following cheating behaviors the real results $R$: (1) {\em tampered values}: some strings in $R$ do no exist in the original dataset $D$; (2) {\em soundness violation}: the server returns $R' = R \cup FS$, where $FS\subseteq F$; and (3) {\em completeness violation}: the server returns $R' = R - SS$, where $SS\subseteq R$. 

The tampered values can be easily caught by the authentication procedure, as the hash values of the tampered strings are not the same as the original strings. This leads to that the root signature of $MB$-tree  re-constructed by the client differet from the root signature from the data owner. The client can catch the tampered values by Step 2 of the authentication procedure. Next, we mainly focus on the discussion of how to catch soundness and completeness violations. 

\noindent{\bf Soundness.} The server may deal with the $VO$ construction of $R' = R \cup {FS}$ in two different ways: \\
\underline{Case 1.} The server constructs the $VO$ $V$ of the correct result $R$, and returns $\{R', V\}$ to the client;

\noindent\underline{Case 2.} the server constructs the $VO$ $V'$ of $R'$, and returns $\{R', V'\}$ to the client.  
Note that the strings in $FS$ can be either NC-strings or C-strings. Next, we discuss how to catch these two types of strings for both Case 1 and 2. 

For Case 1, for each NC-string $s\in FS$, $s$ must fall into a $MF$-tree node in $V$. Thus, there must exist an $MF$-tree node whose associated string range overlaps with $s$. The client can catch $s$ by Step 1 of the authentication procedure. For each C-string $s\in FS$, $s$ must be treated as a C-string in $V$. Thus the client can catch it by computing $DST(s, s_q)$ (i.e., Step 3 of the authentication procedure). 

For Case 2, the C-strings in $FS$ will be caught in the same way as Case 1. Regarding NC-strings in $FS$, they will not be included in any MF-tree in $V'$. Therefore, the client cannot catch them by Step 1 of the authentication procedure (as Case 1). However, as these strings are included in $R'$, the client still can catch these strings by computing the edit distance of $s_q$ and any string in $R'$ (i.e., Step 3 of the authentication procedure). 

\noindent{\bf Completeness.} To deal with the $VO$ construction of $R' = R -  {SS}$, we again consider the two cases as for the discussion of correctness violation. In particular, let $V$ and $V'$ be the $VO$ constructed from the correct result $R$ and the incomplete result $R' = R - SS$ respectively, where $SS\subseteq R$. We discuss how to catch these two cases in details. 

For Case 1 (i.e., the server returns \{$R', V\}$),  any string $s\in SS$ is a C-string. These strings can be caught by re-computing the edit distance between the query string and any C-string (i.e., Step 3 of the authentication procedure). 

For Case 2 (i.e., the server returns \{$R', V'\}$), any string $s\in SS$ is either a NC-string or a C-string in $V'$. For any C-string $s\in SS$, it can be caught by re-computing the edit distance between the query string and the C-strings (i.e., Step 3 of the authentication procedure). 
For any NC-string $s\in SS$, it must be included into a non-candidate MB-tree node. We have the following theorem. \vspace{-0.1in}
\begin{theorem}
\label{theorm:cheat}
{\em 
Given a query string $s_q$ and a non-candidate node $N$ of range $[N_b, N_e]$, including any string $s'$ into $N$ such that $s'\approx s_q$ must change $N$ to be a candidate node. 
}
\end{theorem}
\vspace{-0.1in}
The proof is straightforward. It is easy to see that $s'\not\in[N_b, N_e]$. Therefore, including $s'$ into $N$ must change the range to be either $[s', N_e]$ or $[N_b, s']$, depending on whether $s'<N_b$ or $s'>N_e$. Now it must be true that $DST_{min} (s_q, N)\leq\theta$, as $DST(s', s_q)\leq\theta$. 

Following Theorem \ref{theorm:cheat}, for any string $s\in SS$ that is considered a NC-string in $V'$, the client can easily catch it by verifying whether $DST_{min}(s_q, N)>\theta$, for any non-candidate node (i.e., Step 3 of the authentication procedure). 

\vspace{-0.05in}
\subsection{String Embedding Authentication: E-$VS^2$}
\label{sc:cvss}

One weakness of $VS^2$ is that if there exist a significant number of C-strings, its $VO$ can be of large size. This may bring expensive network communication cost. Furthermore, since the client needs to compute the edit distance $DST(s_q, s)$ for each C-string $s$, too many C-strings may incur expensive verification cost at the client side too. 
Our goal is to shrink $VO$ size with regard to C-strings, so that both network communication cost and verification cost can be reduced. 
We observe that although C-strings are not similar to the query string, they may be similar to each other.
Therefore, we design {\em E-$VS^2$}, a computation-efficient method on top of $VS^2$.
The key idea of E-$VS^2$ is to construct a set of {\em representatives} of C-strings based on their similarity, and only include the representatives of C-strings in $VO$. 
To construct the representatives of C-strings, we 
first apply a similarity-preserving mapping function on C-strings, and transform them into the Euclidean space, so that the similar strings are mapped to the close points in the Euclidean space. 
Then C-strings are organized into a small number of groups called {\em distant bounding hyper-rectangles (DBHs)}. DBHs are the representatives of C-strings in $VO$. In the verification phase, the client only needs to calculate the Euclidean distance between $s_q$ and DBHs. 
Since the number of DBHs is much smaller than the number of C-strings, and Euclidean distance calculation is much faster than that of edit distance, the verification cost of the E-$VS^2$ approach is much cheaper than that of $VS^2$. 
Next, we explain the details of the E-$VS^2$ approach. Similar to the $VS^2$ approach, E-$VS^2$ consists of three phases: (1) the {\em pre-processing } phase at the data owner side; (2) the {\em query processing} phase at the server side; and (3) the {\em verification} phase at the client side. Next, we discuss the three phases in details. 

\vspace{-.05in}
\subsubsection{Pre-Processing}
Before outsourcing the dataset $D$ to the server, similar to $VS^2$, the data owner constructs the $MB$-tree $T$ on $D$. In addition, the data owner maps $D$ to the Euclidean space $E$ via a similarity-preserving embedding  function $f: D\rightarrow E$ denote the embedding function. 
%There is a  methods \cite{jin2003efficient, wang2000index, hjaltason2003properties} to map string values into a multi-dimensional Euclidean space such that the mapped space preserves the original string distances. 
We use SparseMap \cite{hjaltason2003properties} as the embedding function due to its contractive property
 (Section \ref{sc:pre}). 
%(Equation \ref{eq:contractive}).  
The complexity of the embedding is $O(cdn^2)$, where $c$ is a constant value between 0 and 1, $d$ is the number of dimensions of the Euclidean space, and $n$ is the number of strings of $D$. We agree that the complexity of string embedding is comparable to the complexity of similarity search over $D$. This naturally fits into the amortized model for outsourced computation \cite{gennaro2010non}: the data owner performs a one-time computationally expensive phase (in our case constructing the embedding space), whose cost is amortized over the authentication of all the future query executions. 

The data owner sends $D$, $T$ and the embedding function $f$ to the server. The server constructs the embedded space of $D$ by using the embedding function $f$. The function $f$ will also be available to the client for result authentication. 

\vspace{-0.05in}
\subsubsection{VO Construction}
\label{sc:evss-vo}
Given the query $Q(s_q, \theta)$ from the client, the server applies the embedding function $f$ on $s_q$, and finds its corresponding node $P_q$ in the Euclidean space. Then the server finds the result set $R$ of $s_q$. To prove the soundness and completeness of $R$, the server builds a verification object $VO$. First, similar to $VS^2$, the server searches the $MB$-tree to build $MFs$ of NC-strings. 
For the C-strings, the server constructs a set of {\em distant bounding hyper-rectangles} (DBHs) from their embedded nodes in the Euclidean space. 
%The distance between the point $P_q$ of the query string $s_q$ {\bf QUESTION: SO THE SERVER NEEDS TO CONSTRUCT THE EUCLIDEAN SPACE OF $D\CUP\{s_q\}$?} and the DBHs suffices to demonstrate that the C-strings are in the false hits $F$ of $s_q$. 
Before we define DBH, first, we define the minimum distance between an Euclidean point and a hyper-rectangle. Given a set of points $\cal P$ $\{P_1, \dots, P_t\}$ in a $d$-dimensional Euclidean space, a hyper-rectangle $R(<l_1, u_1>, \dots, <l_d, u_d>)$ is the {\em minimum bounding hyper-rectangle} (MBH) of $\cal P$ if $l_i=min_{k=1}^{t}(P_k[i])$ and $u_i=max_{k=1}^{t}(P_k[i])$, for $1\leq i\leq d$, where $P_k[i]$ is the $i$-dimensional value of $P_k$. 
For any point $P$ and any hyper-rectangle $R(<l_1,u_1>, \dots, <l_d, u_d>)$, the minimum Euclidean distance between $P$ and $R$ is 
$dst_{min}(P, R) = \sqrt{\sum_{1\leq i\leq d} m[i]^2}$, 
where $m[i]=max\{l_i-p[i],0,p[i]-u_i$. 
Intuitively, if the node $P$ is inside $R$, the minimum distance between $P$ and $R$ is 0. Otherwise, we pick the length of the shortest path that starts from $P$ to reach $R$. We have: 
\vspace{-0.05in}
\begin{lemma}
\label{them:mdis}
{\em 
Given a point $P$ and a hyper-rectangle $R$, for any point $P'\in R$, the Euclidean distance $dst(P', P)> dst_{min}(P, R)$. 
}
\end{lemma}
\vspace{-0.05in}
The proof of Lemma \ref{them:mdis} is trivial. We omit the details due to the space limit.

Now we are ready to define {\em distant bounding hyper-rectangles} (DBHs). 
Given a query string $s_q$, let $P_q$ be its embedded point in the Euclidean space. For any hyper-rectangle $R$ in the same space, $R$ is a {\em distant bounding hyper-rectangle} (DBH) of $P_q$ if  $dst_{min}(P_q, R)>\theta$. 

Given a DBH $R$, Lemma \ref{them:mdis} guarantees that $dst(P_q, P)>\theta$ for any point $P\in R$. Recalling the contractive property of the SparseMap method, we have $dst(P_i, P_j) \leq DST(s_i, s_j)$ for any string pair $s_i$, $s_j$ and their embedded points $P_i$ and $P_j$. Thus we have the following theorem:
\vspace{-0.05in}
\begin{theorem}
{\em 
Given a query string $s_q$, let $P_q$ be its embedded point. Then for any string $s$, $s$ must be dissimilar to $s_q$ if there exists a DBH $R$ of $P_q$ such that $P\in R$, where $P$ is the embedded point of $s$. 
\label{th:DBH}
}
\end{theorem}
\vspace{-0.05in}
%DBH from a set of points
Based on Theorem \ref{th:DBH}, to prove that the C-strings are dissimilar to the query string $s_q$, the server can build a number of DBHs to the embedded Euclidean points of these C-strings. We must note that not all C-strings can be included into DBHs. This is because the embedding function may introduce {\em false positives}, i.e., there may exist a false hit string $s$ of $s_q$ whose embedded point $P$ becomes close to $P_q$. 
Given a query string $s_q$, we say a C-string $s$ of $s_q$ is an {\em FP-string} if $dst(P, P_q)\leq\theta$, where $P$ and $P_q$ are the embedded Euclidean points of $s$ and $s_q$. Otherwise (i.e. $dst(P, P_q)>\theta$), we call $s$ a {\em DBH-string}. We have: 
\vspace{-0.05in}
\begin{theorem}
\label{them:dbhs}
{\em
Given a query string $s_q$, for any DBH-string $s$, its embedded point $P$ must belong to a DBH. 
}
\end{theorem}
\vspace{-0.05in}
The proof of Theorem \ref{them:dbhs} is straightforward. For any DBH-string whose embedded point cannot be included into a DBH with other points, it constructs a hyper-rectangle $H$ that only consists of one point. Obviously $H$ is a DBH. 

Therefore, given a query string $s_q$ and a set of C-strings $C$, first, the server classifies $C$ into FP- and DBH-strings, based on the Euclidean distance between their embedded points. 
Apparently the embedded points of FP-strings cannot be put into any DBH. Therefore, the server only consider DBH-strings and tries builds DBHs of DBH-strings. The VO of DBH-strings will be computed from DBHs. Therefore, in order to minimize the verification cost at the client side, the server aims to minimize the number of DBHs. Formally, 

%Before we formally define the problem of constructing a minimal number of DBHs, we define the following concepts. Given an Euclidean point $P\{p_1, \dots, p_d\}$, we say $P$ is {\em inside} the hyper-rectangle $R$, denoted as $P\in R$, if $\forall i\in[1, d]$, $p_i\in[l_i, u_i]$. Given two hyper-rectangles $R_i(<l_1^i, u_1^i>, \dots, <l_d^i, u_d^i>)$ and $R_j(<l_1^j, u_1^j>, \dots, <l_d^j, u_d^j>)$, we say $R_i$ and $R_j$ are {\em non-overlapping} if $\forall k\in[1, d]$, $[l_k^i, u_k^i]$ and $[l_k^j, u_k^j]$ do not overlap (i.e., $u_k^i<l_k^j$) . 

\nop{
Prior to discussing the method to build DBHs, we give the definition of {\em minimum bounding hyper-rectangle} first.
\vspace{-0.05in}
%MBH
\begin{definition}
Given a set of points $P=\{p_1, \dots, p_t\}$ in a $d$-dimensional Euclidean space, a hyper-rectangle $R(<l[1], u[1]>, \dots, <l[d], u[d]>)$ is the {\em minimum bounding hyper-rectangle} (MBH) of $P$ if $l[i]=min\{p_1[i], \dots, p_t[i]\}$ and $u[i]=max\{p_1[i], \dots, p_t[i]\}$ for $1\leq i\leq d$.
\label{def:MBH}
\end{definition}
\vspace{-0.05in}
The MBH of $P$ is the minimum hyper-rectangle to cover all points in $P$. 
It is worth noting that a MBH may only  the point of one DBH-string. In that case, the MBH is exactly the point.}

%how to construct DBHs from TP

%Now we are ready to define the problem of constructing a minimal number of DBHs (MDBH). 

{\em $MDBH$ Problem}: Given a set of DBH-strings $\{s_1, \dots, s_t\}$, let $\cal P$ $\{P_1, \dots, P_t\}$ be their embedded points. Construct a minimum number of DBHs $\cal R$ = $\{R_1, \dots, R_k\}$ such that: (1) $\forall R_i, R_j\in\cal R$, $R_i$ and $R_j$ do not overlap; and (2) $\forall P_i\in\cal P$, there exists a DBH $R\in\cal R$ such that $P_i\in R$. 

Next, we present the solution to the $MDBH$ problem. We first present the simple case for a 2-D dimension (i.e. $d=2$). Then we discuss the scenario when $d>2$. For both settings, consider the same input that includes a query point $P_q$ and set of Euclidean points $\cal P$ $\{P_1, \dots, P_t\}$ which are the embedded points of a query string $s_q$ and its DBH-strings respectively. 

\begin{figure*}[!ht]
	\centering
	\begin{tabular}{cc}
	    \vspace{-0.3in}
	    \includegraphics[width=0.7\textwidth]{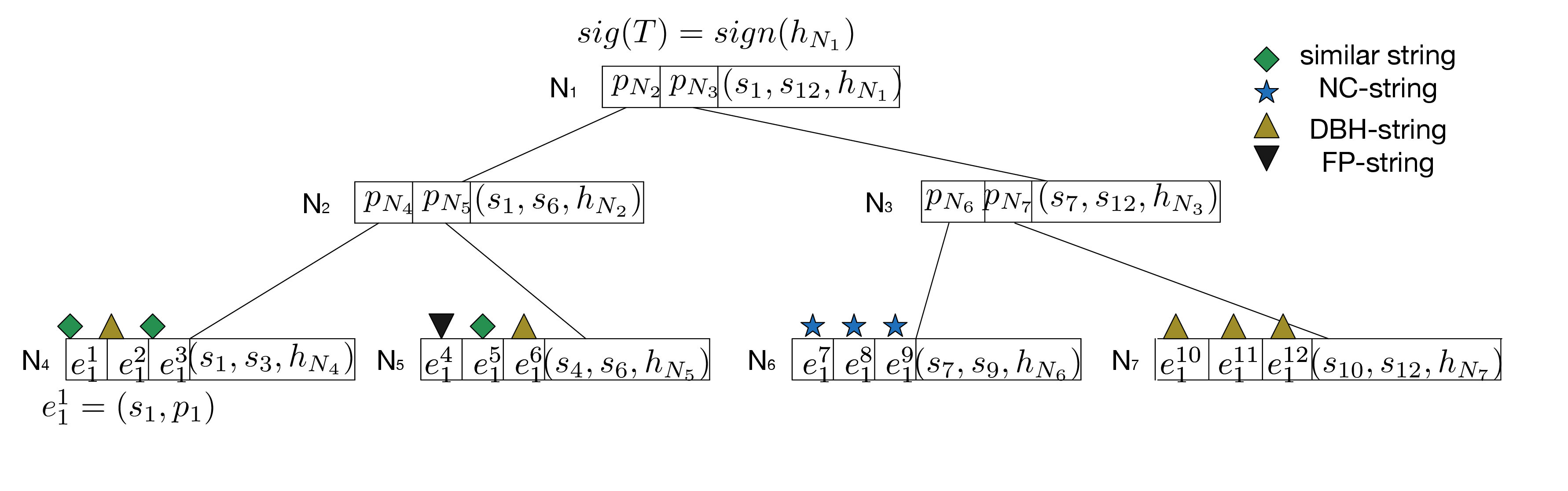}
	    &
	    \includegraphics[width=0.3\textwidth]{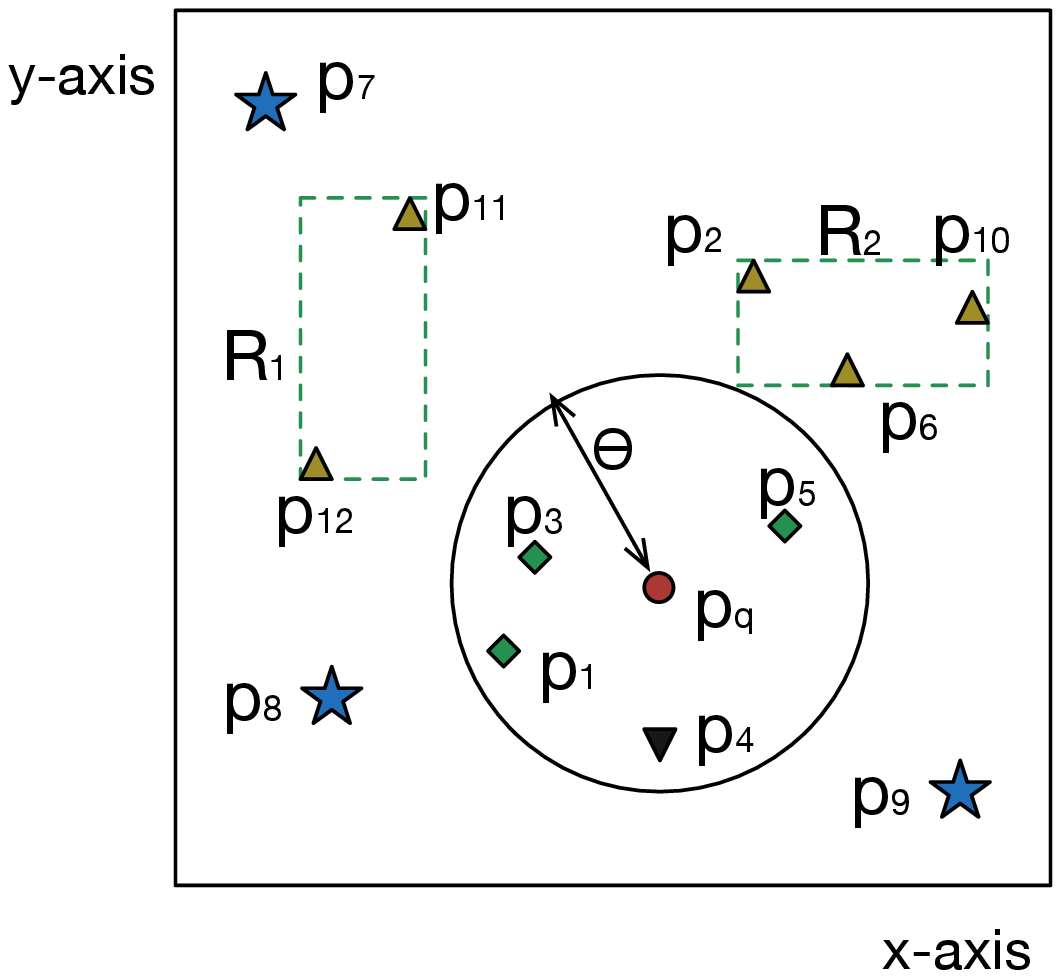}
	    \\
        (a) The $MB$-tree
        & 
        (b) The embedded Euclidean space
    \end{tabular}
    \caption{An example of VO construction by E-$VS^2$ method}
	\label{fig:exampleevs2}
	\vspace{-0.1in}
\end{figure*}

\nop{
\begin{figure*}[!ht]
	\centering
	 \begin{subfigure}[t]{0.6\textwidth}
        \centering
        \includegraphics[width=\textwidth]{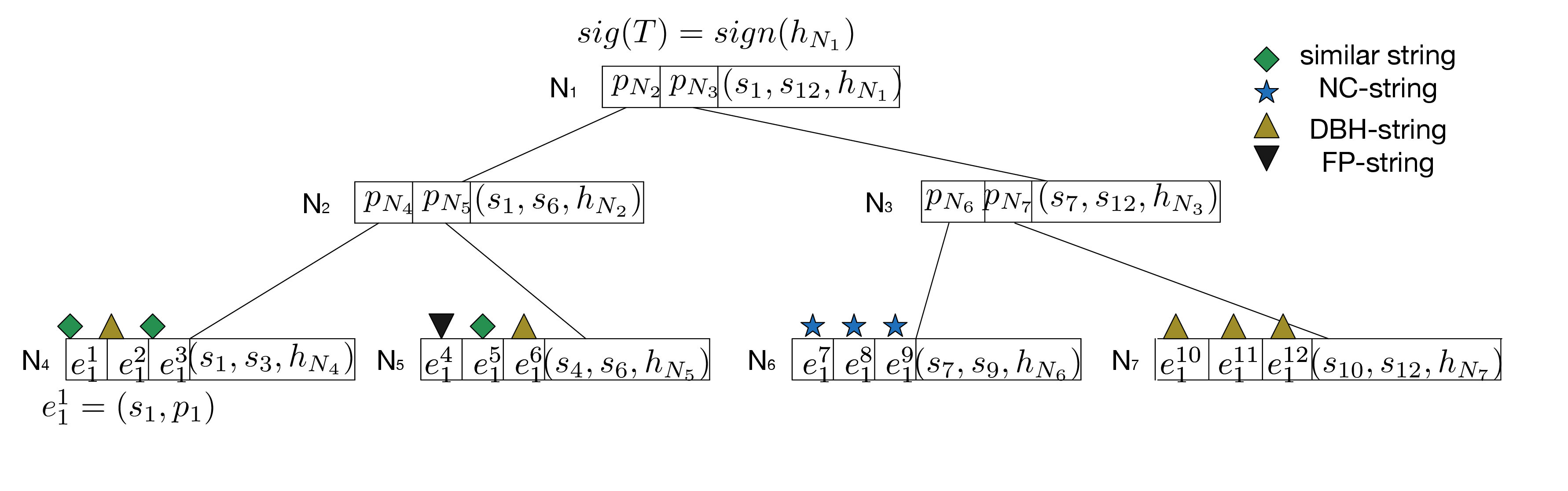}
        \vspace{-0.15in}
        \caption{The $MB$-tree}
    \end{subfigure}%
    ~ 
    \begin{subfigure}[t]{0.3\textwidth}
        \centering
        \includegraphics[width=\textwidth]{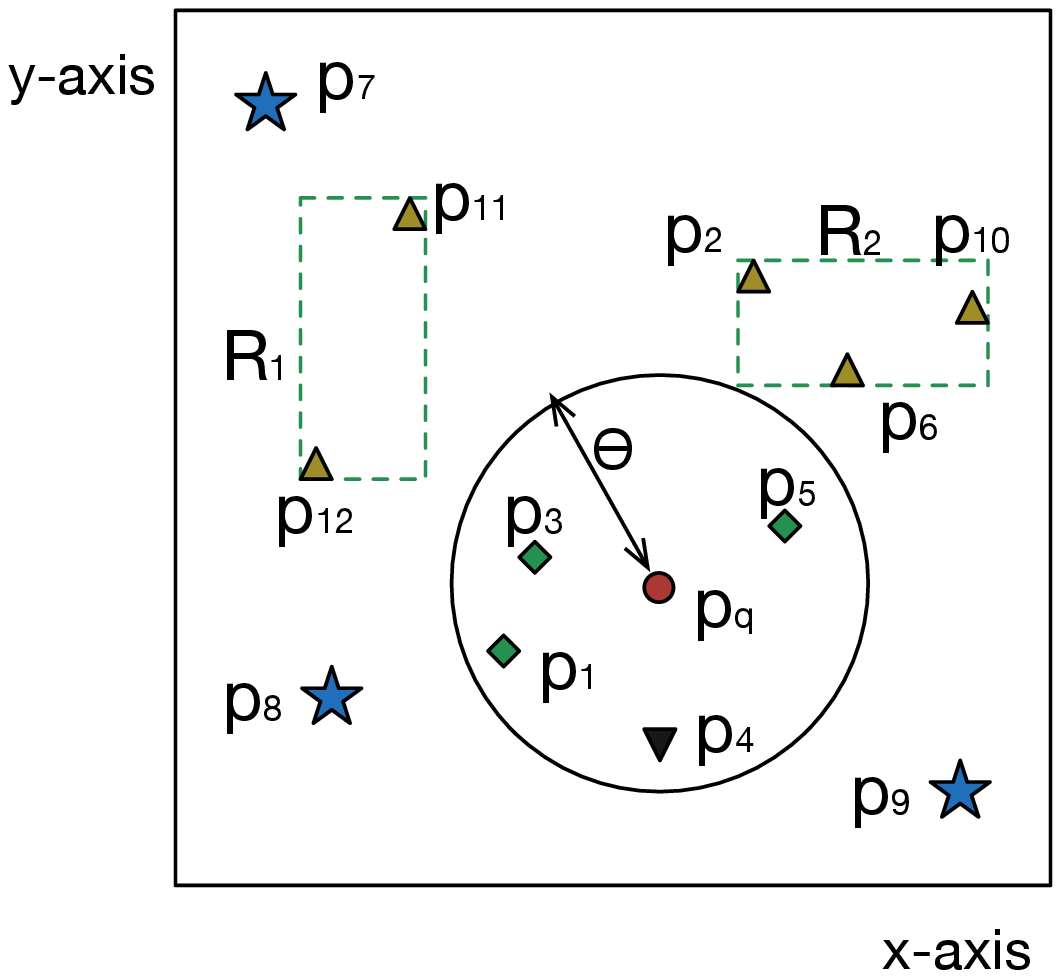}
        \caption{The embedded Euclidean space}
    \end{subfigure}
    \vspace{-0.15in}
    \caption{An example of VO construction by E-$VS^2$ method}
	\label{fig:exampleevs2}
	\vspace{-0.1in}
\end{figure*}
}
%d=2
\noindent{\bf When $d=2$.} We construct a graph $G=(V,E)$ such that for each point $P_i \in\cal P$, it corresponds to a vertex $v_i\in V$. For any two vertices $v_i$ and $v_j$ that correspond to two points $P_i$ and $P_j$, there is an edge $(v_i, v_j)\in E$ if $dst_{min}(P_q, R)>\theta$, where $R$ is the MBH of $P_i$ and $P_j$. We have:
\vspace{-0.05in}
%clique theorem
\begin{theorem}
\label{th:clique}
{\em Given the graph $G=(V, E)$ constructed as above, for any clique $C$ in $G$, let $R$ be the MBH constructed from the points corresponding to the vertice in $C$. Then $R$ must be a DBH.}
\end{theorem}
\vspace{-0.05in}
The proof of Theorem $\ref{th:clique}$ is in Appendix.
%minimum clique partition
Based on Theorem \ref{th:clique}, the {\em MDBH} problem is equivalent to the well-known {\em clique partition problem}, which is to find the smallest number of cliques in a graph such that every vertex in the graph belongs to exactly one clique. 
The clique partition problem is NP-complete. 
Thus, we design the heuristic solution to our {\em MDBH} problem.
Our heuristic algorithm is based on the concept of {\em maximal cliques}. 
%algorithm
Formally, a clique is maximal if it cannot include one more adjacent vertex. The maximal cliques can be constructed in polynomial time \cite{eidenbenz2000maximum}. It is shown that every maximal clique is part of some optimal clique-partition \cite{eidenbenz2000maximum}. Based on this, finding a minimal number of cliques is equivalent to finding a number of maximal cliques.
%a heuristic algorithm of finding a minimal number of cliques is to find a number of maximal cliques. 
Thus we construct maximal cliques of $G$ iteratively, until all the vertices belong to at least one clique. 

\nop{
We have the pseudocode in Algorithm \ref{alg:partition}. To solve the $MDBH$ problem, we first construct the graph $G$ (Line 1). Then we initialize a clique with a single node whose degree is the minimum (Line 4-5). This is because the vertex of small degree is difficult to put into a large clique. Thus we start with the most difficult vertex and build a maximal clique for it. In Line 6 and 7, we define $U_1$ and $U_2$, the union of which is the set of vertices that are not in any partition yet. Specifically, $U_1$ contains the set of vertices that have the potential to be added to the current clique $C$, because every vertex in $U_1$ is connected with all the nodes in $C$ in the graph $G$. In Line 21, from all the vertices in $U_1$, we pick the one of smallest $deg_{U_2}$ as it is the one that is most difficult to be grouped into other cliques. After we extend the clique $C$ by one vertex, we update the set $U_1$ and $U_2$ accordingly (Line 24). We repeatedly add vertices to $C$ until it is maximal, i.e., not extendable. We repeat the steps to find maximal cliques until all vertices are in a clique. The DBHs can be constructed by building MBHs for the points of the vertices in every clique. }

%special simple case
There is a special case where the $MDBH$ problem can be solved in polynomial time: when the embedded points of all DBH-strings lie on a single line, we can construct a minimal number of DBHs in the complexity of $O(\ell$), where $\ell$ is the number of DBH-strings. Due to space limit, the details of DBH construction can be found in Appendix. 

\noindent{\bf When $d>2$.} 
%clique does not hold
Unfortunately, Theorem \ref{th:clique} can not be extended to the case of $d>2$. 
%is shown in Figure \ref{fig:3d}, in which 
We found an example in which the MBHs of the pairs $(v_i, v_j)$, $(v_i, v_k)$, and $(v_j, v_k)$ are DBHs. However, the MBH of the pair $(v_i, v_j, v_k)$ is not a DBH, as it includes a point $w$ such that $w$ is not inside $R(v_i, v_j)$, $R(v_i, v_k)$, and $R(v_j, v_k)$, but $dst(P_q, w)<\theta$. 

\nop{
\begin{figure}
\begin{center}
	\includegraphics[width=0.45\textwidth]{./figs/rect4.eps}
	\caption{An example of 3-d space}
	\label{fig:3d}
\end{center}
\end{figure}
}

%solution
To construct the DBHs for the case $d>2$, we slightly modify the clique-based construction algorithm for  the case $d=2$. In particular, when we extend a clique $C$ by adding an adjacent vertex $v$, we check if the MBH of the extended clique $C'=C\cup \{v\}$ is a DBH. If not, we delete the edges $(u,v)$ from $G$ for all $u\in C$. 
%(Line 9-20 of Algorithm \ref{alg:partition}). 
This step ensures that if we merge any vertex in $U_1$ to $C$, the MBH of the newly generated clique is still a DBH.
%complexity

For both cases $d=2$ and $d>2$, the complexity of constructing DBHs from  DBH-strings is $O(n_{DS}^3)$, where $n_{DS}$ is the number of DBH-strings.

\nop{
 A {\em minimum bounding hyper-rectangle} (MBH) $R$ in $\cal E$ is defined as 
$R (<l_1, u_1>,  \dots, <l_d, u_d>) = \{(p_1, p_2, \dots, p_d), \forall i \in[1, d], p_i \in [l_i, u_i]\}$. For any point $P (p_1, \dots, p_d)$, the minimum distance between $P$ and $R$ is defined as: 
\[ MinD(P, R) = \sqrt{\sum_{1\leq i\leq d} max^2(p_i-l_i, 0, u_i-p_i)}.\] 

that correspondd to $LFs$. The key idea of E-$VS^2$ is to first apply a similarity-preserving mapping function on the strings to transform them into the Euclidean space, so that the similar strings are mapped to the close points in the Euclidean space. Then the $VO$ is constructed from the Euclidean space, with the points of similar strings grouped into {\em distant bounding hyper-rectangles} (DBHs).  The $VO$ of $LF$ is then constructed from DBH instead of individual strings. Next, we discuss the details of the $E$-$VS^2$ method.

Given the query string $s_q$, let $L$ be its $LF$-strings. Obviously, $L$ is a set of dissimilar strings of $s_q$.  
Let $\cal E$ be the Euclidean space constructed by applying a similarity-preserving string embedding method on $L\cup \{s_q\}$, and $\cal P$ be the corresponding points of $L$ in $\cal E$. 
For simplicity, we say $\cal P$ is the {\em $LF$-points} of $s_q$.
Formally, given a $d$-dimensional Euclidean space $\cal E$, a point 
$P = (p_1, \dots, p_d)$ defines a point in $\cal E$. A {\em minimum bounding hyper-rectangle} (MBH) $R$ in $\cal E$ is defined as 
$R (<l_1, u_1>,  \dots, <l_d, u_d>) = \{(p_1, p_2, \dots, p_d), \forall i \in[1, d], p_i \in [l_i, u_i]\}$. For any point $P (p_1, \dots, p_d)$, the minimum distance between $P$ and $R$ is defined as: 
\[ MinD(P, R) = \sqrt{\sum_{1\leq i\leq d} max^2(p_i-l_i, 0, u_i-p_i)}.\] 
{\bf DOUBLE CHECK CORRECTNESS}

Now we are ready to define the {\em distant bounding hyper-rectangle (DBH)}.
\begin{definition} 
Given a query string $s$, let $P$ be its corresponding point in the embedded Euclidean space. Given a set of MBHs $\cal R$, a MBH  $R\in\cal R$ is {\em distant} if $MinD(P, R) > \theta$. We require that each DBH must contain at least two points.
\end{definition}  
{\bf WHY DO WE NEED TO MENTION Single-point DBH?}

Due to the contractiveness property of the embedding methods, for any two embedded points $P_i$ and $P_j$ that are $\theta$-dissimilar (i.e. $dst(P_i, P_j)\geq \theta$, where  $dst()$ is the distance function of the Euclidean space),  it must be true that their original strings $s_i$ and $s_j$ must be  $\theta$-dissimilar. 
Therefore, constructing $VO$ from  DBHs enables the client to verify  whether the dissimilar strings are  indeed distant to the query string $s_q$ by checking whether these strings reside into any DBH. Note that there may exist some $LF$-strings that cannot be assigned to any DBH. We will discuss how to construct VO of these strings in Section  \ref{sc:cvss-vo}. 

Consider the $MB$-tree in Figure \ref{fig:mbtree-example}, {\bf WHAT IS $s_q$, and the leaf maximum subtrees $L$? NEED TO REVISE THE EXAMPLE}
{\em The query string is $s_1$, the set of dissimilar strings of $s_1$ $L=\{s_2, s_6, s_{10}, s_{11}, s_{12}\}$.}
the client does not need to compute the edit distance between $s_1$ and $s_3$, $s_6$. Instead, she only needs to check the minimum distance between $p_1$ and the hyper-rectangle in green color in Figure \ref{fig:example} (b). The hyper-rectangle $rect_1$ includes the points $p_{11}$ and $p_{12}$ of string $s_{11}$ and $s_{12}$. If the client calculates the minimum distance between $p_1$ and $rect_1$ is larger than $\theta$, then the client is sure that $p_{11}$ and $p_{12}$ are not similar to $p_1$, so are $s_{11}$ and $s_{12}$. The same procedure applies to $rect_2$.

The number of DBHs decides the size of $VO$. Therefore, for a given query string $s_q$, its $LF$-strings $L$, and the corresponding $LF$-points $\cal P$, we aim to construct a minimal number of DBHs that cover $\cal P$. Next, we first discuss how to construct DBHs.

\vspace{-.1in}
\subsubsection{Construction of DBHs}
The DBH construction procedure is designed for two scenarios: (1) the dimension of the Euclidean space $d = 2$; and (2) $d>2$. 

\noindent{\bf When d=2.} Given a set of points $p_1(x_1, y_1), \dots, p_k(x_k, y_k)$ ($k\geq 2$), we say $p_1, \dots, p_k$ defines a minimal bounding hyper-rectangle (MBH) $R(<l_1, u_1>, <l_2, u_2>)$, where $l_1 = min(x_1, \dots, x_k)$, $l_2 = max(x_1, \dots, x_k)$, $u_1 = min(y_1, \dots, y_k)$, and $u_2 = max(y_1, \dots, y_k)$. It is possible that all points exist on a single line.  We consider the line as a special case. 

Consider a MBH $R$ and a point $P$ in the two-dimension space. We define the {\em minimum distance} between $P$ and $R$: 
\begin{equation}
DST_{min}(P, R)=\sqrt{\sum_{i=1}^2 max\{P[i]-R[i].min, 0, R[i].max-P[i]\}^2},
\label{eq:dstmin}
\end{equation}
where $R[i].max$ ($R[i].min$) is the max (min) value on the $i$-th dimension of $R$. {\bf WHY DO WE DEFINE $DST_{MIN}$? IS IT EQUIVALENT TO MinD(P, R)?}
{\em Yes. They are equivalent. Originally, we only show one of them.}

For any query string $s_q$, let $L$ be its $LF$-strings, and $\cal P$ be its $LF$-points. To construct a minimal number of DBHs of $\cal P$, first, we construct an undirected graph $G=(V, E)$. Specifically, each string $s'\in L$ corresponds to a vertex $v$ in $V$. For any two strings $s_i$, $s_j\in L$, let $v_i$ and $v_j$ be their corresponding points in $G$, and  $P_i$ and $P_j$ be their corresponding points in the embedded Euclidean space. Let $R$ be the MBH that is defined by $P_i$ and $P_j$. Then there is an edge between the nodes $v_i$ and $v_j$ if $DST_{min}(P_q, R)>\theta$, where $P_q$ is the point of string $s_q$ in the embedded Euclidean space. We have the following lemma.

\begin{theorem}
\label{th:clique}
Given a string $s_q$, its $LF$-strings $L$, and the graph $G=(V, E)$ constructed from $L$, for any clique $C$ in $G$, let $R$ be the MBH defined by the points in $C$. Then $R$ must be a DBH. 
\end{theorem}
{\bf PROOF?}
{\em The proof is in Appendix.}

%Based on Theorem \ref{th:clique},
Based on Lemma \ref{lemma:clique}, the problem of finding a minimal number of DBHs is equivalent to
the problem of finding the {\em clique covering number} of a given graph $G$. The clique covering number of a graph $G$ is the minimum number of cliques in G needed to cover the vertex set of $G$. Since it involves the minimum number of cliques, only 
maximal cliques need be considered. It is shown that every maximal clique is part of some optimal clique-partition \cite{eidenbenz2000maximum}.
Therefore, the problem of finding a minimal number of DBHs is equivalent to
the well-known {\em clique partition problem}, which is  NP-complete.  
%, and the fact it is not approximatable within $n^\epsilon$ for any $\epsilon>0$, we 
Therefore, we can adapt any heuristic method of the clique partition problem to our problem to  find the DBHs. For example, we can use the dynamic ordering algorithm \cite{pattillo2011clique} to construct the cliques. The idea is to initially construct a clique that only contains one vertex and repeatedly add vertex into it, until there is no other vertex that can fit in the clique. 

It is possible that all $LF$-points lie on a single line. Then finding a minimal number of DBHs is not a NP-complete problem any more. Specifically, let $P_q$ be the embedding point of the query string, and $\cal L$ be the line that $LF$-points form. A polynomial algorithm is to draw a perpendicular line from $P_q$ to $\cal L$.  {\bf WHAT IF $P_q$ lies on $\cal L$?}
{\em In a special case that $P_q$ lies on $\cal L$, $P_q$ splits the $LF$-points into two subsets, each subset containing the points that reside in one side of $P_q$.}
The perpendicular line splits $\cal L$ into two DBH: the points on each side of the perpendicular line forms a DBH \ref{fig:line}. 
{\bf CAN YOU PROVE THE TWO hyper-rectangleS MUST BE DBHs? IS THE SINGLE hyper-rectangle THAT CONTAINS ALL $LF$-POINTS GOOD ENOUGH? }
{\em Let the $LF$-points be $\{P_1, \dots, P_k\}$ and $v$ be the unit vector of $\cal L$. For each point $P\in \{P_1, \dots, P_k, P_q\}$, we can calculate the scalar product between $P$ and $v$, which represents the projection length of $P$ on $\cal L$. Based on the scalar product result, we can order $\{P_1, \dots, P_k, P_q\}$. Suppose the ordering based on the scalar product is $\{P_{\phi(1)}, P_{\phi(1)}, \dots, P_{\phi(x)}, P_q, P_{\phi(x+1)}, \dots, P_{\phi(k)}\}$. We have the two hyper-rectangles $R_1=rect(P_{\phi(1)}, P_{\phi(1)}, \dots, P_{\phi(x)})$ and $R_2=rect(P_{\phi(x+1)}, \dots, P_{\phi(k)})$. It is straightforward that $DST_{min}(P_q, R_1)=DST(P_q, P_{\phi(x)})$ and $DST_{min}(P_q, R_2)=DST(P_q, P_{\phi(x+1)})$. Because for any $P\in \{P_1, \dots, P_k\}$, $DST(P_q, P)>\theta$, $R_1$ and $R_2$ are DBHs.

We need exactly two DBHs.
}

\begin{figure}
\begin{center}
	\includegraphics[width=0.45\textwidth]{./figs/line.eps}
	\caption{How to construct DBHs When the $LF$-points lie on one single line}
	\label{fig:line}
\end{center}
\end{figure} 

\noindent{\bf When d>2.} Unfortunately, Theorem \ref{th:clique} can not be extended to the case of $d>2$. An example is shown in Figure \ref{fig:3d}, in which the MBHs $R(v_i, v_j)$ (i.e., the MBH of $v_i$ and $v_j$),  $R(v_i, v_k)$, and $R(v_j, v_k)$ are DBHs. However, the MBH   $R(v_i, v_j, v_k)$ is not a DBH, as it includes a point $w$, with $DST(p, w)<\theta$. 
{\bf WHAT'S THE CLIQUE ACCORDING TO THEOREM 4.1?}
{\em The clique according to THEOREM 4.1 contains $v_i, v_j, v_k$.}

\begin{figure}
\begin{center}
	\includegraphics[width=0.45\textwidth]{./figs/rect4.eps}
	\caption{An example of 3-d space}
	\label{fig:3d}
\end{center}
\end{figure}

For the Euclidean space of $d>2$  dimensions, we build the graph $G$ in the same way as for the $d=2$ case. However, the clique construction method is slight different. In particular, when constructing the cliques by using the same heuristic  algorithm for the $d=2$ case,   for each clique $C$, and for each candidate point $P$ that is going to be added to $C$, we check if $MinD(P, C') > \theta$, where $C'$ is the MBH decided by the points $C\cup \{P\}$ {\bf DOUBLE CHECK}. If $MinD(P, C') \leq \theta$, we delete the edges  between $P$ and any point in $C$ from $G$. Otherwise, we keep $C'$ and move on to the remaining points in $G$. This guarantees that MBHs constructed from the cliques must be DBHs.  

%As the problem is very close to the clique partition problem, even in the multidimensional space, we focus our attention on the solutions of clique partition. 

\nop{
The heuristic algorithms for clique partition can be put into four categories. We briefly introduce them.

{\bf Category 1: Linear programming.} The graph partition problem on $G$ is equivalent to the graph coloring problem on $\bar{G}$, where $\bar{G}$ is the complement graph. The graph coloring problem can be expressed by a integer programming formulation. Obviously, this is a non-convex problem. By relaxing the condition to linear programming problem, we can find the coloring number. However, to solve the semidefinite programming problem on a medium-sized graph, the time complexity could be prohibitive \cite{dukanovic2008semidefinite}. 

{\bf Category 2: Local search.} The idea is to get an initial local optimal solution, and among the neighboring solutions, find the extension which gives the best result. Two vertex permutations are neighbors if they do not conflict in graph coloring \cite{dukanovic2008semidefinite}. 

{\bf Category 3: Metaheuristics.} Use techniques such as backtracking to prevent from being trapped into a poor initial solution. There are mainly two types of logic: {\em annealing} \cite{homer1996experiments} and {\em tabu} \cite{hertz1987using}.

{\bf Category 4: Decide the order of vertex to put into clique.} This solution is based on the following theorem. The idea is to define a sorting metric to arrange the vertices such that by following the order we repeatedly increase the size of a clique until it is maximal \cite{tseng1986automated}. In that case, we create a new clique and do the same operation to it. There are algorithms based on static ordering and dynamic ordering. 
}

Below we explain the details of DBH construction when $d>2$. 
For each string $s_q$ and its $LF$-strings $L$, we firstly construct the graph $G=(V,E)$ from $L$. Then we initialize each clique with a single node whose degree is the minimum. {\bf AS YOU SAID EACH CLIQUE, DO YOU KNOW HOW MANY CLIQUES ARE CONSTRUCTED?} This is because the vertex of small degree tend to be difficult to be put into one clique. Thus, we tackle with the difficult ones first. In Line 6 and 7, we define $U_1$ and $U_2$, the union of which is the set of vertices that are not in any partition yet. Specifically, $U_1$ contains the set of vertices that have the potential to be added to the current clique $C$, because every vertex in $U_1$ is connected with all the nodes in $C$ in the graph $G$. If the clique's size is larger than 1, Theorem \ref{th:clique} may not be valid. To fix the problem, we need to delete some edges from $G$ and update $U_1$ as in Line 11 to 18. This step ensures that if we merge any vertex in $U_1$ to $C$, the newly generated hyper-rectangle is still distant from $p$. In Line 19, from all the vertices in $U_1$, we pick the one of smallest $deg_{U_2}$ as it is the one that is most difficult to be grouped into other cliques. {\bf THIS BASICALLY EXPLAIN THE PARAGRAPH BEFORE.}

We have $O(b^3)$ Algorithm \ref{alg:partition} to group $FH_2(s)$ into a small number of hyper-rectangles, where $b=|FH_2(s)|$.

\nop{
\noindent{\bf Step 1.} $\mathcal{C}=\emptyset$. Create a clique $C$ with the node of minimum degree.

\noindent{\bf Step 2.} If the clique $C$'s size is greater than or equal to $2$ and the dimensionality $m > 2$, Theorem \ref{th:clique} is not valid. To fix the problem, we need to delete some edges from $G$. We define $U_1$ and $U_2$ as two sets of vertices. Formally,
\begin{equation}
U_1=\{u|\forall u\in V \text{ } \forall C \in \bar{C} u\not\in C \text{ and } \forall v\in V \text{ } \exists C\in\bar{C} \text{ } s.t. \text{ } (u,v)\in E \}.
\label{eq:u1}
\end{equation}
\begin{equation}
U_2=\{u|\forall u\in V \text{ } \forall C \in \bar{C} u\not\in C \text{ and } \exists v\in V \text{ } \exists C\in\bar{C} \text{ } s.t. \text{ } (u,v)\not\in E \}.
\label{eq:u2}
\end{equation}
For any $u\in U_1$, we check if we update $C'=C\cup\{u\}$, $DST_{min}(rect(C'),p)\leq \theta$. If it is the case, we need to remove all edges $(u,v)$ from $E$ for all $v \in C$. 

After that, among all vertices in $U_1$, we select $u$ with minimum $d_{U_2}(u)$, where 
\begin{equation}
d_{U_2}(u)=|\{v|v\in U_2 \text{ and } (u,v)\in E\}|.
\end{equation}
We update $C=C\cup\{u\}$ accordingly.

\noindent{\bf Step 3.} If $U_1$ is empty, which means that there is no more possible vertex that we can add to the current clique $C$, we add $C$ to $\cal{C}$. We repeat Step 1 and 2 on the vertices that are not included in any clique, until every vertex is contained in a clique. Then the hyper-rectangles defined by the points in the cliques are the hyper-rectangles that we use to minimize the client's verification cost. Now we can minimize the verification complexity of $FH_2(s)$ by finding minimum number of cliques to cover all vertices in $G$. 
}
}

\nop{
\begin{algorithm}
\caption{Build a small number of DBHs to cover $TP$-strings}
\label{alg:partition}
\begin{algorithmic}[1]
	\REQUIRE The query point $P_q$, all the $TP$-points $\{p_1, \dots, p_t\}$
	\ENSURE The set of DBHs $\mathcal{R}$
	\STATE Construct $G=(V, E)$ from $\{p_1, \dots, p_t\}$
	\STATE $\mathcal{C}=\emptyset$	
	\WHILE{$V\neq \emptyset$}
		\STATE Find the node $v$ with smallest degree $deg(v)$
		\STATE Set a new clique $C=\{v\}$
		\STATE $U_1=\{u|\forall u\in V \text{, } \forall C \in \mathcal{C} \text{, } u\not\in C \text{ and } \forall v\in C \text{ } (u,v)\in E \}$
		\STATE $U_2=\{u|\forall u\in V \text{, } \forall C \in \mathcal{C} \text{, } u\not\in C \text{ and } \exists v\in C \text{ } (u,v)\notin E \}$
		\WHILE{$U_1 \neq \emptyset$}
		    \IF{$d > 2$}
			\IF{$|C|\geq 2$}
				\FORALL{$u \in U_1$}
					\STATE $C'=C\cup \{u\}$
					\IF{$DST_{min}(rect(C'), p) \leq \theta$}
						\STATE $U_1=U_1-\{u\}$
						\STATE $U_2=U_2\cup \{u\}$
						\STATE Remove $(u, v)$ from $E$ for each $v\in C$
					\ENDIF
				\ENDFOR
			\ENDIF 
			\ENDIF
			\STATE Pick $u \in U_1$ with smallest $deg_{U_2}(u)=|\{v|v\in U_2 \text{ and } (u, v)\in E\}|$
			\STATE $V=V-\{u\}$
			\STATE $C=C\cup \{u\}$
			\STATE Update $U_1$ and $U_2$
		\ENDWHILE
		\STATE $\mathcal{C}=\mathcal{C}\cup C$
	\ENDWHILE
	\STATE $\mathcal{R}(s)=\{MBH(C)|C\in \mathcal{C}\}$	
	\RETURN $\mathcal{R}(s)$
\end{algorithmic}
\end{algorithm}
}

\nop{
\subsubsection{Handling False Positives by Embedding}
We denote the set of false hits that are covered by any DBH as $FH^{DBH}$.
Due to the fact that the embedding methods may introduce false positives (i.e., the embedding points of dissimilar strings may become close in the Euclidean space), there may exist false hit strings whose embedding points do not exist in any DBH. These residue false hit strings are put into a set, denoted as $FH^{FP}$. Therefore, for a given string $s$, its false hit strings are resided in a non-leaf maximal false hit tree, in a DBH, or in $FH^{FP}$. 
{\bf IS THIS FINISHED?} 
{\em The strings in $FH^{FP}$ can not be verified based on the $MFT$s or DBHs. The server can only put them in $VO$ and let the client compute the pair-wise edit distane between $s_q$ and the strings in $FH^{FP}$ to check the completeness.}
}

Now we are ready to describe $VO$ construction by the E-$VS^2$ approach.  Given a dataset $D$ and a query string $s_q$, 
let $R$ and $F$ be the similar strings and false hits of $s_q$ respectively. $VS^2$ approach groups $F$ into C-strings and NC-strings.  E-$VS^2$ approach further groups C-strings into FP-strings and DBH-strings. Then E-$VS^2$ constructs $VO$ from $R$, NC-strings, FP-strings, and DBH-strings. Formally, 
%definition of VO
\vspace{-0.05in}
\begin{definition}
{\em 
Given a query string $s_q$, let $R$ be the returned similar strings of $s_q$. Let NC be the NC-strings, and DS the DBH-strings.  Let $T$ be the $MB$-tree, $MF$ be the maximum false hit trees of NC. Let $\cal R$ be the set of DBH constructed from DBH. Then the $VO$ of $s_q$ consists of: (i) string $s$, for each $s\in D - NC - DS$; 
(ii) a pair $(N, h^{1\rightarrow f})$ for each non-leaf $MF$ that is rooted at node $N$, where $N$ takes the format of $[N_b, N_e]$, with $[N_b, N_e]$ the string range associated with $N$, and $h^{1\rightarrow f}=h(h_{C_1}||\dots||h_{C_f})$, with $C_1, \dots, C_f$ being the children of $N$; 
(iii) $\cal R$; and 
(iv) a pair $(s, p_R)$ for each $s\in DS$, where $p_R$ is the pointer to the DBH in $\cal R$ that covers the Euclidean point of $s$; 
Furthermore, in $VO$, a pair of square bracket is added around the strings  pairs that share the same parent in $T$. \qed }
\label{def:evss-vo}
\end{definition}
\vspace{-0.1in}
\begin{example}
\label{exp:evs}
{\em 
To continue with our running example in Example \ref{exp:vs}, recall that the query string is $s_1$. The similar strings $R=\{s_1,s_3, s_5\}$. The NC-strings $NC =\{s_7, s_8, s_9\}$. The C-strings $C = \{s_2, s_4, s_6, s_{10}, s_{11}, s_{12}\}$.
Consider the embedded Euclidean space shown in Figure \ref{fig:exampleevs2} (b). Apparently $s_4$ is a FP-string as $dst(p_q, p_4)<\theta$. So the DBH-strings are \{$s_2$, $s_6$, $s_{10}$, $s_{11}$, $s_{12}\}$. The DBHs of these DBH-strings are shown in the rectangles in Figure \ref{fig:exampleevs2} (b). The $VO$ of query string $s_q$ is
\begin{eqnarray*}
VO(s_q)&=&\{(((s_1, (s_2, p_{R_2}), s_3), (s_4, s_5, (s_6, p_{R_2}))),([s_7, s_9], h^{7\rightarrow 9}),\\
&&((s_{10}, p_{R_2}), (s_{11}, p_{R_1}), (s_{12}, p_{R_1}))), \{R_1, R_2\}\},
\end{eqnarray*}
where $h^{7\rightarrow 9}=h(h(s_7)||h(s_8)||h(s_9)))$. \qed
}
\end{example}
\vspace{-0.05in}
\begin{table*}
\begin{center}
\begin{small}
\begin{tabular}{|c|c|c|c|}
\hline
Phase & Measurement & $VS^2$ & E-$VS^2$ \\\hline
\multirow{2}{*}{Pre-processing}& Time & $O(n)$ &  $O(cdn^2)$ \\\cline{2-4}
& Space & $O(n)$ & $O(n)$ \\\hline \hline 
\multirow{2}{*}{VO construction}& time & $O(n)$ & $O(n+n_{DS}^3)$ \\\cline{2-4}
& $VO$ Size & $(n_R+n_C)\sigma_S+n_{MF}\sigma_M$ & $(n_R+n_F)\sigma_S + n_{MF}\sigma_M+n_{DBH}\sigma_D$ \\\hline\hline
Verification & Time & $O((n_R+n_{MF}+n_C)C_{Ed})$ & $O((n_R+n_{MF}+n_F)C_{Ed}+n_{DBH}C_{El})$ \\\hline
\end{tabular}
\end{small}
\vspace{-0.05in}
\caption{Complexity comparison between $VS^2$ and E-$VS^2$\\
(
$n$: \# of strings in $D$; 
$c$: a constant in [0, 1]; 
$d$: \# of dimensions of Euclidean space; 
$\sigma_S$: the average length of the string; 
$\sigma_M$: Avg. size of a $MB$-tree node; 
$\sigma_D$: Avg. size of a DBH; 
$n_R$: \# of strings in $R$;  
$n_C$: \#  of C-strings; \\
$n_F$: \#  of FP-strings; 
$n_{DS}$: \#  of DBH-strings; 
$n_{DBH}$: \# of DBHs; 
$n_{MF}$: \# of $MF$ nodes;\\ 
$C_{Ed}$: the complexity of an edit distance computation; 
$C_{El}$: the complexity of Euclidean distance calculation.)
}
\vspace{-0.05in}
\label{tb:complexity}
\end{center}
\vspace{-0.3in}
\end{table*}
\vspace{-0.1in}
\subsubsection{VO-based Authentication}
After receiving $(R,VO)$ from the server, the client uses $VO$ to verify if $R$ is sound and complete. The verification of E-$VS^2$ consists of four steps. The first three steps are similar to the three steps of the $VS^2$ approach. The fourth step is to re-compute a set of Euclidean distance. Next, we discuss the four steps in details.

\noindent{\bf Step 1 \& 2}: these two steps are exactly the same as Step 1 \& 2 of $VS^2$.

\noindent{\bf Step 3: Re-computing necessary edit distance:} Similar to $VS^2$, first, for each $s\in R$, the client verifies $DST(s, s_q)\leq\theta$.  Second, for each range $[N_b, N_e]\in VO$, the client verifies whether $DST_{min}(s_q, N)>\theta$, where $N$ is the corresponding $MB$-tree node associated with the range $[N_b, N_e]$. 
The only difference of the E-$VS^2$ approach is that for each FP-string $s$, the client verifies if $DST(s_q, s)>\theta$. If not, the client concludes that $R$ is incomplete.
%In E-$VS^2$, the number of FP-strings is much smaller than the number of C-strings in $VS^2$. Thus the required number of edit distance computation is much smaller. In addition to the FP-strings, for each $s\in R$, the client checks if $DST(s_q, s)\leq \theta$. If it is, the client concludes that $R$ is sound.

\noindent{\bf Step 4: Re-computing of necessary Euclidean distance:} Step 3 only verifies the dissimilarity of FP- and NC-strings. In this step, the client verifies the dissimilarity of DBH-strings. 
First, for each pair $(s, p_R)\in VO$, the client checks if $P_s\in R$, where $P_s$ is the embedded point of $s$, and $R$ is the DBH that $p_R$ points to. If all pairs pass the verification, the client ensures that the DBHs in $VO$ covers the embedded points of all the DBH-strings. Second, for each DBH $R\in VO$, the client checks if $dst_{min}(P_q, R)>\theta$. If it is not, the client concludes that the returned results are not correct. 
Otherwise, third, for each similar string $s\in R$, the client checks if there exists any DBH that includes $P_s$, where $P_s$ is the embedded point of $s$. If there does, the client concludes that the results violate soundness. 

Note that we do not require to re-compute the edit distance between any DBH-string and the query string. Instead we only require the computation of the Euclidean distance between a set of DBHs and the embedded points of the query string. Since Euclidean computation is much faster than that of the edit distance. Therefore E-$VS^2$ saves much verification cost compared with $VS^2$. More comparison of $VS^2$ and E-$VS^2$ can be found in Section \ref{sc:vs2vsevs2}. 
%If it is, the client knows that all the points of the DBH-strings are dissimilar to $s_q$. Based on the contractive property, the client is assured that all DBH-strings are $\theta$-dissimilar to $s_q$.
\vspace{-0.05in}
\begin{example}
\label{exp:evs2}
{\em
Following the running example in Example \ref{exp:evs}, after calculating the root signature $sig$' from $VO$ and compares it with the signature $sig$ received from the data owner, the client performs the following computations: 
(1) for $R=\{s_1, s_3, s_5\}$, compute the edit distances between $s_1$ and any string in $R$; 
(2) for NC-strings $NC=\{s_7, s_8, s_9\}$, compute $DST_{min}(s_q, N_3)$; 
(3) for FP-strings $FP=\{s_4\}$, compute the edit distance $DST(s_q, s_4)$; and 
(4) for $DBR$-strings $DS = \{s_2, s_6, s_{10}, s_{11}, s_{12}\}$, compute $dst_{min}(P_q, R_1)$ and $dst_{min}(P_q, R_2)$.
Compared with the $VS^2$ approach in Example \ref{exp:vs}, which computes 9 edit distances, E-$VS^2$ computes 4 edit distances, and 2 Euclidean distances. Recall that the computation of Euclidean distance is much cheaper than that of the edit distance. \qed
}
\end{example}
\vspace{-0.05in}
\vspace{-0.1in}
\subsection{Security Analysis}
Similar to the security discussion for the $VS^2$ approach (Sec. \ref{sc:vssrobust}), the server may perform three types of cheating behaviors, i.e., tampered values, soundness violation, and completeness violation. E-$VS^2$ can catch the cheating behaviors of tampered values by re-computing the root signature of $MB$-tree (i.e., Step 2 of the authentication procedure). Next, we mainly focus on how to catch the correctness and completeness violation by the E-$VS^2$ approach. 

\noindent{\bf Soundness.} To violate soundess, the server returns $R' = R \cup FS$, where $FS\subseteq F$ (i.e., $FS$ is a subset of false hits). 
We consider two possible ways that the server constructs $VO$: 
(Case 1.) the server constructs the $VO$ $V$ of the correct result $R$, and returns $\{R', V\}$ to the client; and 
Case 2.) the server constructs the $VO$ $V'$ of $R'$, and returns $\{R', V'\}$ to the client. 
Note that the strings in $FS$ can be NC-strings, FP-strings, and DBH-strings. Next, we discuss how to catch these three types of strings for both cases. 

For Case 1, for any NC-string $s\in FS$, $s$ can be caught in the same way as by $VS^2$ approach (i.e., Step 1 of authentication). For any FP-string $s\in FS$, $s$ can be caught by re-computing the edit distance $DST(s, s_q)$ (i.e., Step 3 of authentication).
For any DBH-string $s\in FS$, let $P_s$ be the embedded point of $s$. Since the VO $V$ is constructed from the correct $R$ and $F$, there must exist a DBH in $V$ that includes $P_s$. Therefore,  $s$ can be caught by  verifying whether there exist any DBH that includes the embedded points of $s$ (Step 4 of the authentication procedure). 

For Case 2, the NC-strings and FP-strings in $FS$ can be  caught in the same way as in Case 1. The DBH-strings  cannot be caught by Step 4 now, as: (1) the DBHs constructed from a subset of DBH-strings are still DBHs, and (2) no string in $FS$ is included in a DBH in the VO $V'$.  However, these DBH-strings are treated as FP-strings (i.e., not included in any DBH), and thus can be caught by Step 3. 

\noindent{\bf Completeness.} To violate the completeness requirements, the server returns returns $R' = R - SS$, where $SS\subseteq R$. Let $V$ and $V'$ be the VO constructed from $R$ and $R'$ respectively. We again consider the two cases as for the discussion of soundness violation. 

For Case 1 (i.e.,  the server returns \{$R', V\}$), any string $s\in SS$ is a FP-string, as $s$ is not included in $R$. Then by calculating $DST(s, s_q)$, (i.e., Step 3 of the authentication procedure), the client discovers that $s$ is indeed similar to $s_q$ and thus catch the incomplete results. 

For Case 2 (i.e,  the server returns $\{R',V'\}$), any string $s\in SS$ is either a FP-string or a DBH-string. If $s$ is treated as a FP-string, it can be caught by recomputing of edit distance (Step 3 of the authentication procedure). If $s$ is a DBH-string, then its Euclidean point $P_s$ must be included in a DBH $R$. We have the following theorem.
\vspace{-0.05in}
\begin{theorem}
{\em
Given a query string $s_q$ and a DBH $R$ (i.e., $dst_{min}(P_q, R)>\theta$), then for any string $s$ such that $s\approx s_q$, adding the embedded point of $s$ to $R$ must change $R$ to be a non-DBH.
}
\label{th:evss_security}
\end{theorem}
\vspace{-0.05in}
The proof of Theorem \ref{th:evss_security} is straightforward. 
Let $R'$ be the hyper-rectangle after adding $P_s$ to $R$. It must be true that $dst_{min}(P_q, R)\leq dst(P_q, P) \leq dst(s_q, s) \leq \theta$. Then $R'$ cannot be a DBH. 

Following Theorem \ref{th:evss_security}, for any string $s\in SS$, including its embedded point into any DBH $R$ will change $R$ to be a non-DBH. Then the client can easily catch it by re-computing the euclidean distance (Step 4 of verification).

\subsubsection{$VS^2$ Versus E-$VS^2$}
\label{sc:vs2vsevs2}

In this section, we compare $VS^2$ and E-$VS^2$ approaches in terms of the time and space of the pre-processing, $VO$ construction, and verification phases. The comparison results are summarized in Table \ref{tb:complexity}. 
Regarding the $VO$ construction overhead at the server side, as shown in our empirical study, $n_{DS}<<n$, thus the overhead $O(n+n_{DS}^3)$ of the E-$VS^2$ approach is comparable to $O(n)$ of the $VS^2$ approach. 

Regarding the VO size, the VO size of the $VS^2$ approach is calculated as the sum of two parts: (1) the total size of the similar strings and C-strings (in string format), and (2) the size of $MF$ nodes. Note that $\sigma_M = 2\sigma_S + |h|$, where $|h|$ is the size of a hash value. In our experiments, it turned out that $\sigma_M /\sigma_S \approx 10$. 
The VO size of the E-$VS^2$ approach is calculated as the sum of three parts: (1) the total size of the similar strings and FP-strings (in string format), (2) the size of $MF$ nodes, and (3) the size of DBHs. 
Our experimental results show that $\sigma_D >>\sigma_S, \sigma_M$.  

Regarding the complexity of verification time, 
note that $n_C=n_F+n_{DS}$, where $n_C$, $n_F$ and $n_{DS}$ are the number of C-strings, FP-strings, and DBH-strings respectively. 
Usually, $n_{DBH}<n_{DS}$ as a single DBH can cover the Euclidean points of a large number of DBH-strings. 
Also note that $C_{Ed}$ (i.e., complexity of an edit distance computation) is much more expensive than $C_{El}$ (i.e, the complexity of Euclidean distance calculation). Our experiments show that the time to compute one single edit distance can be 20 times of computing one Euclidean distance. 
Therefore, compared with $VS^2$, E-$VS^2$ significantly reduces the verification overhead at the client side. We admit that it increases the overhead of pre-processing at the data owner side and the $VO$ construction at the server side. 
We argue that, as the pre-processing phase is a one-time operation, the cost of constructing the embedding function can be amortized by a large number of queries from the client.

\section{Multi-string Similarity Search}
\label{sc:mstring}

So far we discussed the authentication of single-string similarity search queries. To authenticate a multi-string query $Q (S, \theta)$ that contains multiple unique search strings $S = \{s_1, \dots, s_\ell$\}, a straightforward solution is to create VO for each string $s_i\in S$ and its similarity result $R_i$. Apparently this solution may lead to $VO$ of large sizes in total. Thus, we aim to reduce the size of VOs. 
Our $VO$ optimization method consists of two main strategies: 
(1) optimization by triangle inequality; and 
(2) optimization by overlapping dissimilar strings. 

\subsection{Optimization by Triangle Inequality}

It is well known that the string edit distance satisfies the {\em triangle inequality}, i.e. $|DST(s_i,s_k)-DST(s_j,s_k)|\leq DST(s_i, s_j)\leq DST(s_i,s_k)+DST(s_j,s_k)$. 
%Therefore, given a query string $s_q$, let $F$ be the false hits of $s_q$. Then for any two dissimilar strings $s_i, s_j\in F$, if there exists a string $s_k\in F$ such that  $|DST(s_i,s_k)-DST(s_j,s_k)|>\theta$,
Therefore, consider two query strings $s_{q_1}$ and $s_{q_2}$, assume the the server has executed the similarity search of $s_{q_1}$ and prepared the VO of the results. Then consider the authentication of the search results of string $s_{q_2}$,  for any string $s\in D$ such that   $DST(s,s_{q_1})- DST(s_{q_1},s_{q_2})> \theta$,
there is no need to prepare the proof of $s$ showing that it is a false hit for $s_{q_2}$. A straightforward method is to remove $s$ from the $MB$-tree $T$ for $VO$ construction for query string $s_{q_2}$. Now the question is whether removing $s$ always lead to $VO$ of smaller sizes. 
%{\bf IS IT possible? AN EXAMPLE to show that removing one string from $T$ may result in more $NLFs$?} One straightforward solution is that the server constructs two $MB$-trees $T_1$ and $T_2$, such that $T_1$ contains $s_k$ but $T_2$ does not. The server then constructs VOs from $T_1$ and $T_2$ separately, and picks the one that delivers the $VO$ of smaller size. Though correct, this may bring significant cost. Therefore, we aim to decide whether to include a string $s_k$ in the $MB$-tree $T$ for $VO$ construction. 
We have the following theorem. 
\vspace{-0.05in}
\begin{theorem}
\label{theorem:ti}
{\em
Given a string $s$ and a $MB$-tree $T$, let $N\in T$ be the corresponding node of $s$, and $N_P$ be the parent of $N$ in $T$. Let $C(N_P)$ be the set of children of $P$ in $T$, and $N_P'$ be the node constructed from $C(N_P)\backslash N$ (i.e., the children nodes of $N_p$ excluding $N$). Then if $N_P$ is a non-candidate, it must be true that $N_P'$ must be a non-candidate. 
}
\end{theorem}

Let $[N_b, N_e]$ be the range of $N_P$. The proof of Theorem  \ref{theorem:ti} considers two possible cases: (1) 
$s\neq N_b$ and $s\neq N_e$. Then removing $s$ will not change $N_b$ and $N_e$, which results in that both $N_p'$ and $N_p$ have the same range $[N_b, N_e]$.
(2)  $s= N_b$ or  $s =N_e$. Then removing $s$ from $C(N_P)$ changes the range of $N_P$ to be $[N_b', N_e']$. Note that it must be true $[N_b', N_e']\subseteq [N_b, N_e]$. Therefore, $N_P'$ must be a non-candidate. 

Based on Theorem \ref{theorem:ti}, removing the $MB$-tree nodes that correspond to any dissimilar string  indeed does not change the structure of the $MB$-tree. Therefore, we can optimize the VO construction procedure by removing those strings covered by the triangle inequality from the $MB$-tree. 

Another possible optimization by triangle inequality is that for any two query strings $s_{q_1}$ and $s_{q_2}$, for any string $s\in D$ such that   $DST(s,s_{q_1})+DST(s_{q_1},s_{q_2})\leq \theta$, then the client does not need to re-compute $DST(s, s_{q_2})$ proving that $s_j$ is a similar string in the results. We omit the details due to the space limit. 

%In Algorithm \ref{alg:multi_construct} we show the pseudocode  of VO constuction at the server side. Intuitively, $M[i,j]=k$ if and only if $DST(s_i,s_k)+DST(s_j,s_k) \leq \theta$, while $N[i,j]=k$ if and only if $|DST(s_i,s_k)-DST(s_j,s_k)| > \theta$. $R^{TRI}(s_i)$ ($FH^{TRI}(s_i)$ resp.) includes the set of similar (dissimilar resp.) strings to $s_i$ which can be verified by the triangle inequality. From Line 9 to 20, the server constructs $R^{TRI}(s_i)$ and $FH^{TRI}(s_i)$ based on the matrix $M$ and $N$. In Line 21 to 32, the server updates $M$ and $N$ using the similarity search result of $s_i$. In Algorithm \ref{alg:multi_verify}, we display the pseudocode for the client to do verification. Basically, for any string $s_i$, the client does not need to check the verification objects for $R^{TRI}(s_i)$ and $FH^{TRI}(s_i)$. Instead, the verification can be accomplished by using an easy arithmetic based on the triangle inequality.

\subsection{Optimization by Overlapped Dissimilar Strings}

Given multiple-string search query, the key optimization idea is to merge the VOs of individual query strings. 
This is motivated by the fact that any two query strings $s_i$ and $s_j$ may share a number of dissimilar strings. These shared dissimilar strings can enable to merge the VOs of $s_i$ and $s_j$. 
Note that simply merging all similar strings of $s_i$ and $s_j$ into one set and constructing $MB$-tree of the merged set is not correct, as the resulting $MB$-tree may deliver non-leaf $MFs$ that are candidates to both $s_i$ and $s_j$. Therefore, given two query strings  $s_i$ and $s_j$ such that their false hits overlap, let $NC_i$ ($NC_j$, resp.) and $DBH_i$ ($DBH_j$, resp.) be the $NC$-strings and $DBH$-strings of $s_i$ ($s_j$ resp.), the server finds the overlap of $NC_i$ and $NC_j$, as well as the overlap between $DBH_i$ and $DBH_j$. Then the server  constructs $VO$ that shares the same data  structure on these overlapping strings. In particular, first, given the overlap $O_1 = NC_i \cap NC_j$, the server constructs the non-leaf $MFs$ from $O_1$, and include the constructed $MFs$ in the VOs of both $s_i$ and $s_j$. Second, given the overlap $O_2 = DBH_i \cap DBH_j$, the server constructs the $DBHs$ from $O_2$, and include the constructed $DBHs$ in the VOs of both $s_i$ and $s_j$.

%two strings $s_i, s_j \in Q$ share the same part of VO, either $MFT$ or $DBR$. That is to say, there may exist a $MFT$ node $N$ whose lower bound distances to both $s_i$ and $s_j$ are greater than $\theta$. One way for the server to prepare the VOs for $s_i$ and $s_j$ is to include $N$'s triple $(N_{min}, N_{max}, h_N)$ in $VO(s_i)$ and $VO(s_j)$, and then send them to the client. Similarly, there may also exist a $DBR$ $R$ which is similar to neither $p_i$ nor $p_j$. However, the repeated transmission of the same object lead to unnecessary bandwidth consumption and storage space. In order to resolve this issue, we propose that in the multiple string case, the server only needs to prepare a repository of unique verification objects (unique $MFT$s and $DBR$s). If two strings $s_i$ and $s_j$ share the same object, the server only needs to include a pointer to the common object in both $VO(s_i)$ and $VO(s_j)$.

\nop{
{\bf DO WE MERGE VOs ON-FLY DURING CONSTRUCTION OR AFTER THEY ARE CONSTRUCTED?}

Apparently for any two strings that have the same VOs, they can share the VOs for the authentication. Next, we discuss how to deal with the overlapping VOs. First, we formally define the overlapping MHTs and DBRs.

\begin{definition}
Given two maximal false hit trees ($MHTs$) rooted at node $N_i$ and $N_j$, we say these trees overlap if $cover(N_i) \cap cover(N_j)\neq \oslash$. 
\end{definition}

\begin{definition}
Given two distant minimum bounding rectangles ($DBRs$) $R_i(l_1^i, u_1^i,  \dots, l_d^i, u_d^i)$ and $R_j(l_1^j, u_1^j,  \dots, l_d^j, u_d^j)$, we say $R_i$ and $R_j$ overlap if $\exists k\in[1, d]$, $l_k^i\in[l_k^j, u_k^j]$ or $u_k^i\in[l_k^j, u_k^j]$. 
\end{definition}
{\bf DOUBLE CHECK!}

For any two strings $s_i$ and $s_j$, we say their VOs $VO_i$ and $VO_j$ {\em overlap} if they contain overlapping MHTs or overlapping DBRs. Next, we discuss how to merge overlapping VOs.

First,  due to the construction procedure of $MB^{ed}$-trees, any two overlapping $MHTs$ indeed follow the strict hierarchical relationship, i.e., the node $N_i$ must be the ancestor of the node $N_j$. Therefore, for any two VOs that contain overlapping $MHTs$, there is no merge of these $MHTs$. 

Now let us consider the overlapping $DBRs$. Given two query strings $s_i$ and $s_j$, let $R_i(l_1^i, u_1^i,  \dots, l_d^i, u_d^i)$ and $R_j(l_1^j, u_1^j,  \dots, l_d^j, u_d^j)$ be the $DBRs$ that belongs to the VO of $s_i$ and $s_j$ respectively. 
The server constructs $R^{ij} (l_1, u_1,  \dots, l_d, u_d)$ where $\forall k\in[1, d]$, $l_k=min(l_k^i, l_k^j)$, and  $u_k=max(u_k^i, u_k^j)$. Then the server checks whether $R^{ij}$ is a DBR to $s_i$ and $s_j$. {\bf IS MERGE DESTINED TO FAIL?}. If it does, then the server replaces $R_i$ and $R_j$ with $R^{ij}$ for both $s_i$ and $s_j$. 
The server keeps looking for all overlapping $DBRs$, merge them if they can, until all $DBRs$ are not mergable. Then the server constructs the VOs from the merging results.

{\bf ANY DATA STRUCTURE FOR SHARED MHT \& DBR?}
}

\nop{
\begin{algorithm}
\begin{small}
\caption{ $Multi\_VO\_Construct(Q, D, \theta)$)
\label{alg:multi_construct}
}
\begin{algorithmic}[1]
\REQUIRE{the set of search strings $Q=\{s_1, s_2, \dots, s_x\}$, the string dataset $D=\{s_1, s_2, \dots, s_n\}$, the similarity threshold value $\theta$}
\ENSURE{The set of verification objects for $Q$}
%\STATE

\STATE{let $M,N$ be two $x*n$ matrix}
\FOR{$i \in [1,x]$}
  \FOR{$j \in [1,n]$}
    \STATE{$M[i,j]=-1$}
    \STATE{$N[i,j]=-1$}
  \ENDFOR
\ENDFOR
\FORALL{$S_i \in Q$}
  \FORALL{$S_j \in R(S_i)$}
    \IF{$M[i,j] \neq -1$}
      \STATE{$R^{TRI}(s_i)=R^{TRI}(s_i)\cup \{(S_j, M[i,j])\}$}
      \STATE{$R(s_i)=R(s_i)-\{s_j\}$}
    \ENDIF
  \ENDFOR
  \FORALL{$S_j \in FH(s_i)$}
    \IF{$N[i,j] \neq -1$} 
      \STATE{$FH^{TRI}(s_i)=FH^{TRI}(s_i)\cup \{(S_j, N[i,j])\}$}
      \STATE{$FH(s_i)=FH(s_i)-\{s_j\}$}
    \ENDIF
  \ENDFOR
  
  \FORALL{$s_j, s_k \in R(s_i)$ and $j \neq k$} 
    \IF{$M[j,k]=-1$ AND $DST(s_i,s_j)+DST(s_i,s_k)\leq \theta$}
      \STATE{$M[j,k]=i$}
      \STATE{$M[j,k]=i$}
    \ENDIF
  \ENDFOR  
  \FORALL{$s_j \in FH(s_i)$, $s_k\in FH(s_i)\cup R(s_i)$}
    \IF{$N[j,k]=-1$ AND $|DST(s_i,s_j)-DST(s_i,s_k)|> \theta$}
      \STATE{$N[j,k]=i$}
      \STATE{$N[k,j]=i$}
    \ENDIF
  \ENDFOR
  \STATE{construct $VO(S_i)$ for $R(s_i)$ and $FH(s_i)$}
  \STATE{$VO(s_i)=VO(s_i)\cup R^{TRI}(s_i) \cup FH^{TRI}(s_i)$}
\ENDFOR

\RETURN{the set of VOs for $Q$}

\end{algorithmic}
\end{small}
\end{algorithm}

\begin{algorithm}
\begin{small}
\caption{ $Multi\_VO\_Verify(Q, \mathcal{VO}, \theta)$)
\label{alg:multi_verify}
}
\begin{algorithmic}[1]
\REQUIRE{the set of search strings $Q=\{s_1, s_2, \dots, s_x\}$, the set of VOs $\mathcal{VO}=\{VO(s_1), VO(s_2), \dots, VO(s_x)\}$, the similarity threshold value $\theta$}
\ENSURE{the correctness of the result}
%\STATE

\FORALL{$s_i \in Q$}
  \FORALL{$(s_j, k)\in R^{TRI}(s_i)$}
    \IF{$DST(s_i, s_k)+DST(s_j,s_k) > \theta$}
      \STATE{catch server's soundness cheating}
    \ENDIF
  \ENDFOR
  \FORALL{$(s_j, k)\in FH^{TRI}(s_i)$}
    \IF{$|DST(s_i, s_k)-DST(s_j, s_k)| \leq \theta$}
      \STATE{catch server's completeness cheating}
    \ENDIF
  \ENDFOR
\ENDFOR

\RETURN{the set of VOs for $Q$}

\end{algorithmic}
\end{small}
\end{algorithm}
}
\vspace{-0.1in}
\section{Authentication of Top-k Results}
\label{sc:ranking}

In this section, we discuss how to authenticate the top-$k$ similarity results. 
Formally, given a query string $s_q$ and a threshold value $\theta>0$, the query is to
find the top-$k$ similar strings of $s$, which are sorted by their distance to the query string $s_q$ in ascending order. 
In other words, the query returns a set of strings $R=\{s_1, s_2, \dots, s_k\}$, where $DST(s, s_i)\leq \theta$ and $DST(s, s_i)\leq DST(s, s_{i+1}), \forall i\in[1, k-1]$. 
Let $c$ be the number of strings that are similar to $s$ in terms of $\theta$. We consider two cases: (1) $k = c$ (i.e., all similar strings are ranked); and (2) $k<c$ (i.e., only a subset of similar strings are returned).

\noindent{\bf k=c.} Besides verifying the soundness and completeness of the returned strings, the authentication procedure checks if the ranking is correct. Both the soundness and completeness verification can be achieved by our $VS^2$ or E-$VS^2$ approach. A straightforward solution to ranking authentication is to calculate the pairwise distance $DST(s_q, s)$ for any $s\in R$, re-sort strings in $R$ based on their distances, and compare the ranking results with $R$. Since the calculation of pairwise distance is required for soundness verification (for both $VS^2$ and E-$VS^2$), the authentication can be done by one additional sorting step. 

\noindent{\bf k<c.} Given a query string $s_q$ and the top-$k$ similar strings $R=\{s_1, s_2, \dots, s_k\}$, the client needs to verify:

{\em Requirement 1.} No returned strings are tampered with, i.e., $\forall s_i\in R, i\in [1, k]$, $s_i\in D$; 

{\em Requirement 2.} $\forall i \in [1, k-1]$,  $DST(s_i, s_q)\leq\theta$, and $DST(s_i, s_q) \leq DST(s_{i+1}, s_q)$; 

{\em Requirement 3.} No genuine top-$k$ results are missing, i.e.,  $\forall s_i\not\in R$, $DST(s_i, s_q) \geq DST(s_k, s_q)$. 

Requirement 1 can be easily verified by restoring the digests of the root signature of the $MB$-tree. Requirement 2 can be verified by two steps: (1) verifying the soundness of $R$ by either our $VS^2$ or E-$VS^2$ approach; (2) re-compute $DST(s_i, s_q)$ for each $s_i\in R$, and sort the strings of $R$ by their distance in an ascending order. 
Requirement 3 can be verified by checking whether for all the false fits in the VO, their distance to $s_q$ is longer than $DST(s_k, s_q)$. This is equivalent to checking whether the server returns sound the complete results with regard to the similarity threshold $\theta= DST(s_k, s_q)$. Therefore, we can use $VS^2$ and E-$VS^2$ approaches to construct VO and do verification, by using the similarity threshold as $DST(s_k, s_q)$. 
%We find that the verificaton for KNN similarity search can be an extension to our $VSS$ and $C-VSS$ approaches. When the server finishes searching for $R(s)$ of the top-k similar strings, he can calculate a threshold value $\theta=max\{DST(s,s')|s'\in R(s)\}$ and prepare the verification object using $\theta$. After receiving the VO, the client first checks if $\theta$ is correctly set by the server by calculating the distance between $s$ and $R(s)$. If $\theta$ is correct, then the client can check the result correctness of $R(s)$. If the test is passed, the client can be sure that $R(s)$ is correct as no other strings is has a distance value smaller than $\theta$, which is the maximum distance for a string to be $s$'s k nearest neighbor.

\vspace{-0.1in}
\section{Experiments}
\label{sc:exp}
%In order to evaluate the efficiency of our string similarity search authentication approaches, we run an extensive set of experiments on real-world datasets. 
In this section, we report the experiment results.

\subsection{Experiment Setup}
%machine
\noindent{\bf Datasets and queries.} We use two real-world datasets collected by {\em US Census Bureau} in 1990\footnote{\url{http://www.census.gov/topics/population/genealogy/data/1990_census/1990_census_namefiles.html}}: (1) the {\em LastName} dataset that contains 88799 last names. The maximum length of a name is 13, while the average is 6.83; 
and (2) the {\em FemaleName} dataset including 4475 actor names. The maximum length of a name is 11, while the average length is 6.03. 
We designed ten single-string similarity search queries. For the following results, we report the average of the ten queries. 

\noindent{\bf Parameter setup.} 
The parameters include: (1) string edit distance threshold $\theta$, (2) dimension $d$ of the embedding space, and (3) the fanout $f$ of $MB$-tree nodes  (i.e., the number of entries that each node contains). The details of the parameter settings can be found in Table \ref{table:psetting} in Appendix. We also include the details of the query selectivity in Table \ref{tb:selectivity} in Appendix. 
%The threshold $\theta$ is valued from 2 to 6. Each threshold value leads to a different number of similar strings. More details of the thresholds can be found in Table \ref{tb:selectivity} in Appendix. Regarding the dimension $d$, the {\em LastName} dataset is embedded to the Euclidean spaces of dimension $5, 10, 15, 20$ and $25$. {\bf Due to the large size of the {\em Actor} dataset, we only embed it to spaces with $2, 4, 6, 8$ and $10$ dimensions.(MAY NEED TO CHANGE)} Regarding the fanout value $f$, we use $f$= 10, 15, 20, 25, 30, 35, and 40.  

Due to space limits, the details of experimental environments are in Appendix. For the following discussions, we use the following notations:
(1) $n$: the number of strings in the dataset, (2) $n_R$: the number of similar strings, (3) $n_{MF}$: the number of $MFs$, (4) $n_C$: the number of C-strings, (5) $n_F$: the number of FP-strings, (6) $n_{DBH}$: the number of DBHs, and (7) $n_{DS}$: the number of DBH-strings. We use $\sigma_S$ to indicate the average string size, and $\sigma_M$ ($\sigma_S$, resp.) as the size of $MB$-tree node (DBH, resp.). 

\subsection{VO Construction Time}
We measure the VO construction time at the server side by both $VS^2$ and E-$VS^2$ methods. 

\begin{figure*}[!ht]
\centering
\begin{tabular}{ccc}
 \includegraphics[width=0.29\textwidth]{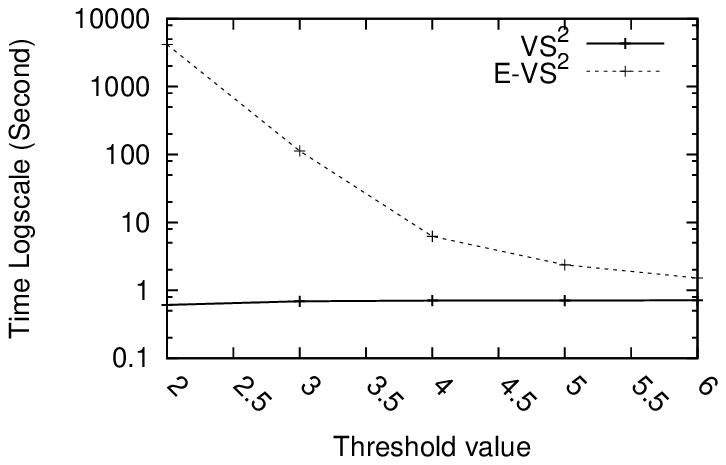}
 &
\includegraphics[width=0.29\textwidth]{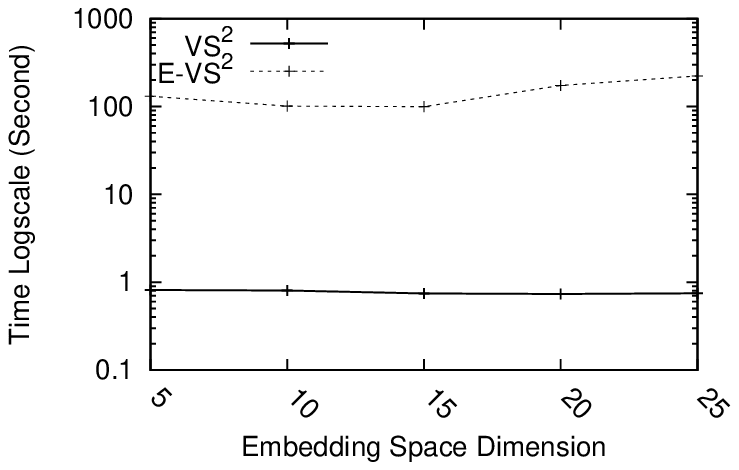}
&
\includegraphics[width=0.29\textwidth]{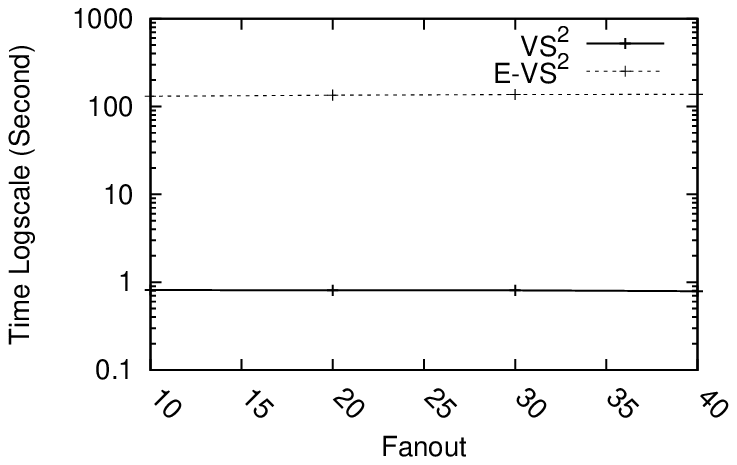}
 \\
 {\scriptsize (a) Various $\theta$ ($d=5, f=10$)}
 & 
 {\scriptsize (b) Various $d$ ($\theta=3, f=10$)}
 &
 {\scriptsize (c) Various $f$ ($\theta=3, d=5$)}
\end{tabular}
\vspace{-0.05in}
    \caption{VO construction time ({\em LastName} dataset) }
    \label{fig:ctime}
    \vspace{-0.1in}
\end{figure*}

\noindent{\bf The impact of $\theta$.} In Figure \ref{fig:ctime} (a), we show the VO construction time  with regard to different $\theta$ values. 
First, we observe that there is an insignificant growth of VO construction time by $VS^2$ when $\theta$ increases. For example, on the {\em LastName} dataset, the time increases from 0.7085 seconds to 0.7745 seconds when $\theta$ changes from 2 to 6. This is because the number of $MF$s reduces with the increase of $\theta$. For example, when $\theta=2$, $n_{MF}=390$, while $\theta=6$, $n_{MF}=0$. Consequently, $VS^2$ visits more $MB$-tree nodes to construct the VO. However, compared with $n$, the increase of $n_{MF}$ is not significant. Therefore, the VO construction time of $VS^2$ slightly increases with the growth of $\theta$.  
Second, we observe the dramatic decrease in VO construction of the E-$VS^2$ when $\theta$ increases. This is because with the increase of $\theta$ value, $n_R$ rises, while $n_{DS}$  reduces sharply (e.g., on {\em LastName} dataset, when $\theta=2$, $n_R=136$, $n_{DS}=13928$; when $\theta=3$, $n_R=1104$, $n_{DS}=3568.6$. Since the complexity of VO construction is cubic to $n_{DS}$, the total VO construction time decreases intensively when $n_{DS}$ decreases. 

\nop{
\begin{figure}[!ht]
\centering
\begin{tabular}{cc}
 \includegraphics[width=0.25\textwidth]{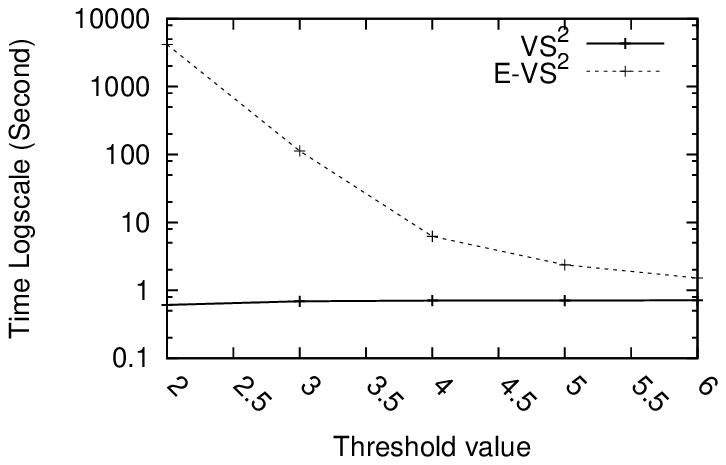}
 &
 \includegraphics[width=0.25\textwidth]{./exp//exp/lastname//proof_preparation_time_vs_threshold.eps}
 \\
 {\scriptsize (a) {\em LastName} dataset}
 & 
 {\scriptsize (b) {\em /exp/lastname/} dataset}
\end{tabular}
    \caption{VO construction time v.s. $\theta$ ($d=5, f=10$)}
    \label{fig:construction_vs_theta}
\end{figure}
}

\noindent{\bf The impact of $d$.} We change the dimension $d$ of the embedding space and observe its impact on VO construction time. The results are displayed in Figure \ref{fig:ctime} (b). On both datasets, $d$ has no effect on $VS^2$, because $VS^2$ does not interact with the embedding space. However, the time performance of E-$VS^2$ increases with the $d$ value. Intuitively, larger dimension leads to smaller $n_{F}$, and thus larger $n_{DS}$. As the VO construction time of E-$VS^2$ is cubic to $n_{DS}$, the VO construction time increases when $d$ increases.

\nop{
\begin{figure}[!ht]
\centering
\begin{tabular}{cc}
 \includegraphics[width=0.25\textwidth]{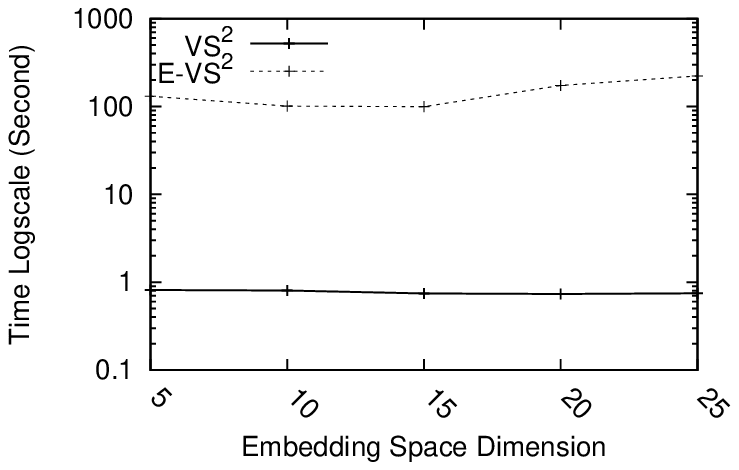}
 &
 \includegraphics[width=0.25\textwidth]{./exp//exp/lastname//proof_preparation_time_vs_k.eps}
 \\
 {\scriptsize (a) {\em LastName} dataset}
 & 
 {\scriptsize (b) {\em /exp/lastname/} dataset}
\end{tabular}
    \caption{VO construction time v.s. $d$ ($\theta=3, f=10$)}
    \label{fig:construction_vs_d}
\end{figure}
}

\noindent{\bf The impact of $f$.} For both approaches, we observe that the VO construction time is stable for various $f$ fanout values (Figure \ref{fig:ctime} (c)). This is because the complexity of VO construction is decided by the number of strings in the dataset and the number of DBH-strings. Both numbers do not change by $f$ values. 
%We use different fanout in the $MB$-tree and get relatively stable VO construction time, which is shown in Figure \ref{fig:construction_vs_f}. For $VS^2$, we observe (1) the total number of nodes in the $MB$-tree decreases with larger fanout values, and (2) the decrease in the number of $MF$s (when $f=10$, $n_{MF}=147.9$; when $f=40$, $n_{MF}=0$.). These two factors counteract and lead to a small decrease in the total VO construction time. While for E-$VS^2$, it is the opposite. With a node having more children, the number of $MF$s becomes smaller. As a consequence, the number of DBH-strings increases. But the increment in the number of DBH-strings is not intensive (when $f=10$, $n_{DS}=3528$; when $f=40$, $n_{DS}=3586$). Therefore, the VO construction does not increase much.

\nop{
\begin{figure}[!ht]
\centering
\begin{tabular}{cc}
 \includegraphics[width=0.25\textwidth]{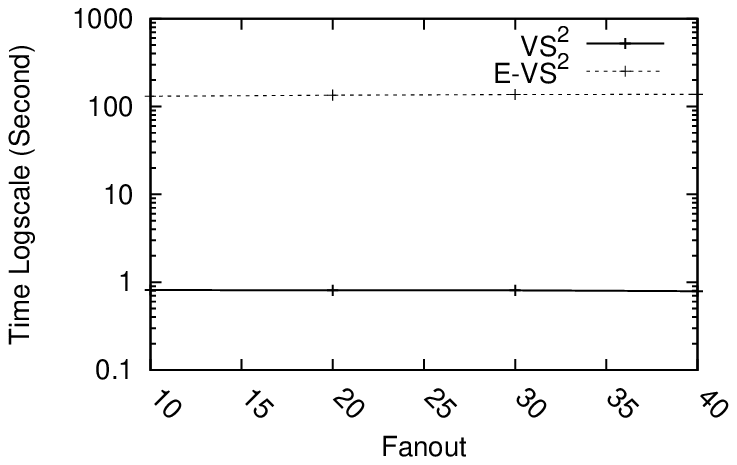}
 &
 \includegraphics[width=0.25\textwidth]{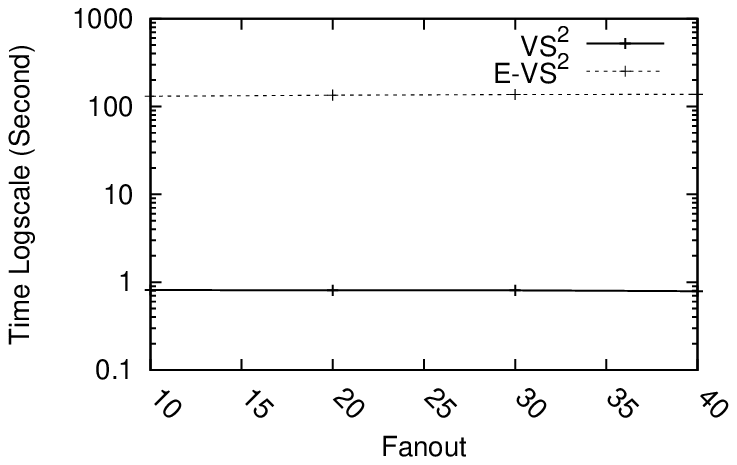}
 \\
 {\scriptsize (a) {\em LastName} dataset}
 & 
 {\scriptsize (b) {\em /exp/lastname/} dataset}
\end{tabular}
    \caption{VO construction time v.s. $f$ ($\theta=3, d=5$)}
    \label{fig:construction_vs_f}
\end{figure}
}

\subsection{VO Size}
We measure the size of the VO constructed by the $VS^2$ and E-$VS^2$ approaches. 

\begin{figure*}[!ht]
\centering
\begin{tabular}{ccc}
 \includegraphics[width=0.29\textwidth]{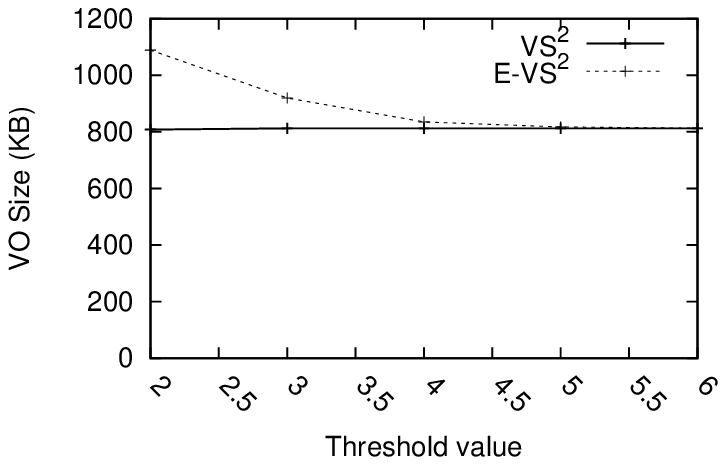}
 &
\includegraphics[width=0.29\textwidth]{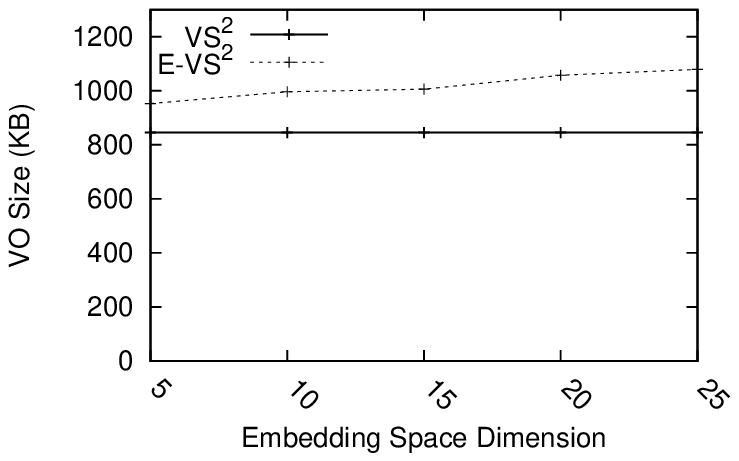}
&
\includegraphics[width=0.29\textwidth]{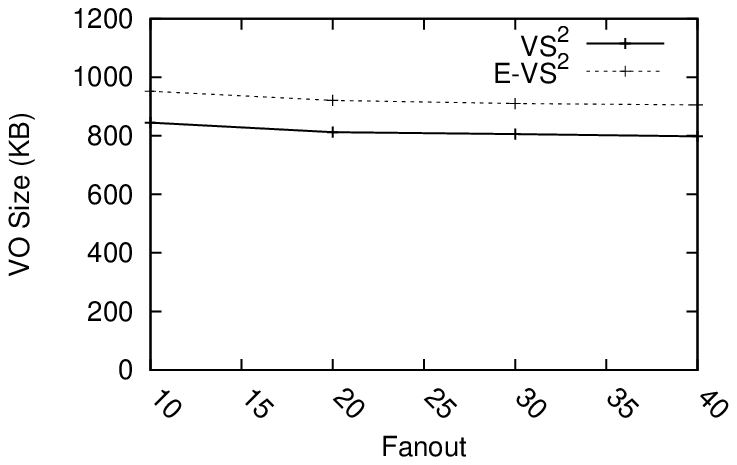}
 \\
 {\scriptsize (a) Various $\theta$ ($d=5, f=10$)}
 & 
 {\scriptsize (b) Various $d$ ($\theta=3, f=10$)}
 &
 {\scriptsize (c) Various $f$ ($\theta=3, d=5$)}
\end{tabular}
\vspace{-0.05in}
    \caption{VO size ({\em LastName} dataset) }
    \label{fig:vosize}
    \vspace{-0.1in}
\end{figure*}

\noindent{\bf The impact of $\theta$.} The results are shown in Figure \ref{fig:vosize} (a). The first observation is that the VO size of $VS^2$  with $\theta$ value. Apparently, larger $\theta$ values lead to more similar strings (i.e., larger $n_R$), fewer $MF$s (i.e., smaller $n_{MF}$), and fewer C-strings (i.e., smaller $n_{C}$). As the VO size is decided by $(n_R+n_C)\sigma_S+n_{MF}\sigma_M$, where  $\sigma_M /\sigma_S \approx 10$, and the increase of $(n_R+n_C)\sigma_S$ is cancelled out by the decrease of $n_{MF}\sigma_M$, the VO size of $VS^2$ approach stays relatively stable. 
On the contrary, for the E-$VS^2$ approach, since $\sigma_D>>\sigma_S, \sigma_M$, the VO size is dominately decided by $n_{DBH}$. We observe that the slight increase of $\theta$ values lead to sharp decrease of $n_{DS}$ and thus $n_{DBH}$ (e.g., when $\theta=2$, $n_{DBH}=1272$; when $\theta=6$, $n_{DBH}=5$). Thus the VO size decreases significantly for larger $\theta$.
When $\theta \geq 4$, the VO size of E-$VS^2$ is very close to that of $VS^2$.

\nop{
\begin{figure}[!ht]
\centering
\begin{tabular}{cc}
 \includegraphics[width=0.25\textwidth]{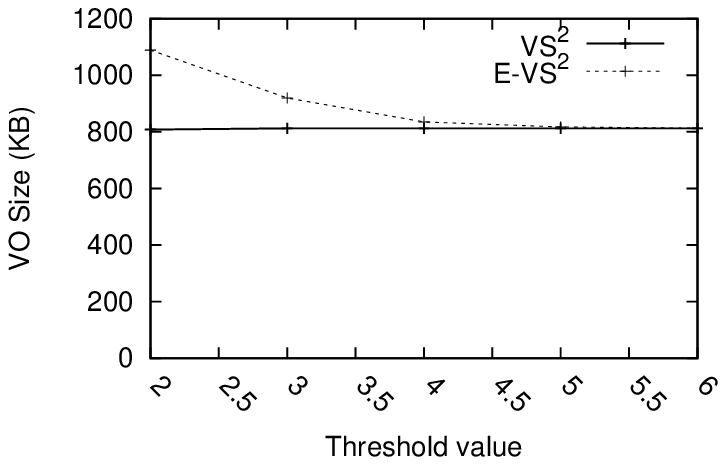}
 &
 \includegraphics[width=0.25\textwidth]{./exp//exp/lastname//proof_size_vs_threshold.eps}
 \\
 {\scriptsize (a) {\em LastName} dataset}
 & 
 {\scriptsize (b) {\em /exp/lastname/} dataset}
\end{tabular}
    \caption{VO size v.s. $\theta$ ($d=5, f=10$)}
    \label{fig:size_vs_theta}
\end{figure}
}

\noindent{\bf The impact of $d$.} From the results reported in Figure \ref{fig:vosize} (b), we observe that VO size of $VS^2$ is not affected by various $d$ values. This is straightforward as $VS^2$ does not rely on embedding. On the other hand, the VO size of E-$VS^2$ increases with larger dimension value, since larger dimension leads to smaller $n_{F}$, larger $n_{DS}$, and thus larger $n_{DBH}$. Furthermore, when $d$ increases, the average $DBH$ size increases too. These two factors contribute to the growth of VO size for the E-$VS^2$.

\nop{
\begin{figure}[!ht]
\centering
\begin{tabular}{cc}
 \includegraphics[width=0.25\textwidth]{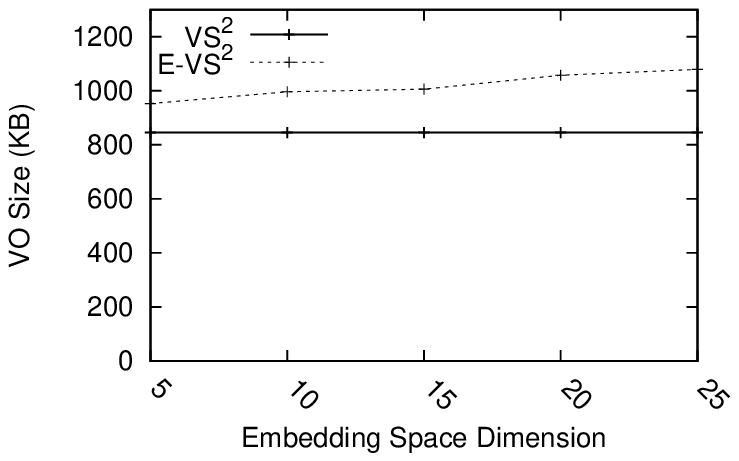}
 &
 \includegraphics[width=0.25\textwidth]{./exp//exp/lastname//proof_size_vs_k.eps}
 \\
 {\scriptsize (a) {\em LastName} dataset}
 & 
 {\scriptsize (b) {\em /exp/lastname/} dataset}
\end{tabular}
    \caption{VO size v.s. $d$ ($\theta=3, f=10$)}
    \label{fig:size_vs_d}
\end{figure}
}

\noindent{\bf The impact of $f$.} As shown in Figure \ref{fig:vosize} (c), the VO size decreases with the growth of the fanout $f$. 
First, for $VS^2$, recall that its $VO$ size is calculated as $(n_R+n_C)\sigma_S+n_{MF}\sigma_M$. 
When $f$ increases, $n_R$ is unchanged. 
Meanwhile, $n_C$ slightly grows with $f$ (e.g., when $f=10$, $n_C=86,216$, while when $f=40$, $n_C=87,695$).
Furthermore, when $f$ increases, $n_{MF}$ decreases (when $f=10$, $n_{MF}=148$, while when $f=40$, $n_{MF}=0$). Also $\sigma_M /\sigma_S \approx 10$. Therefore, the decrease of $n_{MF}$ leads to smaller VO size. 
For E-$VS^2$, by which the VO size is measured $(n_R+n_F)\sigma_S + n_{MF}\sigma_M+n_{DBH}\sigma_D$, again $n_R$ is unchanged, while $n_{MF}$ decreases for larger $f$. However, 
$n_F$ and $n_{DBH}$ keep relatively stable with different $f$ (e.g., when $f=10$, $n_F=82688$, $n_{DBH}=80$; when $f=40$, $n_F=84108$, $n_{DBH}=80$). Therefore, 
the VO size by E-$VS^2$ approach decreases with the growth of $f$. 
\nop{
\begin{figure}[!ht]
\centering
\begin{tabular}{cc}
 \includegraphics[width=0.25\textwidth]{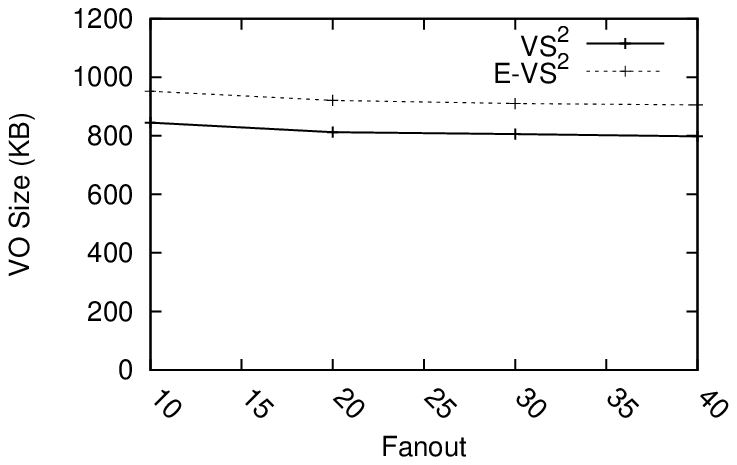}
 &
 \includegraphics[width=0.25\textwidth]{./exp//exp/lastname//proof_size_vs_fanout.eps}
 \\
 {\scriptsize (a) {\em LastName} dataset}
 & 
 {\scriptsize (b) {\em /exp/lastname/} dataset}
\end{tabular}
    \caption{VO size v.s. $f$ ($\theta=3, d=5$)}
    \label{fig:size_vs_f}
\end{figure}
}

Another observation is that the VO size of $VS^2$ is always larger than that of E-$VS^2$. This is straightforward as E-$VS^2$ has to include $DBH$s in VO, which contributes to a substantial portion of VO in terms of its size. Nevertheless, as $n_{DBH}$ is small in most cases (always smaller than 100), the additional VO size  required by E-$VS^2$ is not substantial compared with the total VO size. 

\subsection{VO Verification Time}

\begin{figure*}[!ht]
\centering
\begin{tabular}{ccc}
 \includegraphics[width=0.32\textwidth]{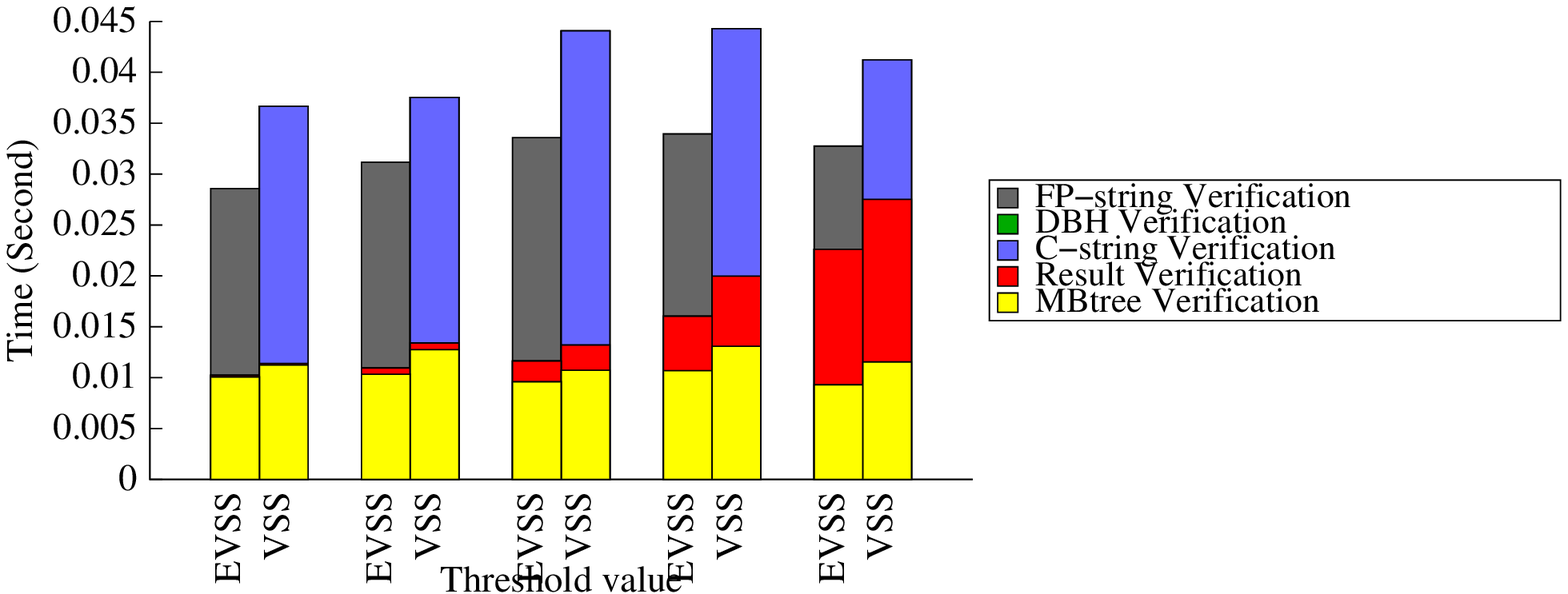}
 &
\includegraphics[width=0.32\textwidth]{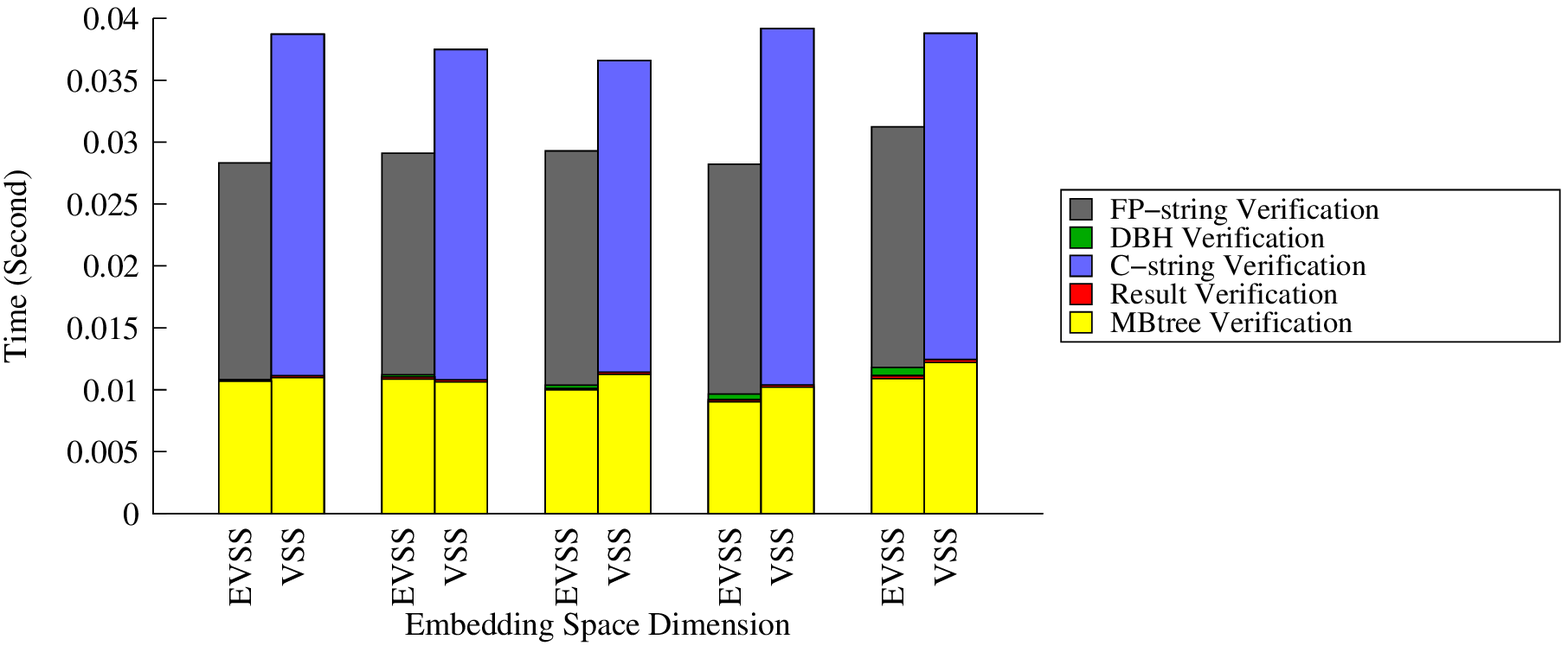}
&
\includegraphics[width=0.32\textwidth]{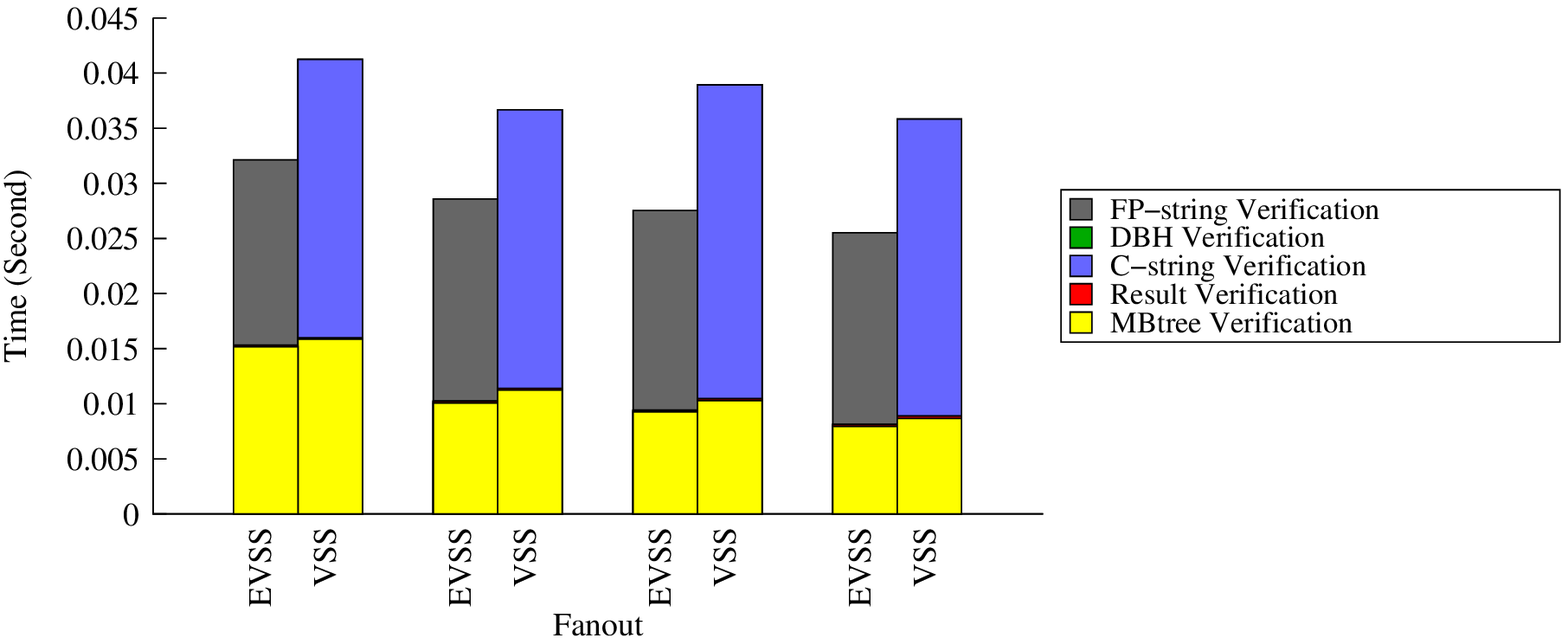}
 \\
 {\scriptsize (a) Various $\theta$ ($d=5, f=10$)}
 & 
 {\scriptsize (b) Various $d$ ($\theta=2, f=10$)}
 &
 {\scriptsize (c) Various $f$ ($\theta=2, d=5$)}
\end{tabular}
\vspace{-0.05in}
    \caption{VO verification time ({\em FemaleName} dataset) }
    \label{fig:vtime}
    \vspace{-0.1in}
\end{figure*}

In this section, we measure the VO verification time at the client side. We split the verification into five components: (1) time to verify the similarity of the returned similar strings (Result Verification), (2) time to re-construct the root signature of MB-tree and verifying NC-strings (MBtree Verification), (3) time to compute edit distance for C-strings (C-string Verification), (4) time to calculate the Euclidean distance for DBH-strings (DBH-string Verification), and (5) time to compute the edit distance for FP-strings (FP-string Verification). We use the {\em FemaleName} dataset and report these five components in details. 

Before we discuss specific parameters, an important observation is that the verification time of E-$VS^2$ can be as small as 75\% of $VS^2$. This proves that E-$VS^2$ can save the verification time at the client side significantly.

\noindent{\bf The impact of $\theta$.} From Figure \ref{fig:vtime} (a), we observe that, first, the MBtree Verification time keeps stable. This is because the number of $MBH$s is small (always smaller than 10). Even though the increase of $\theta$ decreases the number of $MBH$s, the time to re-construct the root signature does not increase much. 
Second, the Result Verification time (0.4\% - 32\% of the total verification time) increases sharply with $\theta$ as the number of similar strings increases fast with $\theta$. For example, when $\theta=2$, $n_R=31$, while when $\theta=6$, $n_R=2385$. 
Third, the C-string Verification time decreases when $\theta$ grows, since the number of C-strings drops fast with the rapid growth of similar strings. 
For $VS^2$, there is no $DBH-$ or $FP$-string, leading to zero DBH-string and FP-string Verification time. 
Regarding the E-$VS^2$ approach, we only discuss  DBH-string Verification time and FP-string Verification time, as the other components are the same as the $VS^2$ approach. Due to the efficient Euclidean distance calculation, the DBH-string Verification time is very small (smaller than 0.2\% of the total verification time). The FP-string Verification time decreases when $\theta$ grows, as a large portion of FP-strings become similar when $\theta$ increases. Overall, the total verification time of $E-VS^2$ increases when $\theta$ changes from 2 to 4, but keeps stable after that. This is because when $\theta$ increases from 2 to 4, $n_{DS}$ drops very fast (from 1349 to 45). The time saved by verifying the $DBH$s thus shrinks. When $\theta>4$, $n_{DS}$ does not change much (from 45 to 4). Thus the total verification time keeps  stable.

\nop{
\begin{figure}[!ht]
\centering
\begin{tabular}{cc}
 \includegraphics[width=0.25\textwidth]{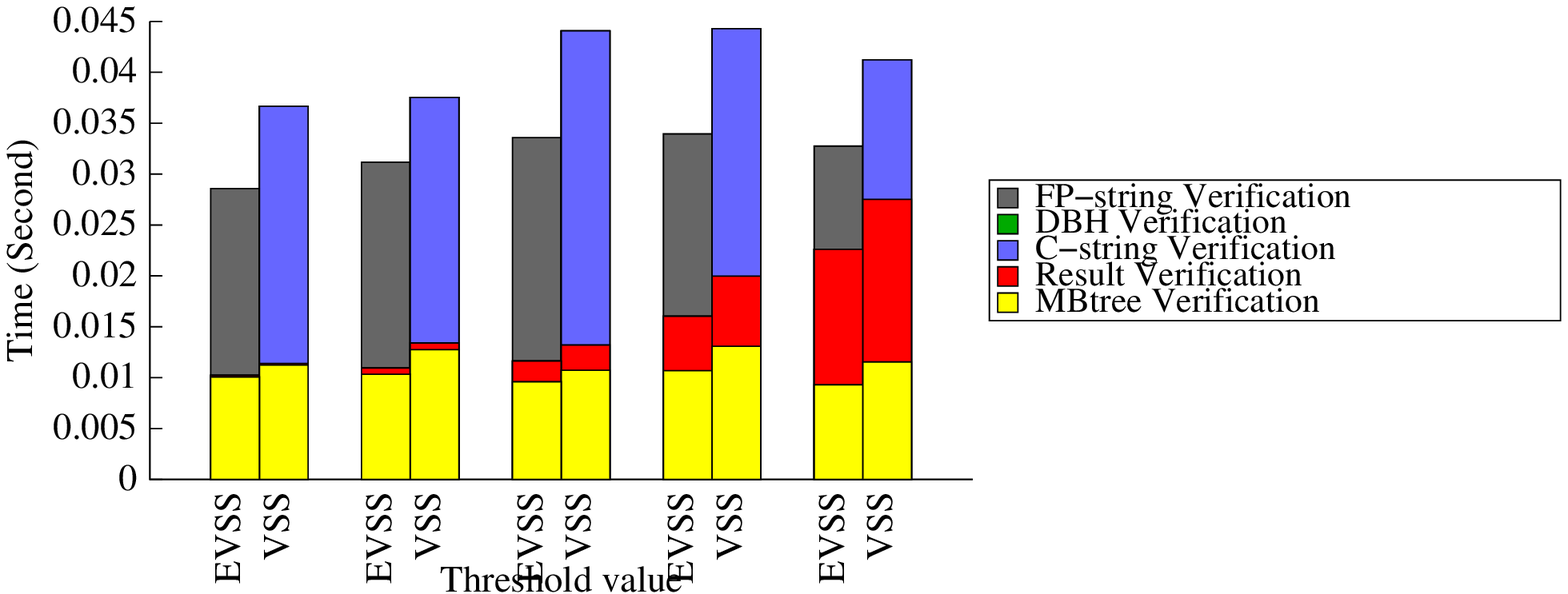}
 &
 \includegraphics[width=0.25\textwidth]{./exp//exp/lastname//proof_verification_time_vs_threshold.eps}
 \\
 {\scriptsize (a) {\em LastName} dataset}
 & 
 {\scriptsize (b) {\em /exp/lastname/} dataset}
\end{tabular}
    \caption{VO verification time v.s. $\theta$ ($d=5, f=10$)}
    \label{fig:verification_vs_theta}
\end{figure}
}
\noindent{\bf The impact of $d$.} We only discuss the verification time of E-$VS^2$ as the verification time of $VS^2$ does not rely on the dimension of the embedding space. According to the results shown in Figure \ref{fig:vtime} (b), for E-$VS^2$, the total verification time keeps stable with the increase of $d$. The reason is that the number of DBH-strings varies little much with the increase of $d$. When $d=5$, $n_{DS}=1349$; when $d=25$, $n_{DS}=1357$. This shows that $E-VS^2$ is efficient even for the high-dimension embedding space.
%In all circumstances, E-$VS^2$ is faster than $VS^2$ in terms of VO verification. Besides, the advantage of E-$VS^2$ is higher with larger $d$ value.

\nop{
\begin{figure}[!ht]
\centering
\begin{tabular}{cc}
 \includegraphics[width=0.25\textwidth]{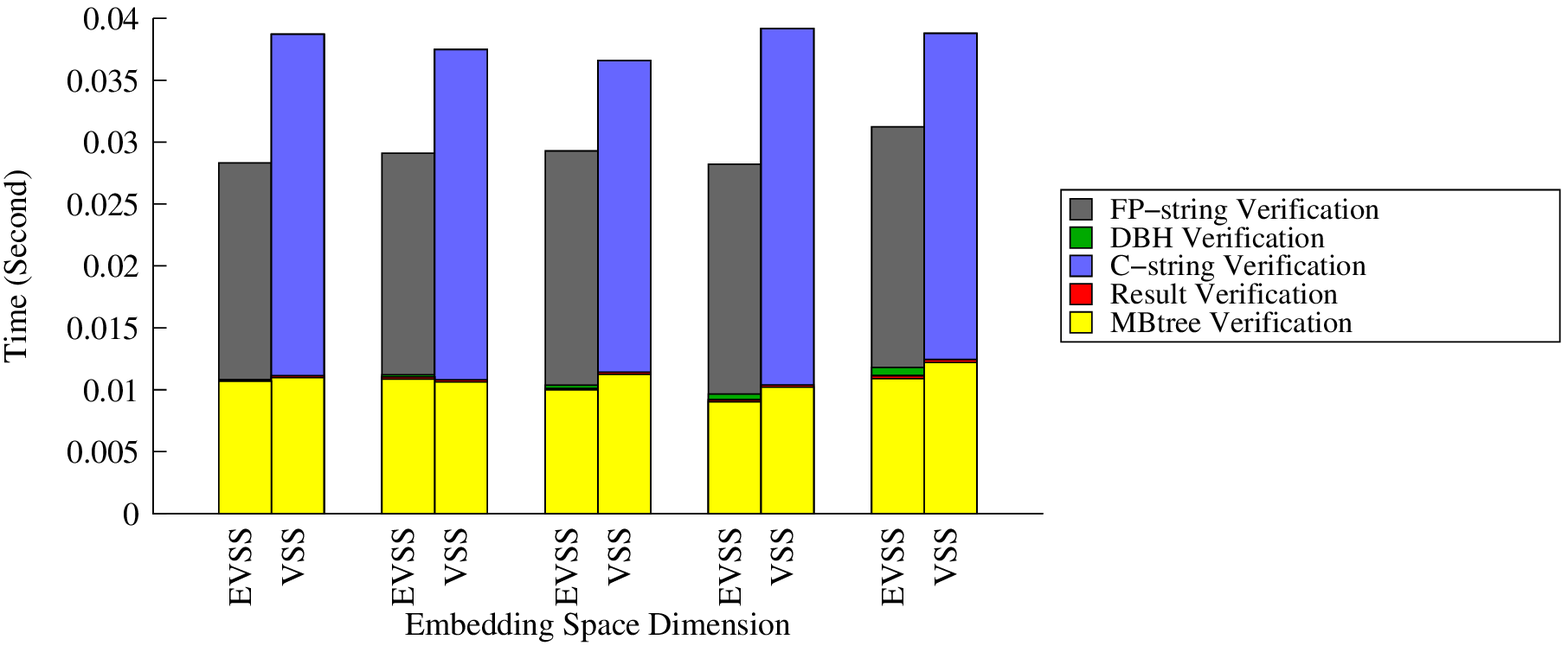}
 &
 \includegraphics[width=0.25\textwidth]{./exp//exp/lastname//proof_verification_time_vs_k.eps}
 \\
 {\scriptsize (a) {\em LastName} dataset}
 & 
 {\scriptsize (b) {\em /exp/lastname/} dataset}
\end{tabular}
    \caption{VO verification time v.s. $d$ ($\theta=3, f=10$)}
    \label{fig:verification_vs_d}
\end{figure}
}

\noindent{\bf The impact of $f$.} According to the results shown in Figure \ref{fig:vtime} (c), larger $f$ value results in shorter verification time for both approaches. The reason is that larger $f$ leads to a smaller $MB$-tree, and  thus small time to re-compute the tree's root signature. 
%While as the number of DBH-strings and $DBH$s keeps stable with different $f$ values, the time difference between $VS^2$ and E-$VS^2$ is also stable.

\nop{
\begin{figure}[!ht]
\centering
\begin{tabular}{cc}
 \includegraphics[width=0.25\textwidth]{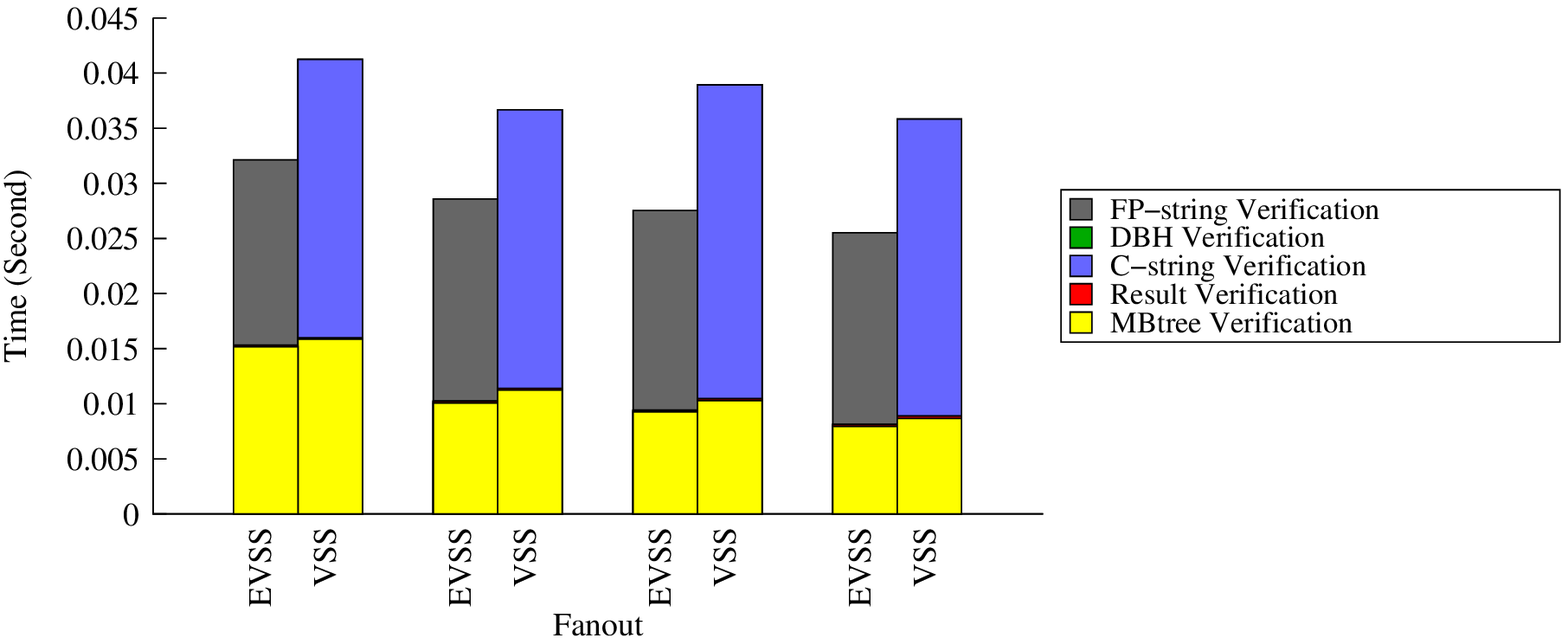}
 &
 \includegraphics[width=0.25\textwidth]{./exp//exp/lastname//proof_verification_time_vs_fanout.eps}
 \\
 {\scriptsize (a) {\em LastName} dataset}
 & 
 {\scriptsize (b) {\em /exp/lastname/} dataset}
\end{tabular}
    \caption{VO verification time v.s. $f$ ($\theta=3, d=5$)}
    \label{fig:verification_vs_f}
\end{figure}
}

%%%%%%%%%%%%%%%%%%%%journal version
%\subsection{Verification VS Local Search}

%\input{discussion}
%\input{exp}
\vspace{-0.1in}
\section{Conclusion}
\label{sc:concl}
In this paper, we designed two efficient authentication methods, namely $VS^2$ and E-$VS^2$, for outsourced string similarity search. Both $VS^2$ and E-$VS^2$ approaches are based on a novel authentication data structure named $MB$-tree that integrates both $B^{ed}$-tree and Merkle hash tree. The E-$VS^2$ approach further applies string embedding methods to merge dissimilar strings into smaller $VO$. Experimental results show that our methods authenticate similarity query searches efficiently. 
%and outperform the baseline method by a wide margin. In general, VA has a better communication efﬁciency and MA has a better computation efﬁciency. 
In the future, we will investigate how to design authentication methods for database with updates. We also plan to study how to authenticate the correctness of {\em privacy-preserving} string similarity search.
\section{Appendix}

\begin{figure*} [t!]
	\begin{center}
		\begin{tabular}{ccc}
		\includegraphics[width=0.3\textwidth]{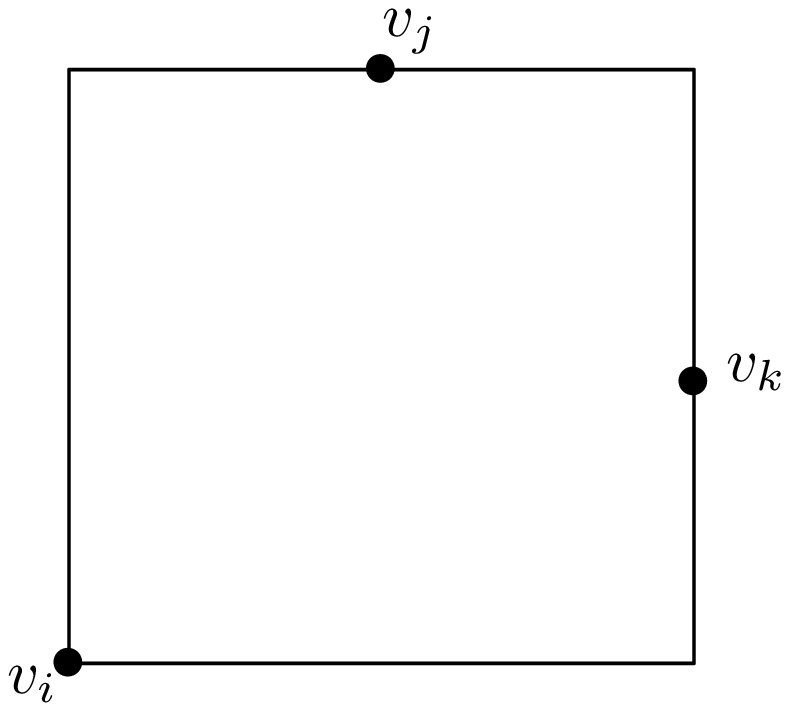}
		&
		\includegraphics[width=0.3\textwidth]{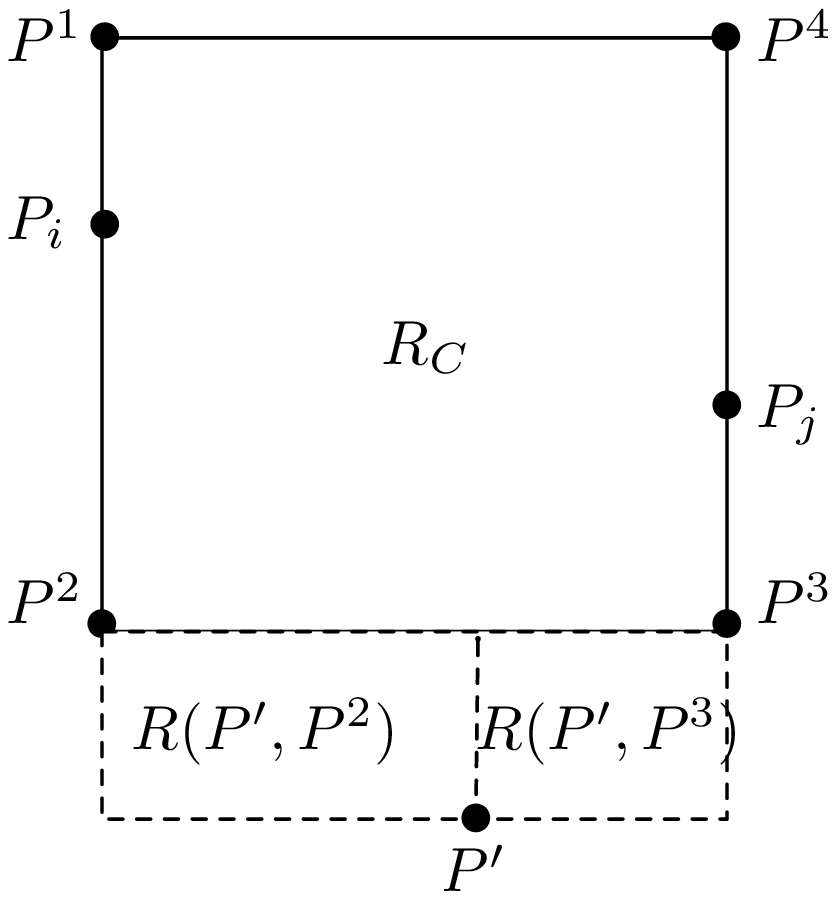}
		&
		\includegraphics[width=0.3\textwidth]{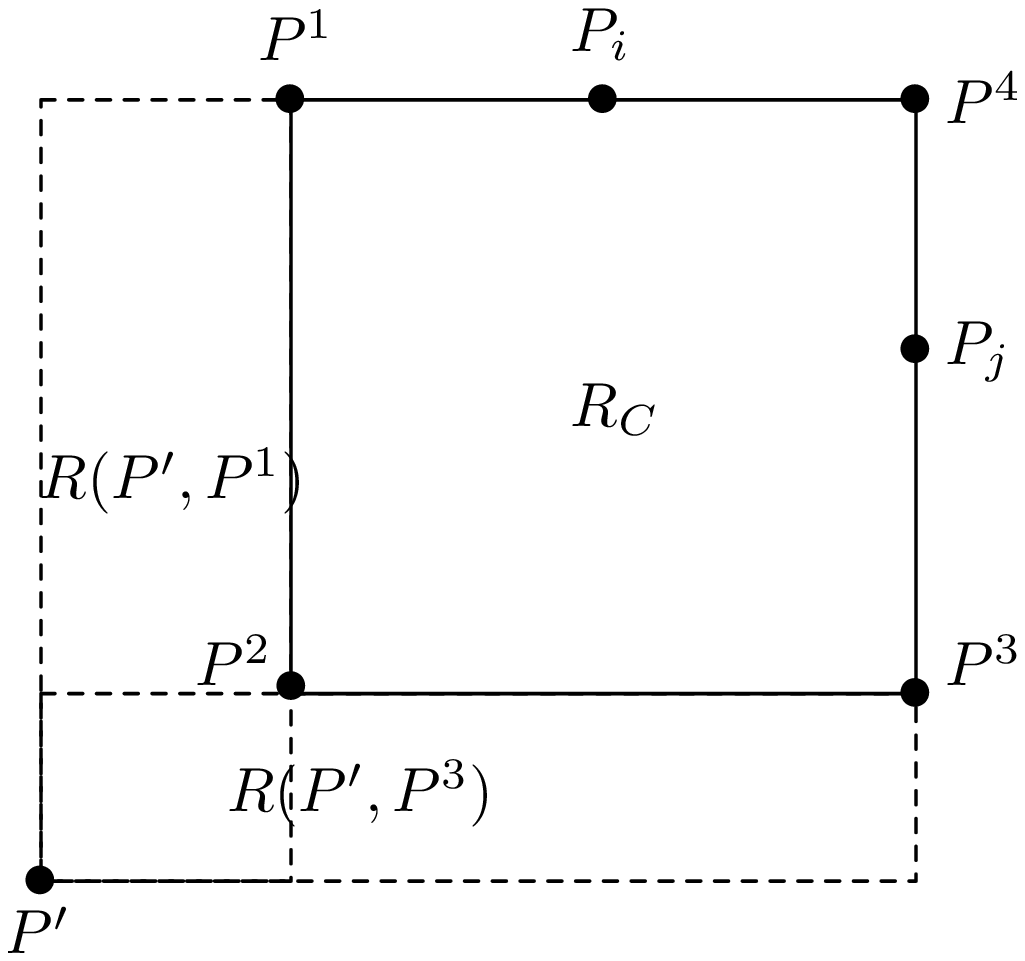}
		\\
		(a) Base case
		&
		(b) Case 2 of induction
		&
		(c) Case 3 of induction
		\end{tabular}
		\caption{Illustration: Proof of Theorem 5.4}
				\label{fig:rect}
	\end{center}	
	\end{figure*}

\subsection{Proof of Theorem 5.4}

%In the $2D$ space, given a string $s$, its false hits $LH(s)$, and the graph $G=(V, E)$ constructed from $LH(s)$, for any clique $C$ in $G$, let $R$ denote the MBH defined by the points in $C$. Then it must be true that $MinD(P, R) > \theta$. 

We construct a graph $G=(V,E)$ such that for each point $P_i \in\cal P$, it corresponds to a vertex $v_i\in V$. For any two vertices $v_i$ and $v_j$ that correspond to two points $P_i$ and $P_j$, there is an edge $(v_i, v_j)\in E$ if $dst_{min}(P_q, R)>\theta$, where $R$ is the MBH of $P_i$ and $P_j$. We have the following theorem: given the graph $G=(V, E)$ constructed as above, for any clique $C$ in $G$, let $R$ be the MBH constructed from the points corresponding to the vertice in $C$. Then $R$ must be a DBH.

\noindent{\bf Proof.} We prove it by induction. 
Let $|C|$ denote the number of vertices in the clique. It is trivial to show that the theorem holds if $|C|<3$. Next, we mainly discuss $|C|\geq 3$.

	\noindent{\bf Base case.} When $|C|=3$, let $v_i$, $v_j$ and $v_k$ denote the vertices in the clique $C$. Let $R_{ij}$, $R_{jk}$, and $R_{ik}$ be the MBHs constructed from the pairs $(v_i, v_j)$, $(v_j, v_k)$, and $(v_i, v_k)$ respectively. 
	Let $R_{ijk}$ be the MBH constructed from $v_i$,  $v_j$, and $v_k$. Apparently $R_{ijk} = R_{ij} \cup R_{jk} \cup R_{ik}$. Given the fact that $dst_{min}(p, R_{ij}) > \theta$, $dst_{min}(p, R_{ik}) > \theta$, and $dst_{min}(p, R_{jk}) > \theta$, it must be true that \\$dst_{min}(p, R_{ijk}) > \theta$. Therefore, $R$ must be a DBH. 
	
	\noindent{\bf Induction step.} 	If we add $v'$ into $C$, we get a new clique $C'$. Let $R_{C'}$ be the MBH constructed from  $C'$. Next, we prove that $R_{C'}$ is always a DBH. We prove this for three cases: (1) $P' \in R_C$, (2) $P'$ falls out of the range of $R_C$ at one dimension, and (3) $P'$ falls out of the range of $R_C$ at both dimensions.  
	
	{\bf Case 1.} $P' \in R_C$. This case is trivial as it is easy to see that $R_{C'}=R_C$. So $R_{C'}$ must be a DBH.
	
	{\bf Case 2.} At exactly one dimension, $P'$ falls out of $R_C$. 
	Then it must be true that either $P'[i]<l_i^C$ or $P'[i]>u_i^C$, for either $i = 1$ or $i = 2$. 
	%We depict this case in Figure \ref{fig:rect} (b). 
	Without loss of generality, we define the four boundary nodes of $R_C$ as $P^1$, $P^2$, $P^3$, and $P^4$ (as shown in Figure \ref{fig:rect} (b)). 
	Also we assume that $P'[1]\in[l_1^C, u_1^C]$ and $P'[2]<l_2^C$. It is easy to see that $R_{C'}=(<l_1^C, u_1^C>, <P'[2], u_2^C>)=R_C\cup R(P', P^2) \cup R(P', P^3)$.  Apparently, $R_C$ is covered by $R_{C'}$. 
	
	Before we prove that $R_{C'}$ is a DBH, we present a lemma. 
	
	\begin{lemma}
	Given two rectangles $R_1(<l_1^1, u_1^1>, \dots, <l_d^1, u_d^1>)$ and $R_2(<l_1^2, u_1^2>, \dots, <l_d^2, u_d^2>)$ in the same Euclidean space, if $R_1$ is covered by $R_2$, i.e. $l_i^2 \leq l_i^1 \leq u_i^1 \leq u_i^2$ for any $i=1, \dots, d$, then $dst_{min}(P, R_1)\geq dst_{min}(P, R_2)$ for any point $P$.
	\label{lm:containment}
	\end{lemma}
	
	Lemma \ref{lm:containment} states that if $R_1$ is covered by $R_2$, for any point $P$, its minimum distance to $R_1$ is no less than the minimum distance to $R_2$.
	
	Next, let's consider $R_{C'}$ that is constructed from adding the point $P'$ to the existing DBH $R_C$. We pick a point $P_i\in R_{C'}$ ($1\leq i \leq t$) s.t. $P_i[1]=l_1^C$ and $P_i[2]\in [l_1^C, u_1^C]$. Apparently, $R(P', P^2)$ is covered by $R(P', P_i)$. Because there is an edge between $v'$ and $v_i$ in the graph $G$, it must be true that $dst_{min}(P_q, R(P', P_i))>\theta$. Following Lemma \ref{lm:containment}, we can infer that $dst_{min}(P_q, R(P', P^2))>\theta$.
	Similarly, there must be a point $P_j$ with $P_j[1]=u_1^C$ and $P_j[2]\in [l_1^C, u_1^C]$.
	Because $R(P', P^3)$ is covered by $R(P', P_j)$ and $dst_{min}(P_q, R(P', P_j))>\theta$, we can prove that $dst_{min}(P_q, R(P', P^3))>\theta$. Thus, we prove that $dst_{min}(P_q, R_{C'})>\theta$ and $R_{C'}$ is a DBH.
	
	%Consider a clique $C$ in $G$. Let $R_C$ be the MBH constructed from the nodes of $C$. Suppose $R_C(<l_1^C, u_1^C>, <l_2^C, u_2^C>)$ is a DBH (i.e.,  $dst_{min}(P_q, R_C) > \theta$) that covers the embedded points $P_1, P_2, \dots, P_t$. The vertices that $P_1, \dots, P_t$ correspond to in the graph $G$ are $v_1, \dots, v_t$. Assume there is a node $v'$ that is  connected with every vertex in $v_1, \dots, v_t$. 	In other words, for every minimum rectangle $R(P', P_i)$ that covers $P'$ and $P_i$, where $i\in \{1, \dots, t\}$, $dst_{min}(P_q, R(P', P_i))>\theta$.

	{\bf Case 3.} On both dimensions, $P'$ falls out of $R_C$. 
	Formally, either $p'[i]<l_i^C$ or $P'[i]>u_i^C$, for both $i = 1, 2$. 
	Without loss of generality, we assume that $P'[0]\in[l_1^C, u_1^C]$ and $P'[2]<l_2^C$. It is easy to see that $R_{C'}=(<P'[1], u_1^C>, <P'[2], u_2^C>)=R_C\cup R(P', P^1) \cup R(P', P^3)$. %, if we include a new node  clique's size from $|C|$ to $|C|+1$, it also holds that MBH of the new clique does not intersect query area.
	There must exists a $P_i$($1\leq i \leq t$) s.t. $P_i[2]=u_2^C$ and $P_i[1]\in [l_1^C, u_1^C]$. In other words, $R(P', P^1)$ is covered by $R(P', P_i)$. As $dst_{min}(P_q, R(P', P_i))>\theta$, it must be true that $dst_{min}(P_q, R(P', P^1))>\theta$.
	Similar to Case 2, we can prove that $dst_{min}(P_q, R(P', P^3))>\theta$ based on $P_j$. Thus, we prove that $dst_{min}(P_q, R_{C'})>\theta$ and $R_{C'}$ is a DBH.
	
%	\noindent{\bf Conclusion:} The distance between $p$ and he MBH defined by cliques in $G$ is always larger than $\theta$. 

\subsection{Special Case of $MDBH$ Problem}
There is a special case where the $MDBH$ problem can be solved in polynomial time. In particular, when the embedded points of all DBH-strings lie on a single line, we can construct a minimal number of DBHs in the complexity of $O(\ell$), where $\ell$ is the number of DBH-strings. 
Let $\cal L$ be the line that the embedded points of DBH-strings lie on. We draw a perpendicular line from $P_q$ to $\cal L$. Let $dst(P_q, \mathcal{L})$ be the  distance between $P_q$ and $\cal L$. 
Depending on the relationship between $\theta$ and $dst(P_q, \mathcal{L})$, there are two cases:

{\em Case 1: $dst(P_q, \mathcal{L})>\theta$.}  
%The distance between $P_q$ and any point on $\cal L$ is larger than $\theta$. Therefore 
We construct the MBH of all the embedded points of DBH-strings. 
\nop{
{\bf WHAT IF $P_q$ lies on $\cal L$?}
{\em In a special case that $P_q$ lies on $\cal L$, $P_q$ splits the $LF$-points into two subsets, each subset containing the points that reside in one side of $P_q$.}
}

{\em Case 2: $dst(P_q, \mathcal{L})\leq \theta$.}
The perpendicular line splits all points of DBH strings into two subsets, $P_L$ and $P_R$, where $P_L$ includes the embedded points that are at one side of $\cal L$, and $P_R$ be the points at the other side. A special case is that $P_q$ lies on $\cal L$. For this case, $P_q$ still splits all points on $\cal L$ into two subsets, $P_L$ and $P_R$. It is possible that $P_L$ or $P_R$ is empty. 
For each non-empty $P_L$ or $P_R$, we construct a corresponding MBH. 

We have the following theorem.
\vspace{-0.05in}
\label{theorem:mbhs}
\begin{theorem}

\noindent{\bf Proof} For Case 1, if $dst(P_q, \mathcal{L}) > \theta$, for any point $P$ located on $\cal L$, it must be true that $dst(P_q, P)\geq dst(P_q, \mathcal{L}) > \theta$. 
For Case 2, if $P_L$ is non-empty, then $dst_{min}(P_q, P_L)\geq min\{dst(P_q, P)|P\in P_L\} > \theta$. So $P_L$ must be a $DBH$. The same reasoning holds for $P_R$. 
\end{theorem} 
\vspace{-0.05in}

\subsection{Experiments: Experimental environment}
We implement both $VS^2$ and E-$VS^2$ approaches in C++. The hash function we use is the {\em SHA256} function from the OpenSSL library. We execute the experiments on a machine with 2.5 GHz CPU and 6 GB RAM, running Mac OS X 10.10. 

\subsection{Experiments: Parameter Settings}

Table \ref{table:psetting} includes the details of the parameter settings of our experiments. 
\begin{table*}[t!]
  \centering
  \begin{tabular}{|c|c|c|}
    \hline
   	Parameter& dataset & setting\\\hline
   	\multirow{2}{*}{Similarity threshold $\theta$} &{\em Lastname}& 2, 2.5, 3. 3.5, 4, 4.5, 5, 5.5, 6\\\cline{2-3}
   	&{\em Femalename}& 2, 3, 4, 5, 6 \\\hline
   	\multirow{2}{*}{Dimension $d$ of embedding space} &{\em Lastname}&5, 10, 15, 20, 25 \\\cline{2-3}
   	&{\em Femalename}&5, 10, 15, 20, 25  \\\hline
   	\multirow{2}{*}{MB-tree fanout $f$} &{\em Lastname}& 10, 15, 20, 25, 30, 35, 40\\\cline{2-3}
   	&{\em Femalename}&5, 10, 15, 20 \\\hline
  \end{tabular}
  \caption{Parameter settings}
  \label{table:psetting}
\end{table*}

\subsection{Experiments: Selectivity of Search Queries}
In Table \ref{tb:selectivity}, we report the selectivity of the threshold values on both datasets, where selectivity is defined as the percentage of similar strings in the dataset. The reported result is the average selectivity of 10 query strings. 
\begin{table}[h]
  \centering
  \begin{tabular}{|c|c|c|c|}
    \hline
   	$\theta$ & {\em Lastname} & {\em Femalename} \\\hline
   	2 & 0.1378 & 0.7277 \\\hline
   	3 & 1.1188 & 3.329 \\\hline
   	4 & 5.214 & 10.61 \\\hline
   	5 & 16.4  & 27.49 \\\hline
   	6 & 39.015 & 55.80 \\\hline
  \end{tabular}
  \caption{Selectivity (\%) of search queries w.r.t. different threshold values}
  \label{tb:selectivity}
\end{table}

%%%%%%%%%%%%%%%%%%%%%journal version %%%%%%%%%%%%%%%%%%%%%%%%
\nop{
\subsection{Experiments: Query Selectivity Analysis}

\begin{table*}[h]
\small
\centering
\begin{tabular}{c c}
    \begin{tabular}{|c|c|c|c|c|c|}
    \hline
     $\theta$ & $n_R$ & $n_{NC}$ & $n_C$ & $n_{DS}$ & $n_F$  \\\hline
     2 & 136 & 7791 & 80872 & 13928 & 50293 \\\hline
     3 & 1104 & [0, 1479] & [86216, 87695] & [2426, 3587] & [82688, 84109] \\\hline
     4 & 5145 & 13 & 83641 & 830 & 82810 \\\hline
     5 & 16181 & 0 & 72618 & 162 & 72456 \\\hline
     6 & 38494 & 0 & 50305 & 12 & 50293  \\\hline
    \end{tabular}
 &  
    \begin{tabular}{|c|c|c|c|c|c|}
    \hline
     $\theta$ & $n_R$ & $n_{NC}$ & $n_C$ & $n_{DS}$ & $n_F$  \\\hline
     2 & 31 & [11, 302] & [3942, 4233] & [1282, 1429] & [2660, 2815] \\\hline
     3 & 142 & 5 & 4128 & 359 & 3769 \\\hline
     4 & 453 & 0 & 3822 & 45 & 3777 \\\hline
     5 & 1175 & 0 & 3100 & 4 & 3096 \\\hline
     6 & 2385 & 0 & 1890 & 0 & 1890  \\\hline
    \end{tabular}
 \\
    (a) {\em Lastname} dataset
 & 
    (b) {\em Femalename} dataset
\end{tabular}
\caption{Query Selectivity}
\label{tab:query_selectivity}
\end{table*}
}

\bibliographystyle{abbrv}
{
\bibliography{bib}  
}
\end{document}